%% file: TDI_JMLR_Main.tex
\documentclass[11pt]{article}
\usepackage[margin=1in]{geometry}
\usepackage{amsmath}
\usepackage{graphicx}
\usepackage{enumerate}
\usepackage{enumitem}
\usepackage{natbib}
\usepackage{url} % not crucial - just used below for the URL 

\usepackage{algorithmic} %For computer algorithm code in LATEX
\usepackage{algorithm} %For algorithm box around the algorithmic
\usepackage{setspace} %For setting spacing (e.g., double spacing, use \doublespacing)

\usepackage{rotating} %For rotating tables (provides sidewaysfigure and sidewaystable)

\usepackage{amsfonts, amssymb} %Either provides Real number and other font'related stuff.
\usepackage{amsthm}

\usepackage{epstopdf} %for adding eps in pdfLatex	
\usepackage{hyperref} %For adding hyperlinks in documents
\usepackage{bm} %Bolds everything when uses \bm as a command
\usepackage{bbm}
\usepackage{multirow} %Have multiple rows in a table
\usepackage{latexsym} %Latex-based math symbols. Generally load this in conjunction with amssymb
\usepackage{rotating} %arbitrary rotate any object
\usepackage{titlesec} %centering section headings
\usepackage{caption} %captioning for tables
\usepackage{mathtools} %ceiling/floor commands
\usepackage{color} %Provides color for text.
\usepackage{diagbox}
\usepackage{pict2e}
\usepackage{etoolbox}
\usepackage[english]{babel}
\usepackage{empheq}

\usepackage{cases}
\usepackage{lato} 

\hypersetup{
colorlinks=true,
linkcolor=blue,
filecolor=blue,
urlcolor=blue,
citecolor=blue
}

\graphicspath{ {plot/} }% plot path

\newtheorem{theorem}{Theorem}[section]
\newtheorem{lemma}[theorem]{Lemma}

\theoremstyle{definition}
\newtheorem{assumption}{Assumption}

\newtheorem{REMARK}{Remark}

\newtheorem{THEOREM}{Theorem}
\graphicspath{ {plot/} }% plot path

\definecolor{mycolor}{RGB}{0,200,200}

\input{Commands.tex}

\definecolor{red1}{RGB}{255,64,64}
\definecolor{blue1}{RGB}{128,255,255}
\definecolor{green1}{RGB}{0,205,0}

\begin{document}

\setlength{\abovedisplayskip}{6pt}
\setlength{\belowdisplayskip}{6pt}
\setlength{\abovedisplayshortskip}{3pt}
\setlength{\belowdisplayshortskip}{3pt}

\title{\vspace*{-1.5cm} Personalized Two-sided Dose Interval}

\author{
Chan Park$^{a*}$, Guanhua Chen$^{b*}$, Menggang Yu$^{c}$\\
\makebox[1cm][c]{{\footnotesize $^{a}$: Department of Statistics, University of Illinois Urbana-Champaign, Champaign}}\\[-0.1cm]
\makebox[1cm][c]{{\footnotesize $^{b}$: Department of Biostatistics and Medical Informatics, University of Wisconsin--Madison}}\\[-0.1cm]
\makebox[1cm][c]{{\footnotesize $^{c}$: Department of Biostatistics, University of Michigan}}\\[-0.1cm]
\makebox[1cm][c]{{\footnotesize $^{*}$: Corresponding Authors}}
}
 \date{}

\maketitle

\begin{abstract}
In fields such as medicine and social sciences, the goal of treatment is often to maintain the outcome of interest within a desirable range rather than to optimize its value. To achieve this, it may be more practical to recommend a treatment dose interval rather than a single fixed level for a study unit. Since individuals may respond differently to the same treatment level, the recommended dose interval should be personalized based on their unique characteristics. Iterative procedures have been proposed to jointly learn the lower and upper bounds of personalized dose intervals, but they lack theoretical justification. To fill this gap, we propose a method to learn personalized two-sided dose intervals based on empirical risk minimization using a novel loss function. The proposed loss function is designed to be well-defined over a tensor product function space, eliminating the need for iterative procedures. In addition, the loss function is doubly-robust to the misspecification of nuisance functions. We establish statistical properties of the estimated dose interval in terms of excess risk by leveraging the reproducing kernel Hilbert space theory. Our simulation study and real-world applications in warfarin dosing and the Job Corps program show that our proposed direct estimation method outperforms competing methods, including indirect regression-based methods. \\[0.4cm]
\end{abstract}

\newpage

\section{Introduction}

Personalized treatment rules, also known as individualized treatment rules or treatment policies, are strategies for assigning treatments, such as medication or social policy, to individuals based on their characteristics. These strategies have gained attention in various fields, such as medicine, education, and social science, due to their superior performance compared to traditional ``one-size-fits-all'' treatment assignment approaches. For example, many studies have shown that personalized warfarin dosing strategies, which take into account patient characteristics, including pharmacogenetics, are more effective than standard dosing based on an empirical protocol \citep{Warfarin_PM_2007, Warfarin_PM_2014}. Most research on personalized treatment rules, including the warfarin example, aims to optimize the individual’s outcome by producing a single recommended treatment level for each study unit under the rule \citep{laber2015tree, Chen2016, kallus2018policy, zhu2020kernel, zhou2021parsimonious, Schulz2021, hua2022personalized, wang2023non}. However, single-level dose recommendations may be overly ambitious or inadequate, as the primary goal of treatment in many real-world applications is not to achieve a specific value, but rather to ensure that a study unit's outcome falls within a favorable range. For instance, warfarin should be prescribed to keep a patient's international normalized ratio, a measure of the time for the blood to clot, within a desired range, usually between 2 and 3 \citep{Warfarin_AHA_2014}. Similarly, major medical associations recommend targeting appropriate ranges for chronic disease management metrics such as hemoglobin level and blood pressure \citep{TDI_Example_2019, TDI_Example_2020}. 

In practice, the relationship between the amount of treatment and the likelihood of favorable outcomes, referred to as the dose-probability curve, can be biphasic, resulting in an inverted U-shaped curve across dose levels; this phenomenon is called hormesis \citep{Hormesis2008}. These hormetic dosage responses are common in many real-world applications, including the above warfarin example \citep{Blann2003}, including clinical trials \citep{Hormesis2008_2}, public health \citep{Hormesis_PublicHealth2006}, biology, toxicology, and medicine \citep{Hormesis2017}. Under hormesis, to have a probability of favorable outcomes greater than a certain level, the treatment dose often falls within a two-sided interval.

As such, interval-based dosing, rooted in the concept of hormesis, is widely utilized in clinical practice. In many cases, it is not only preferred but, quite often, the only feasible option. This is especially evident in the following two key clinical settings.  The first clinical setting involves subgroup-specific dose interval recommendations, which are commonly established from a policy or guideline-making perspective. For example, \citet{Ageno2012_WarfarinSupport2} and \citet{Witt2016_WarfarinSupport1} recommend an initial warfarin dose ranging from 5 to 10 mg for most patients, while a dose of less than 5 mg is often recommended for elderly patients or those with poor nutrition, liver disease, congestive heart failure, or a high risk of bleeding. In prostate cancer radiotherapy, \citet{ProstateCancer2020} recommended that the definitive treatment doses for prostate cancer can increase to 76–80 gray (Gy) when applying 3-dimensional conformal radiotherapy or intensity-modulated radiotherapy with conventional fractionated radiotherapy. Similar types of interval recommendations were given for radiotherapy of thyroid cancer and lung cancer \citep{ThyroidCancer2008, LungCancer2021, ThyroidCancer2023}. In many scenarios of this setting, the absorbed or intaken dose is more important than the administered dose. The second clinical setting is chronic disease management, where blood sugar level and blood pressure level control are of great importance for elderly subjects. Since these metrics are influenced by various factors, such as lifestyle and dietary habits, a single-level dose recommendation is not feasible even with pharmaceutical interventions. For example, the actual blood sugar level for a given dose of a glucose-lowering drug can vary depending on the timing of administration, the patient’s activity level that day, and dietary intake. As a result, the American Diabetes Association (ADA) and American Heart Association only recommend intervals for these two metrics. For example, the ADA recommends a fasting glucose level of 80-130 mg/dL for healthy elderly and 90-150mg/dL for elderly with notable co-morbidities \citep{ADA2023}. A similar guideline exists for blood pressure management \citep{AHA2017}.  Moreover, identifying and estimating the optimal range of variables associated with preferred outcomes is of significant interest in fields beyond medicine, including education, nutrition, neuroscience, management, and social sciences \citep{TDISupport2014_Economics, TDISupport2015_Nutrition, TDISupport2019_Neuroscience, TDISupport2023_Transportation, TDISupport2023_Education}. These examples illustrate the practical application of interval-based decision rules in various settings.
 
Not surprisingly, determining the optimal personalized therapeutic dose range is challenging, as it requires learning both the lower and upper bounds of the unit-specific dose interval. However, existing methods designed for optimal single-level dose rules, which rely on the assumption that the dose-probability curve is unimodal \citep{Chen2016}, or for learning one-sided intervals, which rely on the assumption that the dose-probability curve is monotonic \citep{Chen2022,Park2022_OptTrt}, are not suitable for learning personalized therapeutic dose range under the hormetic dose-probability curve. This is because these methods are suitable for only estimating the lower or upper bounds of two-sided intervals by treating the other bound as fixed. Consequently, to use their approaches, an iterative procedure must be employed in which the lower bound function is obtained while keeping the upper bound function fixed, followed by obtaining the upper bound function while keeping the lower bound fixed, repeated until convergence; see Remark \ref{remark-1} for details. Thus, implementing these approaches can be computationally challenging, and even if successful, they may suffer from convergence issues. In addition, due to the monotonicity constraint on the lower and upper bound estimators imposed in each iteration, the theoretical properties of the resulting estimators are challenging to characterize. In particular, theoretical tools for studying the approximation error of a general pair of functions using functions within unrestricted function spaces, such as product reproducing kernel Hilbert spaces, are no longer applicable to a function space with a restricted domain. 

This paper proposes a new approach to estimate personalized therapeutic dose ranges under the hormetic dose-probability curve. In brief, we solve an empirical risk minimization task using a novel loss function that can accommodate both monotonic and non-monotonic pairs; here, a monotonic pair refers to a pair of which a lower bound argument is less than or equal to the upper bound argument for all covariate values, while a non-monotonic pair refers to a pair which violates the monotonic relationship for some covariate values. The proposed function enables simultaneous estimation of the lower and upper bounds of the dose interval without requiring a strictly monotonic relationship between the bounds. The loss function assigns higher values to non-monotonic intervals than to monotonic intervals, ensuring that the estimated dose interval is monotonic when evaluated in the training data. The loss function also has a double robustness property \citep{Scharfstein1999, Lunceford2004, Bang2005}, making the estimated dose interval robust to the misspecification of the dose-probability curve and the propensity score of treatment. Our simulation study and the warfarin dosing application demonstrate that our methodology is superior to an indirect method discussed in Section \ref{sec:indirect}. 

Although methods exist for designing personalized policies that involve interval-based quantities, our work offers a concrete and distinct contribution. For instance, the method proposed by \citet{Cai2023} can also generate two-sided dose interval recommendations, but its goal is not to identify the personalized therapeutic dose range. In simple terms, their target estimand is defined as follows. First, the dose range is divided into user-specified sub-intervals, which are often infinitesimally small. The optimal dose range is then identified as the sub-interval that maximizes the dose-probability curve when treatment is assigned within it. Their approach is motivated by detecting multiple change points in the dose-probability curve to identify a dose range that maximizes the outcome, which contrasts with our motivation and goal---identifying a treatment range that leads to a preferred outcome.  Additionally, a series of works on Interval Estimation methods have emerged in the context of multi-stage contextual bandit and reinforcement learning problems \citep{Lai1987, Kaebling1993, Strehl2005, StrehlLittman2008, Karamptziakis2020, Hong2022}. While these methods differ in procedural details, the central idea of Interval Estimation is to identify the policy that maximizes the upper bound of the outcome's uncertainty interval. In these approaches, the term interval refers to the uncertainty in the outcome estimate, rather than to the policy itself. In contrast, our framework defines intervals directly in terms of policies rather than outcome estimates; see Section \ref{sec:Preliminary} for a formal definition of our interval-based policy. Therefore, unlike the Interval Estimation approach, our framework focuses on policies that are explicitly defined as intervals.

\section{Preliminary} \label{sec:Preliminary}

% \subsection{Notations}

Let the subscript $i$ denote the $i$th study unit. For each unit $i =1,\ldots,N$, we observe $\bO_i = (Y_i, A_i, \bX_i) \in \mathcal{O}$ which are independent and identically distributed realizations from a distribution $P_{\bO}$. Here, $\bX_i$ is a $d$-dimensional pretreatment covariate with support $\mathcal{X} \subseteq \mathbbm{R}^d$, $A_i$ is the treatment dose taking its value in the interval $[0,1]$, and $Y_i \in \R$ is the observed outcome/reward. For simplicity, we assume that the treatment is transformed so that its range is the unit interval. To define the dose assignment rule, we use the potential outcome \citep{Rubin1974}. Specifically, let $\pot{Y_i}{a}$ denote the potential outcome/reward when the treatment dose is $A_i = a$. In what follows, we suppress the subscript $i$ unless necessary. 

Let $\F$ be a decision space, a set of dose assignment rules, specified by an investigator, where $f \in \F$ is a dose assignment rule that maps $\bX$, the characteristics of a study unit, to $f(\bX) \in [0, 1]$, a dose level. In the context of learning two-sided dose intervals, we define $\Fprod = \big\{ (f_L,f_U) \cond f_L, f_U \in \F \big\} $, a set of pairs of dose levels, as a decision space. Let $\Fmnt = \{ (f_L,f_U) \in \Fprod \mid f_L(\bx) \leq f_U(\bx) \text{ for all } \bx \}$ be the collection of pairs of functions such that the lower bound is smaller than or equal to the upper bound for all $\bx$. Let $\Fnmnt = \{ (f_L,f_U) \in \Fprod \mid f_L(\bx) > f_U(\bx) \text{ for some } \bx \}$ be the collection of pairs of functions such that the lower bound is larger than the upper bound for some $\bx$.

The interval rules can be more useful than the single-level dose assignment rules when one or more of the following scenarios happen: (i) assigning any value in an interval yields almost the same outcome as assigning the optimal treatment rule; (ii) finding the optimal treatment rule is inefficient; (iii) an investigator has subject matter knowledge about the shape of the dose-probability curve. In particular, we consider a $(\alpha,\thr)$-probability dose interval (PDI) introduced in \cite{Li2018} and \cite{Chen2022}: 
\begin{align}					\label{eq-PDIcriterion}
\forall a \in [ f_L(\bX \con \alpha, \thr), f_U(\bX \con \alpha, \thr) ]
\in \Fmnt
\ \Rightarrow \ \Pr \big\{ \pot{Y}{a} \in \thr \cond \bX \big\} 
\geq \alpha
\ .
\end{align}
Hereafter, we omit $(\alpha, \thr)$ in $f_L$ and $f_U$ for notational brevity. 
In words, if a study unit with characteristic $\bX$ is treated with a dose level that belongs to  $[f_L(\bX), f_U(\bX)]$, the study unit's outcome belongs to the desired range $\mathcal{T} \subseteq \mathbbm{R}$ with a probability greater than or equal to $\alpha$. Here, $\alpha$ plays an analogous role as a confidence level in hypothesis testing because the corresponding PDI admits type I error with probability at most $1-\alpha$. As a result, the selection of $\alpha$ depends on the desired confidence level for a specific scientific question in view. In order to ensure meaningful PDIs, it is important not to set $\alpha$ too high, such as $\alpha=0.99$, as this may result in no PDI satisfying the condition, especially when $\Pr \big\{ \pot{Y}{a} \in \thr \cond \bX \big\}$ is small. The selection of $\alpha$ depends on the application and should ideally be guided by domain expertise. If such expertise is not available, we suggest choosing an $\alpha > 0.5$ or using estimated outcome regression to help determine $\alpha$. We remark that $\alpha$ and $\thr$ may depend on $\bX$, i.e., the probability threshold and the desired range can also be personalized. In addition, let $R = \ind(Y \in \thr)$ be the indicator of whether an individual's outcome belongs to the desired range $\thr$, which is referred to as the discretized outcome. The optimal PDI, denoted by $[f_L^*, f_U^*]$, is the PDI that has the longest length and belongs to $\Fmnt$. Formally, for a fixed $(\alpha,\thr)$, $f_L^*$ and $f_U^*$ satisfy $[f_L(\bX), f_U(\bX)] \subseteq [f_L^*(\bX), f_U^*(\bX)]$ for any $\bX$ and any $f_L$ and $f_U$ satisfying \eqref{eq-PDIcriterion}.

Lastly, we introduce additional notation used throughout. For random variables $U$, $V$, and $W$, let $U \indep V \cond W$ denote the conditional independence of $U$ and $V$ given $W$. We use $O_P$ and $o_P$ to denote stochastic boundedness and convergence in probability. For two non-negative sequences $a_N$ and $b_N$, let $a_N \precsim b_N$ denote $a_N \leq C \cdot b_N$ for some constant $C>0$, and let $a_N \asymp b_N$ denote $a_N \precsim b_N$ and $b_N \precsim a_N$. Let $\bm{1}_N$ be the length $N$ vector of ones. Lastly, we denote $(x)_+=\max(x,0)$ and $(x)_-=\max(-x,0)$, satisfying the decomposition $x=(x)_+ - (x)_-$.

% \subsection{Assumptions}

Next, in order to establish identification of the PDI, we make the following assumptions.
\begin{assumption}[Stable Unit Value Treatment Assumption; \cite{Rubin1978}] \label{(A1)}
$Y = \pot{Y}{A}$ almost surely;
\end{assumption}
\begin{assumption}[Unconfoundedness] \label{(A2)}
We have $\pot{Y}{a} \indep A \cond \bX$ for any $a \in [0,1]$;
\end{assumption}
\begin{assumption}[Positivity] \label{(A3)}
The generalized propensity score $e^* (A \cond \bX) = P (A \cond \bX)$ \citep{Imbens2000} is bounded below by a constant $c_e>0$, i.e., $e^* (A \cond \bX) >c_e$ for all $(A,\bX) \in [0,1] \otimes \mathcal{X}$. 
\end{assumption}
We refer readers to \cite{Hirano2004} and \cite{HR2020} for substantive discussions on these assumptions. Under these assumptions, the probability of obtaining favorable outcomes given covariates, i.e., $\Pr \big\{ \pot{Y}{a} \in \thr \cond \bX \big\}$, can be identified from the observed data as $\mu^* (a,\bX) = \Pr \big( R=1 \cond A=a, \bX \big)$ which is the dose-probability curve of $R$ for a unit with characteristic $\bX$. % For simplicity, we refer the study unit's outcome is positive if $R = 1$ and negative if $R=0$.

To guarantee the existence of $f_L^*$ and $f_U^*$, we make the following assumptions on $\mu^*$: 
\begin{assumption}[Hormesis] \label{(A4)}
For any $\bX \in \mathcal{X}$, there exists a constant $c_\mu>0$ such that $\mu^* (0,\bX) < \alpha - c_\mu$, $\mu^* (1,\bX) < \alpha - c_\mu$, and $\mu^* (a_{\bX},\bX) > \alpha + c_\mu$ for some $a_{\bX} \in (0,1)$; 
\end{assumption}
\begin{assumption}[Smoothness and Connected PDI] \label{(A5)}
For any $\bX \in \mathcal{X}$, $\mu^*(a,\bX)$ is Lipschitz continuous with respect to $a$ and a level set $\big\{ a \cond \mu^*(a, \bX) \geq \alpha \big\}$ is a closed interval.
\end{assumption}
Assumption \ref{(A4)} states that moderate dose levels are more likely to result in a favorable outcome compared to low and high doses, also known as hormesis \citep{Hormesis2008}. This phenomenon can be found in the inverted U-shaped dose-probability curve; see Section \ref{sec:Data} for details on warfarin and \cite{Hormesis2008_2} for additional examples of drugs that have hormetic dose-probability curves. If Assumption \ref{(A4)} is violated at $\bX=\bx^{\dagger}$, it may imply that (i) study units having their covariates as $\bx^{\dagger}$ always have favorable or unfavorable outcomes regardless of the treatment dose level, i.e., $\mu^*(a,\bx^{\dagger}) > \alpha$ or $\mu^*(a,\bx^{\dagger}) < \alpha$ for all $a$, and/or that (ii) treatment is not hormetic, say $\mu^*(a,\bx^{\dagger})$ increases monotonically with $a$. The first part of Assumption \ref{(A5)} states that $\mu^*$ is smooth with respect to dose, while the second part states that the therapeutic dose range is a connected interval. Assumptions \ref{(A4)} and \ref{(A5)} guarantee the existence of the optimal PDI, but there are other sufficient conditions; see Section \ref{sec:supp:Existence} of the Appendix for details.

\section{Methodology}						\label{sec:method}

\subsection{Indirect Method} \label{sec:indirect}

We present a simple two-step approach to estimate the optimal PDI, often referred to as the indirect method \citep{Moodie2012, Moodie2013}. In the first step, we estimate the dose-probability curve $\widehat{\mu}$ based on a posited outcome model. For instance, one can fit a logistic regression model of $R$ on $(A, A^2, \bX)$. Alternatively, flexible machine learning classifiers can be used, such as random forest \citep{Breiman2001}, gradient boosting \citep{Friedman2001}, neural net \citep{Ripley1994}, support vector machines \citep{SVM}, and many others; see \cite{ESL} for additional examples. In the second step, we find the range $[\hf_L(\bX), \hf_U(\bX)]$ that satisfies the PDI criterion in \eqref{eq-PDIcriterion}, which can be done by a grid search over the unit square $[0,1]^2$. 

Although the indirect method is simple to implement, it can have notable drawbacks in terms of finite-sample performance as demonstrated in the simulation study and application in Sections \ref{sec:Simulation} and \ref{sec:Data}. Additionally, the estimated PDI using indirect methods may not be available in a closed form unless $\widehat{\mu}$ has a very simple form. Furthermore, since the indirect rule only uses the dose-probability curve, it is more sensitive to model misspecification than methods using both the dose-probability curve and the generalized propensity score. This motivates our direct approach presented in the following Sections.

\subsection{Construction of the Loss and Risk Functions} 	\label{sec:lossft}

We propose a method that directly estimates the optimal PDI from empirical risk minimization (ERM). An essential part of the proposed direct method is to design a risk function that is minimized at $(f_L^*, f_U^*)$. If the risk function does not have $(f_L^*, f_U^*)$ as a minimizer, the estimator obtained from the ERM does not converge to $(f_L^*, f_U^*)$ in probability. We first start by reviewing the loss function in \cite{Chen2022} that has an 
inverse probability-weighted (IPW) form defined over $(f_L,f_U)\in \Fmnt$:
\begin{align*}
\mathcal{L}_{\text{IPW}}(\bO,f_L,f_U \con e)
= 
\frac{ \alpha (1-R) \ind \big\{ A \in [ f_L(\bX), f_U(\bX)] \big\}
+
(1-\alpha) R \ind \big\{ A \notin [ f_L(\bX), f_U(\bX)] \big\} }{ e(A \cond \bX) }
\ .
\end{align*}
The loss function is determined by the following four arguments. The first argument, $\bO$, is the observed data. The second and third arguments, $(f_L, f_U)$, form a PDI candidate with the monotonicity condition, i.e., $f_L(\bx) \leq f_U(\bx)$. The last argument is a user-specified propensity score model. The IPW loss function above penalizes a study unit for either of the following cases: (i) the outcome does not belong to the desired range, i.e., $R=0$, even though the dose level belongs to the PDI candidate, or (ii) the outcome belongs to the desired range, i.e., $R=1$, even though the dose level lies outside of the PDI candidate. These errors can be seen as false positives and false negatives, respectively. The coefficients $\alpha$ and $1-\alpha$ assign different weights to these two errors. When the propensity score model is correctly specified, the IPW loss function can be used to identify $(f_L^*,f_U^*)$ as $ \EXP \big\{ \mathcal{L}_{\text{IPW}}(\bO \con f_L,f_U \con e^*) \}$ is minimized at $(f_L^*,f_U^*)$. 

One drawback of $\mathcal{L}_{\text{IPW}}$ is its sensitivity to misspecification of the propensity score model. Therefore, we can consider the following augmented IPW (AIPW) loss function over $(f_L,f_U)\in \Fmnt$ that is robust to the propensity score model misspecification:
\begin{align*}
\mathcal{L}_{\text{AIPW}}(\bO,f_L,f_U \con \mu, e)
& = 
\alpha \frac{ \{\mu(A,\bX)-R\} \ind \{ A \in [ f_L(\bX), f_U(\bX)] \}
}{ e(A \cond \bX) }
\\
& \hspace*{1cm}
+
\alpha \int \{ 1 - \mu(a,\bX) \} \ind \{ a \in [ f_L(\bX), f_U(\bX)] \} \, da 
\\
& \hspace*{1cm}
+ 
(1-\alpha) 
\frac{ \{ R - \mu(A,\bX) \} \ind \{ A \notin [ f_L(\bX), f_U(\bX)] \} }{ e(A \cond \bX) }
\\
& \hspace*{1cm}
+
(1-\alpha) 
\int \mu(a,\bX) \ind \{ a \notin [ f_L(\bX), f_U(\bX)] \} \, da
\ .
% & =
% \left[
% \begin{array}{l}
% \alpha
% \left[
% \frac{ \{\mu(A,\bX)-R\} \ind \{ A \in [ f_L(\bX), f_U(\bX)] \}
% }{ e(A \cond \bX) }
% + \int \{ 1 - \mu(a,\bX) \} \ind \{ a \in [ f_L(\bX), f_U(\bX)] \} \, da 
% \right]
% \\ 
% + 
% (1-\alpha)
% \left[
% \frac{ \{ R - \mu(A,\bX) \} \ind \{ A \notin [ f_L(\bX), f_U(\bX)] \} }{ e(A \cond \bX) }
% + \int \mu(a,\bX) \ind \{ a \notin [ f_L(\bX), f_U(\bX)] \} \, da \right]
% \end{array}
% \right]
% \ .
\end{align*}
Compared to $\mathcal{L}_{\text{IPW}}$, $\mathcal{L}_{\text{AIPW}}$ has one more argument $\mu$ which is a user-specified model for the dose-probability curve. The terms weighted by $\alpha$ and those weighted by $1-\alpha$ can be seen as functions that penalize false positive and false negative errors, respectively. These terms resemble the efficient influence function of the average treatment effect \citep{Hahn1998}, and, not surprisingly, it has the doubly-robust property in the following manner: the AIPW loss function is minimized at the optimal PDI so long as $\mu$ or $e$, but not necessarily both, is correctly specified; see Lemma \ref{lemma-minimizer} for a related discussion.

The AIPW loss function $\mathcal{L}_{\text{AIPW}}$ has a limitation in that it is defined solely over monotonic intervals $\Fmnt$, i.e., PDI candidates in $\Fnmnt$ are excluded from the loss function. Consequently, if $\mathcal{L}_{\text{AIPW}}$ is used, the optimization domain, denoted by $\mathcal{H}_{\text{mnt}}$, must adhere to the monotonicity requirement between the lower and upper bound candidates, which implies that $\mathcal{H}_{\text{mnt}}$ should be a subset of $\Fmnt$. Moreover, $\mathcal{H}_{\text{mnt}}$ should be rich enough to accurately approximate any function in $\Fmnt$. These two observations indicate that an ideal choice for $\mathcal{H}_{\text{mnt}}$ would be a rich subset of $\Fmnt$. However, achieving both of these goals simultaneously is a significant challenge for practitioners for the following reasons. First, to maintain the monotonicity between the lower and upper bounds, iterative procedures are typically utilized in practice. By doing so, $\mathcal{H}_{\text{mnt}}$ may be restricted in a specific way, casting doubt on its suitability as an approximation subspace of $\Fmnt$. On the other hand, if one naively chooses rich product function spaces, such as a product Reproducing Kernel Hilbert Space (RKHS) as $\mathcal{H}_{\text{mnt}}$, it would cause problems in the optimization procedure because the AIPW loss function is not compatible with non-monotonic intervals. 

The limitation of the AIPW loss function originates from its constrained domain. One way to resolve the limitation is to design a loss function that is well-defined over $\Fprod$ and is minimized at the optimal PDI $(f_L^*,f_U^*)$. We can then estimate $(f_L^*,f_U^*)$ over a product function space that approximates $\Fprod$ without the need to restrict the optimization domain. To this end, we propose a new doubly-robust loss function $\loss: \mathcal{O} \otimes \F^{\otimes 2} \rightarrow \R$, which is defined below:
\begin{align}
&
\loss ( \bO, f_L, f_U \con \mu, e)
=
\left\{
\begin{array}{ll}
\loss^{(1)} ( \bO, f_L, f_U \con \mu, e) & \text{if }f_L(\bX) \leq f_U(\bX) \\
C_\loss
& \text{if } f_L(\bX) > f_U(\bX)
\end{array}
\right.
\ ,
\label{eq-Loss}
\\
&
\loss^{(1)} ( \bO, f_L, f_U \con \mu, e)
=
\left[
\begin{array}{l}
\big\{ \mu(A, \bX) - R \big\} \ind \{ A \in [ f_L(\bX), f_U(\bX)] \} / e(A \cond \bX) 
\\
+
\int \big\{ \alpha - \mu(a, \bX) \big\} \ind \{ a \in [ f_L(\bX), f_U(\bX)] \} \, da 
\end{array}
\right]
\ ,
\nonumber
\end{align}
where $C_{\loss}$ is a sufficiently large constant such as $C_\loss=2 \sup_{\bO, f_L, f_U} \big| \loss^{(1)}(\bO, f_L, f_U \con \mu, e) \big|$. Clearly, $\loss$ allows non-monotonic pairs, but the corresponding loss value $C_\loss$ is higher than that of monotonic pairs. Additionally, from simple algebra, we find that the difference between $\mathcal{L}_{\text{AIPW}}$ and $\mathcal{L}^{(1)}$ does not depend on $(f_L,f_U)$, indicating that the minimizers of $\mathcal{L}_{\text{AIPW}}$ and $\mathcal{L}^{(1)}$ are the same; see Section \ref{sec:supp:TwoLoss} of the Appendix for details. In most observational studies, the true nuisance components $(\mu^*,e^*)$ are unknown and must be estimated from the observed data; see Section \ref{sec:ERM} for details.

The loss function $\loss$ has advantageous properties, which are summarized in the following lemma.
\begin{lemma}				\label{lemma-minimizer}
Suppose that Assumptions \ref{(A1)}-\ref{(A5)} hold. Let $\risk: \Fprod \rightarrow \R$ be the risk function associated with $\loss$, i.e., $ \risk( f_L,f_U \con \mu, e )
=
\EXP \big\{
\loss (\bO, f_L, f_U \con \mu, e )
\big\}$. Then, the optimal PDI is the minimizer of the risk function with the true nuisance functions, i.e.,
\begin{align} \label{eq-minimizer}
(f_L^*, f_U^*) \in \argmin_{(f_L, f_U) \in \Fprod} \risk( f_L, f_U \con \mu^*, e^* ) \ .
\end{align}
Additionally, $(f_L^*,f_U^*)$ is the minimizer of \eqref{eq-minimizer} so long as either $\mu^*$ or $e^*$ is correctly specified, i.e., \eqref{eq-minimizer} holds for $(\mu',e^*)$ and $(\mu^*,e')$ for any $\mu'$ and $e'$.
\end{lemma}
Lemma \ref{lemma-minimizer} states that the optimal PDI is achieved by minimizing the risk function when both nuisance components are correctly specified. This means that the loss function is Fisher consistent in detecting the optimal PDI. Additionally, if either the dose-probability curve or the propensity score is correctly specified, the loss function remains Fisher consistent. This property is referred to as a doubly-robust property \citep{Scharfstein1999, Lunceford2004, Bang2005} and is further discussed in Theorem \ref{thm-ExcessRisk}. 

% 	{\color{red} Comparison to the previous paper $\Rightarrow$ }

% 	In \cite{Chen2022}, a similar loss function is designed so that the optimal PDI is the minimizer of the corresponding risk function. However, their loss function does not allow the lower end $f_L$ to be larger than the upper end $f_U$, i.e., $(f_L,f_U)$ must belong to the ``lower triangle function space'' $\LT{}_{\F}$, and this requires additional iterative procedures in the optimization step; see Section \ref{sec:ERM} for details. Moreover, their loss function only uses the propensity score having a inverse probability-weighted (IPW) form, and not fully exploit the dose-probability curve. As a consequence, their methodology is not DR and requires the propensity score to be correctly specified, making it less robust to the model misspecification. Consequently, the excess risk bound using our DR loss function is tighter than that using the IPW loss function; see Theorem \ref{thm-ExcessRisk} for additional discussion.

%	We defer discussion on how to estimate the nuisance functions until Section \ref{sec:practical}.

\subsection{Empirical Risk Minimization Via the Difference of Convex Functions Algorithm} \label{sec:ERM}

\begin{figure}[!htb]
\centering
\includegraphics[width=1\textwidth]{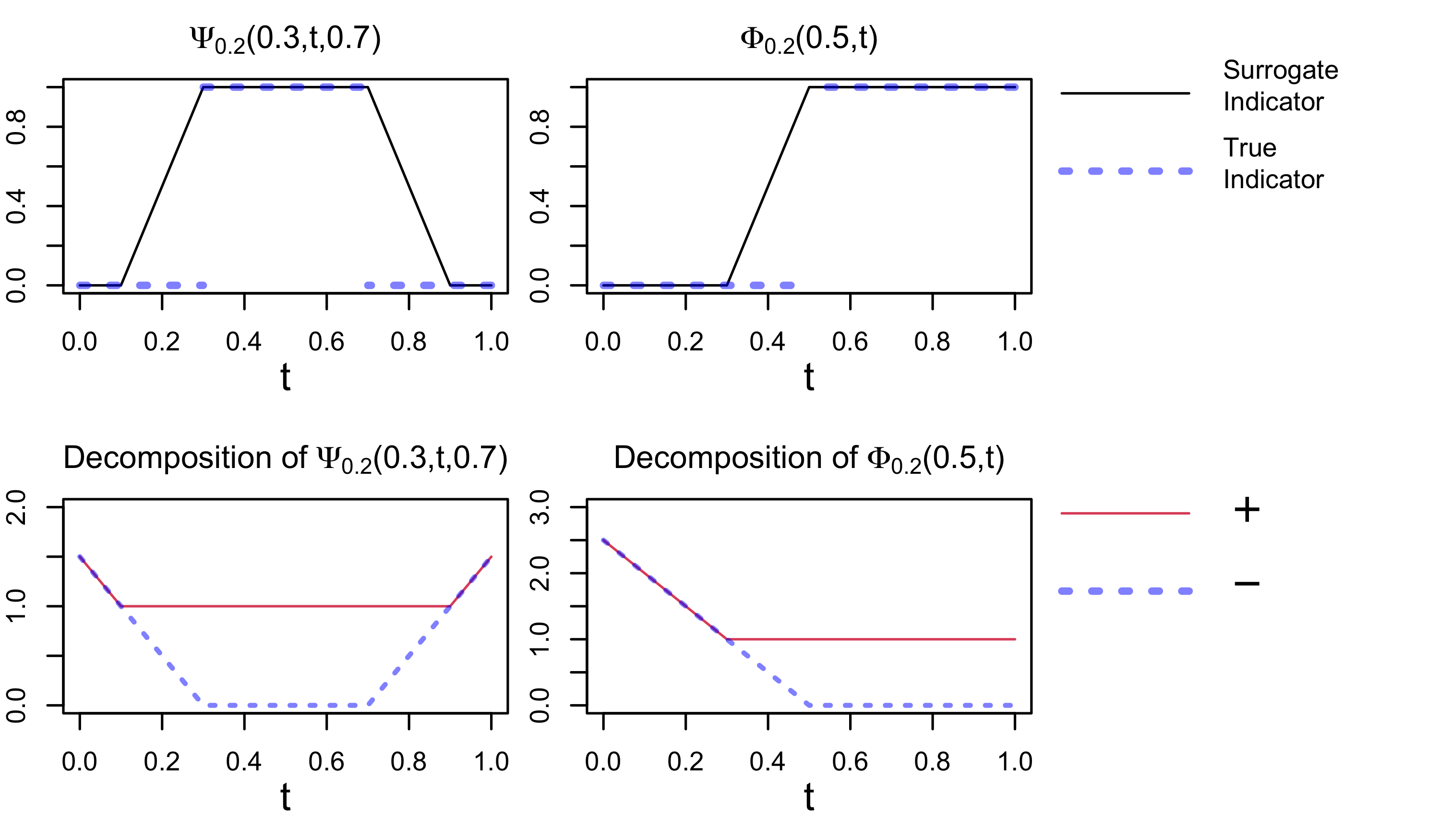}
\caption{Graphical illustrations of $\Psi_\epsilon$ and $\Phi_\epsilon$ with $\epsilon=0.2$, and their convex decomposition.}
\label{Fig-surrogate}
\end{figure}

Despite its theoretical validity, the loss function $\loss$ in \eqref{eq-Loss} itself is difficult to use in ERM because it is not convex or smooth as induced by the indicator functions $\ind \{f_L(\bX) \leq f_U(\bX) \}$ and $\ind\{ A \in [f_L(\bX), f_U(\bX)] \}$. To address the non-smoothness, we design a surrogate loss function $\loss_\sur$ of the loss function $\loss$ by replacing the indicator functions with truncated hinge-type functions as follows:
\begin{align*}
&
\loss_\sur ( \bO, f_L, f_U \con \mu, e )
=
\left\{
\begin{array}{ll}
\loss_\sur^{(1)} ( \bO, f_L, f_U \con \mu, e )
&
\text{ if } 
f_L(\bX) \leq f_U(\bX)
\\
\loss_\sur^{(2)} ( \bO, f_L, f_U \con \mu, e )
&
\text{ if } 
 f_U(\bX) < f_L(\bX) < f_U(\bX) + \epsilon 
\\
C_\loss
&
\text{ if } 
f_U(\bX) + \epsilon \leq f_L(\bX)
\end{array}
\right. \ ,
\\
&
\loss_\sur^{(1)} ( \bO, f_L, f_U \con \mu, e)
=
\left[
\begin{array}{l}
\big\{ \mu(A, \bX) - R \big\} 
\Psi_\epsilon \big( f_L(\bX) , A, f_U(\bX) \big) / e(A \cond \bX)
\\
+
\int \big\{ \alpha - \mu(a, \bX) \big\} \ind \{ a \in [ f_L(\bX), f_U(\bX)] \} 
\, da 
\end{array}
\right] \ ,
\\
&
\loss_\sur^{(2)} ( \bO, f_L, f_U \con \mu, e)
=
\Phi_\epsilon( f_L(\bX), f_U(\bX) ) 
\bigg[
0.5
\bigg\{
\begin{array}{l}
\loss_\sur^{(1)} ( \bO, f_L, f_L \con \mu, e) 
\\
+ \loss_\sur^{(1)} ( \bO, f_U, f_U \con \mu, e) 
\end{array}
\bigg\}
- C_\loss
\bigg] + C_\loss \ ,
\end{align*}	
where $\Psi_\epsilon$ and $\Phi_\epsilon$ are truncated hinge-type functions:
\begin{align*}
&
\Psi_\epsilon (\ell ,t, u)
=
\left\{
\begin{array}{ll}
\frac{t-\ell+\epsilon}{\epsilon}
& \text{if } t \in [\ell-\epsilon, \ell]
\\
1 & \text{if } t \in [\ell , u]
\\
\frac{u+\epsilon-t}{\epsilon} & \text{if } 
t \in [u, u+\epsilon] 
\\
0 & \text{otherwise}
\end{array}
\right.
\ , 
&&
\Phi_\epsilon(\ell, u) 
=
\left\{
\begin{array}{ll}
0 & \text{if } u - \ell \in (-\infty,-\epsilon]
\\
\frac{u-\ell + \epsilon}{\epsilon} & \text{if } u - \ell \in [-\epsilon,0]
\\
1 & \text{if } u - \ell \in [0,\infty)
\end{array} 
\right.
\ .
\end{align*}
Here, $\epsilon>0$ is a bandwidth parameter where $\Psi_\epsilon$ and $\Phi_\epsilon$ converge to the indicator functions as $\epsilon$ goes to zero; an appropriate choice of $\epsilon$ is discussed in Section \ref{sec:theory}. As depicted in the Figure \ref{Fig-surrogate}, $\Psi_{\epsilon}$ and $\Phi_{\epsilon}$ are continuous functions that approximate $\ind \{f_L(\bX) \leq f_U(\bX) \}$ and $\ind\{ A \in [f_L(\bX), f_U(\bX)] \}$, respectively. These continuous approximations enable the use of computationally efficient algorithms, such as gradient-based iterative methods, one of which we implement later in this section. Additionally, the truncated hinge-type functions used in $\loss_\sur$ ensure that the loss function is bounded, making it robust to noisy training data \citep{Wu2007, Chen2016}.  

Using $\loss_\sur$, we obtain a PDI estimator from the ERM with the estimated nuisance components, i.e.,
\begin{align*}
    (\widehat{f}_L,\widehat{f}_U)
    =
    \argmin_{(f_L,f_U) \in \HH^{\otimes 2}}
    \bigg\{ 
    \frac{1}{N}
    \sum_{i=1}^{N}
    \loss_{\sur}
\big( \bO_i , f_L , f_U \con \widehat{\mu} , \widehat{e} \big) 
+
\lambda
\big\| f_L \big\|_{\HH}^2
+
\lambda
\big\| f_U \big\|_{\HH}^2
\bigg\} \ .
\numeq \label{eq-ERM-surloss}
\end{align*}
Here, $\HH$ is a user-specified function space, such as linear functions, splines, RKHS, or deep neural networks, and $\lambda>0$ is a regularization parameter. While other choices are possible in principle, we focus on the case where $\HH$ is an RKHS for the remainder of the paper. Specifically, let $\mathcal{H}$ be an RKHS with the Gaussian kernel function $k(\bx,\bx') = \exp \{ - \| \bx - \bx' \|_2^2 / \gamma^2 \}$ where $\gamma > 0$ is a bandwidth parameter. We remark that other universal kernels can be used with minor modifications. The RKHS is widely recognized for its approximation properties. For example, the Gaussian RKHS is dense in $L_2(P)$, the space of square-integrable functions, whenever $P$ is a finite measure \citep[Theorem 4.63]{SVM}. Moreover, the RKHS is computationally efficient for our problem, as its parameters can be obtained efficiently via an iterative convex optimization procedure. Further details are provided in the remainder of this section.

Under this specification, the ERM \eqref{eq-ERM-surloss} is defined over the tensor product RKHS $\HH^{\otimes 2}$. From the representer theorem \citep{KW1970,SHS2001}, the solution to the ERM, denoted by $(\hf_L,\hf_U)$, is represented as linear combinations of the kernel-transformed covariates, i.e., 
\begin{align*}
    &
    \hf_L (\bx) = \widehat{\xi}_{L,0} + \sum_{i=1}^{N} \widehat{\xi}_{L,i} k(\bx, \bX_i)
    \ , 
    &&
    \hf_U (\bx) = \widehat{\xi}_{U,0} + \sum_{i=1}^{N} \widehat{\xi}_{U,i} k(\bx, \bX_i) \ .
\end{align*}
The coefficients $\widehat{\bxi}_{L} = (\widehat{\xi}_{L,0}, \widehat{\xi}_{L,1},\ldots,\widehat{\xi}_{L,N})\T$ and $\widehat{\bxi}_{U} = (\widehat{\xi}_{U,0}, \widehat{\xi}_{U,1},\ldots,\widehat{\xi}_{U,N})\T$ are obtained from an ERM over an Euclidean space below:
\begin{align*}	
&
\big( \widehat{\bxi}_{L} , \widehat{\bxi}_{U} \big)
=
\argmin_{ \bxi_{L}, \bxi_{U} } 
Q ( \bxi_{L} , \bxi_{U} ) \ ,
\end{align*}
where
\begin{align}						\label{eq-otherERM}
Q ( \bxi_{L} , \bxi_{U} )
& =
\frac{1}{N} \sum_{i=1}^{N}
\loss_{\sur}
\big( \bO_i , \xi_{L,0} + \bm{k}_i \T {\bxi}_{L,(-0)} , \xi_{U,0} + \bm{k}_i \T {\bxi}_{U,(-0)} \con \widehat{\mu} , \widehat{e} \big) 
\nonumber
\\
& \hspace*{1cm}
+
\lambda \big\{
{\bxi}_{L,(-0)}\T K {\bxi}_{L,(-0)}
+
{\bxi}_{U,(-0)}\T K {\bxi}_{U,(-0)}
\big\} \ .
\end{align}
Here, $K = [ k(\bX_i, \bX_j) ]_{i,j} \in \R^{N \times N}$ is the gram matrix, $\bm{k}_i \in \R^{N}$ is the $i$th column of $K$, $\bxi_{L,(-0)} = ({\xi}_{L,1},\ldots,{\xi}_{L,N})\T$, and $\bxi_{U,(-0)} = ({\xi}_{U,1},\ldots,{\xi}_{U,N})\T$. 

    Although \eqref{eq-otherERM} provides a characterization of the PDI estimator, solving it remains challenging due to the nonconvexity of the surrogate loss function. While other nonconvex optimization methods can be applied, we use the Difference of Convex Functions algorithm (DC algorithm; \citealp{DCalgorithm}) due to its ready adaptation to our problem. Briefly, the DC algorithm is implemented as follows. First, we decompose the surrogate loss function $\mathcal{L}_{\text{sur}}$ into a difference between two convex functions, say $\mathcal{L}_{\text{sur}} = \mathcal{L}_{\text{sur},+} - \mathcal{L}_{\text{sur},-}$. We then iteratively solve a sequence of convex optimization problems using $\mathcal{L}_{\text{sur},+}$, $\mathcal{L}_{\text{sur},-}$, and specified initial values. Section 5.4 of \citet{DC2018}, a recent review of the DC algorithm, emphasizes that the stability of the algorithm depends on \HL{(DC-1)}: selecting an appropriate DC decomposition and \HL{(DC-2)}: implementing a strategy to calculate good initial values. Although there is no unified approach to addressing these two factors, we outline how they are handled in our problem below. The proposed specifications demonstrate reasonable performance, as shown in the simulation study in Section \ref{sec:Simulation}.

\begin{itemize}[leftmargin=0cm]
    \item[] \HT{(DC-1)}: \textit{Decomposition of the Surrogate Loss Function}
\end{itemize} 
 
First, we consider the convex decompositions of $\Psi_\epsilon$ and $\Phi_\epsilon$:
\begin{align*}
&
\Psi_{\epsilon, +}(\ell, t, u) = 
\left\{
\begin{array}{ll}
\frac{-t + \ell}{\epsilon} & \text{if } t \leq \ell - \epsilon \\
1 & \text{if } \ell - \epsilon < t < u + \epsilon \\
\frac{t-u}{\epsilon} & \text{if } u + \epsilon \leq t \\
0 & \text{otherwise}
\end{array}
\right.
\quad , \quad
&&
\Psi_{\epsilon, -}(\ell, t, u) = 
\left\{
\begin{array}{ll}
\frac{-t + \ell}{\epsilon} & \text{if } t \leq \ell \\
0 & \text{if } \ell < t < u \\
\frac{t-u}{\epsilon} & \text{if } u \leq t \\
0 & \text{otherwise}
\end{array}
\ ,
\right. 
\\
&
\Phi_{\epsilon,+} ( \ell, u )
=
\left\{
\begin{array}{ll}
\frac{\ell - u }{\epsilon} & \text{if } u - \ell < -\epsilon 
\\
1 & \text{if } -\epsilon \leq u - \ell
\end{array}
\right.
\quad , \quad
&&
\Phi_{\epsilon,-}( \ell, u )
=
\left\{
\begin{array}{ll}
\frac{\ell - u }{\epsilon} & \text{if } u - \ell < 0 
\\
0 & \text{if } 0 \leq u - \ell
\end{array}
\right. \ .
\end{align*}
Note that $\Psi_\epsilon=\Psi_{\epsilon,+} - \Psi_{\epsilon,-}$ and $\Phi_\epsilon=\Phi_{\epsilon,+}-\Phi_{\epsilon,-}$; see Figure \ref{Fig-surrogate} for graphical illustrations of $\Psi_\epsilon$ and $\Phi_\epsilon$ and surrogates of these functions.

Let $\mu_+(a, \bX)$ and $\mu_-(a,\bX)$ be non-negative, non-decreasing, Lipschitz continuous functions in $a$ satisfying $\mu(a,\bX) = \mu_+(a,\bX) - \mu_-(a,\bX)$. Since $\mu(a,\bX)$ is Lipschitz continuous, we can easily find such $\mu_+$ and $\mu_-$. For $0 \leq \ell \leq u \leq 1$, we additionally define $G_+ (\ell, u, \bX)$ and $G_-(\ell, u, \bX)$ as follows:
\begin{align*}
&
G_+ ( \ell, u, \bX ) 
=
\int_{0}^{\ell} \mu_+ (a,\bX) \, da
+
\int_{0}^{u} \mu_- (a,\bX) \, da	
+ \alpha u \ , 
\\
& 
G_- ( \ell, u, \bX )
=
\int_{0}^{u} \mu_+ ( a,\bX) \, da
+
\int_{0}^{\ell} \mu_- (a,\bX) \, da 
+ \alpha \ell \ .
\end{align*}
Using these functions, we define $\loss_{\sur,+}$ and $\loss_{\sur,-}$ as follows:
\begin{itemize}

\item If $f_L(X) \leq f_U(X)$:
\begin{align*}
&\loss_{\sur,+} \big( \bO, f_L, f_U \con \mu, e)
\\
&
=
\frac{\Phi_{\epsilon,+} (f_L(\bX), f_U(\bX) ) }{e(A \cond \bX)}
\left[
\begin{array}{l}
\big\{ \mu (A, \bX) - R \big\}_+ \big[ \Psi_{\epsilon,+} \big( f_L(\bX) , A, f_U(\bX) \big) \big]
\\
-
\big\{ \mu (A, \bX) - R \big\}_- \big[ \Psi_{\epsilon,-} \big( f_L(\bX) , A, f_U(\bX) \big) 
\big]
\end{array} \right]
\\
&
\hspace*{1cm}
+
G_+\big( f_L(\bX), f_U(\bX), \bX \big) 
+ C_\loss	\ ,
\end{align*}
and 
\begin{align*}
& 
\loss_{\sur,-} \big( \bO, f_L, f_U \con \mu, e)
\\
&
= 
\frac{\Phi_{\epsilon,+} (f_L(\bX), f_U(\bX) ) }{e(A \cond \bX)}
\left[
\begin{array}{l}
\big\{ \mu (A, \bX) - R \big\}_+ \big[ \Psi_{\epsilon,-} \big( f_L(\bX) , A, f_U(\bX) \big) \big]
\\
-
\big\{ \mu (A, \bX) - R \big\}_- \big[ \Psi_{\epsilon,+} \big( f_L(\bX) , A, f_U(\bX) \big) \big]
\end{array} \right]
\\
&
\hspace*{1cm}
+ G_-\big( f_L(\bX), f_U(\bX), \bX \big) 
+ C_\loss	\ .
\end{align*}

\item If $f_L(X) > f_U(X)$:
\begin{align*}
&
\loss_{\sur,+} \big( \bO, f_L, f_U \con \mu, e)
\\
&
=
\frac{\Phi_{\epsilon,+} (f_L(\bX), f_U(\bX) ) }{ 2 e(A \cond \bX)}
\left[ 
\begin{array}{l}
\big\{ \mu (A, \bX) - R \big\}_+ \big[ \Psi_{\epsilon,+} \big( f_L(\bX) , A, f_L(\bX) \big) \big]	
\\
+
\big\{ \mu (A, \bX) - R \big\}_+ \big[ \Psi_{\epsilon,+} \big( f_U(\bX) , A, f_U(\bX) \big) \big]
\\
- \big\{ \mu (A, \bX) - R \big\}_- \big[ \Psi_{\epsilon,-} \big( f_L(\bX) , A, f_L(\bX) \big) \big]
\\
- \big\{ \mu (A, \bX) - R \big\}_- \big[ \Psi_{\epsilon,-} \big( f_U(\bX) , A, f_U(\bX) \big) \big]
\end{array}
\right]
\\
&
\hspace*{1cm}
+
\frac{\Phi_{\epsilon,-} (f_L(\bX), f_U(\bX) ) }{ 2 e(A \cond \bX)}
\left[ 
\begin{array}{l}
\big\{ \mu (A, \bX) - R \big\}_+ \big[ \Psi_{\epsilon,-} \big( f_L(\bX) , A, f_L(\bX) \big) \big]	
\\
+
\big\{ \mu (A, \bX) - R \big\}_+ \big[ \Psi_{\epsilon,-} \big( f_U(\bX) , A, f_U(\bX) \big) \big]
\\
- \big\{ \mu (A, \bX) - R \big\}_- \big[ \Psi_{\epsilon,+} \big( f_L(\bX) , A, f_L(\bX) \big) \big]
\\
- \big\{ \mu (A, \bX) - R \big\}_- \big[ \Psi_{\epsilon,+} \big( f_U(\bX) , A, f_U(\bX) \big) \big]
\\
+ 2C_\loss e(A \cond \bX)
\end{array}
\right] 
\\
&
\hspace*{1cm}
+ C_\loss + C_{\text{cvx}} \{ f_L(\bX) - f_U(\bX) \}^2 / \epsilon 	\ ,
\end{align*}
and 
\begin{align*}
& 
\loss_{\sur,-} \big( \bO, f_L, f_U \con \mu, e)
\\
&
= 
\frac{\Phi_{\epsilon,+} (f_L(\bX), f_U(\bX) ) }{ 2e (A \cond \bX)}
\left[ 
\begin{array}{l}
\big\{ \mu (A, \bX) - R \big\}_+ \big[ \Psi_{\epsilon,-} \big( f_L(\bX) , A, f_L(\bX) \big) \big]	
\\
+
\big\{ \mu (A, \bX) - R \big\}_+ \big[ \Psi_{\epsilon,-} \big( f_U(\bX) , A, f_U(\bX) \big) \big]
\\
- \big\{ \mu (A, \bX) - R \big\}_- \big[ \Psi_{\epsilon,+} \big( f_L(\bX) , A, f_L(\bX) \big) \big]
\\
- \big\{ \mu (A, \bX) - R \big\}_- \big[ \Psi_{\epsilon,+} \big( f_U(\bX) , A, f_U(\bX) \big) \big]
\\
+ 2C_\loss e(A \cond \bX)
\end{array}
\right] 
\\
&
\hspace*{1cm} 
+
\frac{\Phi_{\epsilon,-} (f_L(\bX), f_U(\bX) ) }{ 2 e(A \cond \bX)}
\left[ 
\begin{array}{l}
\big\{ \mu (A, \bX) - R \big\}_+ \big[ \Psi_{\epsilon,+} \big( f_L(\bX) , A, f_L(\bX) \big) \big]	
\\
+
\big\{ \mu (A, \bX) - R \big\}_+ \big[ \Psi_{\epsilon,+} \big( f_U(\bX) , A, f_U(\bX) \big) \big]
\\
- \big\{ \mu (A, \bX) - R \big\}_- \big[ \Psi_{\epsilon,-} \big( f_L(\bX) , A, f_L(\bX) \big) \big]
\\
- \big\{ \mu (A, \bX) - R \big\}_- \big[ \Psi_{\epsilon,-} \big( f_U(\bX) , A, f_U(\bX) \big) \big]
\end{array}
\right] 
\\
&
\hspace*{1cm} 
+ C_{\text{cvx}} \{ f_L(\bX) - f_U(\bX) \}^2 / \epsilon \ .
\end{align*}

\end{itemize} 
Here, $C_{\text{cvx}}$ is a constant that guarantees the convexity. Specifically, if $C_{\text{cvx}}$ is large enough, $\loss_{\sur,+}(\bO,\ell,u \con \mu, e)$ and $\loss_{\sur,-}(\bO,\ell,u \con \mu, e) $ are convex in $(\ell,u)$. Therefore, we recommend choosing a large value for $C_{\text{cvx}}$, say $C_{\text{cvx}} = 10^4 C_\mathcal{L}$. 

Using these functions, we finally define the following convex functions $Q_+$ and $Q_-$:
\begin{align} \label{eq-Q-Decomposition}
Q_+( \bxi_{L} , \bxi_{U} )
&		
=
\frac{1}{N} \sum_{i=1}^{N}
\loss_{\sur,+}
\big( \bO_i , \xi_{L,0} + \bm{k}_i \T {\bxi}_{L,(-0)} , \xi_{U,0} + \bm{k}_i \T {\bxi}_{U,(-0)} \con \widehat{\mu} , \widehat{e} \big) 
\nonumber
\\
& \hspace*{1cm}
+
\lambda \big\{
{\bxi}_{L,(-0)}\T K {\bxi}_{L,(-0)}
+
{\bxi}_{U,(-0)}\T K {\bxi}_{U,(-0)}
\big\} \ ,
\nonumber
\\
Q_-( \bxi_{L} , \bxi_{U} )
&		
=
\frac{1}{N} \sum_{i=1}^{N}
\loss_{\sur,-}
\big( \bO_i , \xi_{L,0} + \bm{k}_i \T {\bxi}_{L,(-0)} , \xi_{U,0} + \bm{k}_i \T {\bxi}_{U,(-0)} \con \widehat{\mu} , \widehat{e} \big) \ . 
\end{align}
From straightforward algebra, one can show that $Q_+$ and $Q_-$ are convex decompositions of $Q$ in \eqref{eq-otherERM}, i.e., $Q_+$ and $Q_-$ are convex and satisfy $Q=Q_+-Q_-$. 

\begin{itemize}[leftmargin=0cm]
    \item[] \HT{(DC-2)}: \textit{Computation of Initial Values}
\end{itemize}

The solution to the DC algorithm depends on a $(2N+2)$-dimensional vector $( {\bxi}_{L}^{(0)}, {\bxi}_{U}^{(0)} )$, which can be quite large even for moderate-sized datasets. Consequently, greedy algorithms are not appropriate for searching initial points when the dataset has a moderate to large number of observations. Therefore, we propose an alternative strategy as follows. We focus on the case of $\bxi_{L}^{(0)}$ below, as $\bxi_{U}^{(0)}$ can be selected using a similar procedure. Given the indirect rules $\ell_i$ for $i=1,\ldots,N$, we consider the following internal division points with the internal division ratio parameter $\rho \in [0,1]$:
\begin{align} \label{eq-internaldevision}
& 
\ell_{i}^{(\rho)} = \rho \ell_i + (1-\rho) \overline{\ell}  \ , 
\quad 
i=1,\ldots,N \ ,
&& 
\overline{\ell} = \frac{1}{N} \sum_{i=1}^{N} \ell_i \ .
\end{align} 
In other words, the internal division points $\bm{\ell}^{(\rho)} = (\ell_1^{(\rho)}, \ldots, \ell_N^{(\rho)})\T$ are homogeneous and are close to the average of the indirect rules at small $\rho$. When $\rho=0$, the internal division points for the lower bound are chosen as the average of the indirect rules for all $i$, i.e., $\bm{\ell}^{(\rho)} = \overline{\ell} \bm{1}_N$. On the contrary, when $\rho=1$, the internal division points are chosen as the indirect rules themselves, i.e., $\bm{\ell}^{(\rho)}= (\ell_1,\ldots,\ell_N)\T$. For the given internal division points $\bm{\ell}^{(\rho)}$ and the gram matrix $K$, we find the coefficient vector ${\bxi}_L^{(0)}$ satisfying $\bm{\ell}^{(\rho)}
=
\xi_{L,0}^{(0)} \bm{1}_N + K {\bxi}_{L,(-0)}^{(0)}$. That is, we find the coefficients that are associated with the internal division points. For example, these coefficients can be calculated as follows:
\begin{align*}
    &
    \xi_{L,0}^{(0)} = \frac{1}{N}\sum_{i=1}^{N} \ell_i^{(\rho)}
    \ , 
    &&
    \bxi_{L,(-0)}^{(0)}
    =
    K^{-1} \big\{ \bm{\ell}^{(\rho)} - \xi_{L,0}^{(0)} 
    \bm{1}_N 
    \big\} \ .
\end{align*}
With this choice, we have $\xi_{L,0}^{(0)} = \overline{\ell}$ for all $\rho$. Additionally, $\bxi_{L,(-0)}^{(0)}$ tends to be small and homogeneous for small values of $\rho$, and large and heterogeneous for large values of $\rho$, reflecting the properties of $\bm{\ell}^{(\rho)}$.

Given the convex decomposition and the initial value, the DC algorithm can be implemented as in Algorithm \ref{alg-DC}:
\begin{algorithm}[!htp] 
\caption{DC Algorithm} \label{alg-DC}
\begin{algorithmic}
\REQUIRE Initial coefficients $( {\bxi}_{L}^{(0)}, {\bxi}_{U}^{(0)} )$
\STATE Let $Q_+ $ and $Q_- $ be the convex functions satisfying $Q= Q_+ - Q_- $, i.e., \eqref{eq-Q-Decomposition}.

\STATE Update ${\bxi}_L^{(t)}$ and ${\bxi}_U^{(t)}$ based on the following optimization until convergence:
\begin{align*}
\begin{pmatrix}
{\bxi}_L^{(t+1)} \\ {\bxi}_U^{(t+1)}
\end{pmatrix}																										
=
\argmin_{ {\bxi}_L, {\bxi}_U }												
\left[ Q_+ ( {\bxi}_L, {\bxi}_U )
-
\Big\{ \nabla Q_- ( {\bxi}_L^{(t)}, {\bxi}_U^{(t)} ) \Big\} \T 
\begin{pmatrix}
{\bxi}_L \\ {\bxi}_U 
\end{pmatrix}
\right] \ .
\end{align*}

\RETURN Converged ${\bxi}_L^{(t)}$ and ${\bxi}_U^{(t)}$

\end{algorithmic}
\end{algorithm}

We discuss how to select hyperparameters for implementing the optimization problem \eqref{eq-otherERM}, namely the kernel bandwidth parameter $\gamma$, the regularization parameter $\lambda$, and the internal division ratio parameter $\rho$. For $\gamma$, we recommend using the median heuristic \citep{Garreau2017}, while for $\lambda$ and $\rho$, we suggest using cross-validation. The details of these selection methods are provided in Section \ref{sec:supp:alg-CV} of the Appendix.

Lastly, we consider some remedies to address the invalid intervals; notably, the PDI estimate can be invalid when (i) the bound estimate does not lie over a proper dose range or (ii) the upper bound estimate is not larger than the lower bound estimate, i.e., non-monotonic bounds. For the first violation, we simply winsorize the estimates escaping the dose range; we remark that the winsorization does not affect its statistical property; see Remark \ref{thm-ExcessRisk} for details. To avoid the second violation, in the ERM, we may introduce an additional regularization term penalizing when violations happen. For instance, in the optimization problem in \eqref{eq-otherERM}, we may use the following $\widetilde{Q}_{\kappa}$ having an additional regularization term:
\begin{align}						\label{eq-monotonicity}
\widetilde{Q}_{\kappa} ( {\bxi}_L, {\bxi}_U ) 
=
Q ( {\bxi}_L, {\bxi}_U ) 
+
\kappa \sum_{i=1}^{N} (\xi_{L,i} - \xi_{U,i} )_+ 
\ , \quad 
\kappa\geq 0 
 \ . 
\end{align}
The new regularization term assigns a larger loss function value if the lower bound coefficient is larger than the upper bound coefficient. This helps to ensure that the PDI estimate stays within the specified dose range. Again, the regularization parameter $\kappa$ can be determined through cross-validation. Despite these measures, there may still be cases where violations occur in the test data. In such instances, we resort to the pointwise dose rule by taking the average of $\widehat{f}_L$ and $\widehat{f}_U$. However, our simulations and data analysis indicate that the direct method yields very few violations, even without incorporating the coefficient-regularization term in \eqref{eq-monotonicity}; see the simulations and data applications in Sections \ref{sec:Simulation} and \ref{sec:Data} for empirical evidence. Therefore, we take $\kappa=0$ in the subsequent simulation studies and real-world data analyses.

% see Section \ref{sec:supp:lossft} of the Appendix for details. In addition, in Section \ref{sec:practical} of the Appendixs, we discuss practical considerations such as (i) choices of hyperparameters and initial points for solving \eqref{eq-otherERM}, and (ii) remedies to address invalid intervals.

\begin{REMARK} \label{remark-1}
Instead of learning $f_L$ and $f_U$ jointly, one can use an iterative procedure to optimize the lower or upper bound function at a time while the other is treated as fixed. In particular, the procedure can be summarized as follows: (i) given an estimator of the lower bound at iteration $t$, denoted by $\hf_L^{(t)}$, the upper bound is only estimated, denoted by $\hf_U^{(t+1)}$, with the restriction $\hf_L^{(t)} \leq \hf_U^{(t+1)}$; (ii) given the new upper bound estimator $\hf_U^{(t+1)}$, the lower bound is estimated, denoted by $\hf_L^{(t+1)}$, with the restriction $\hf_L^{(t+1)} \leq \hf_U^{(t+1)}$; and (iii) repeat (i) and (ii) until both bound estimators converge. The procedure can be viewed as an extension of the blockwise optimization method, also known as the nonlinear Gauss-Seidel method; see \cite{Write2015} for details. However, we argue that the iterative procedure is suboptimal for several reasons. First, there is no guarantee of its convergence and it faces a theoretical issue due to the restrictions on the PDI estimators, i.e., $\hf_L^{(t)} \leq \hf_U^{(t+1)}$ and $\hf_L^{(t+1)} \leq \hf_U^{(t+1)}$. These constraints restrict the range of the PDI estimators in a complex way, making the optimization domain smaller than the unrestricted product RKHS $\HH^{\otimes 2}$. Consequently, the approximation theory established under the unrestricted RKHS may not be applicable to the estimator obtained from this iterative procedure and the restricted RKHS. On the other hand, the ERM in \eqref{eq-otherERM} simultaneously uses both lower and upper bound estimators without any restrictions over the candidates, making the computation simpler and the RKHS theory applicable to our setting at the expense of more complicated, yet closed-form surrogate loss function. 
\end{REMARK}

\begin{REMARK} \label{remark-2}
One may restrict the relationship between $f_L^*$ and $f_U^*$ so that the width of the PDI is constant for all $\bX$, i.e., $f_U^*(\bX) = f_L^*(\bX) + w$ for a non-negative constant $w \geq 0$. The estimators of the PDI are represented as $\hf_L(\bx) = \widehat{\xi}_0 + \sum_{i=1}^{N} \widehat{\xi}_i k(\bx,\bX_i)$ and $\hf_U(\bx) = \widehat{\xi}_0 + \widehat{w} + \sum_{i=1}^{N} \widehat{\xi}_i k(\bx,\bX_i)$ where the coefficients $\widehat{\xi}_0$, $\widehat{\bxi}_{(-0)} = (\widehat{\xi}_1,\ldots,\widehat{\xi}_N)\T$, and $\widehat{w} \geq 0$ are obtained from the following ERM:
\begin{align*}
&
\big( \widehat{\xi}_{0}, \widehat{\bxi}_{(-0)}, \widehat{w} \big)
=
\argmin_{ \xi_{0}, \bxi_{(-0)}, w } 
\bigg[
\begin{array}{l}
N^{-1} \sum_{i=1}^{N}
\loss_{\sur}
\big( \bO_i , \xi_{0} + k_i \T \bxi_{(-0)} , \xi_{0} + w + k_i \T \bxi_{(-0)} \con \widehat{\mu} , \widehat{e} \big) 
\\
+
\lambda 
\bxi_{(-0)}\T K \bxi_{(-0)}
\end{array}
\bigg] \ .
\end{align*}
Although the constant-width PDI is easier to interpret and implement than the proposed method, it fails to account for variations in PDI width that may depend on $\bX$. In the context of the warfarin dataset analyzed in Section \ref{sec:Data}, the estimated PDI exhibited significant variations with respect to age and gender in the warfarin dataset; see Table \ref{tab:Data-2} for details. This suggests that the constant-width PDI model may be unreasonable for numerous real-world applications, including the warfarin dataset.	
\end{REMARK}

\subsection{Estimation of the Nuisance Components} \label{sec:NuisanceFt}

% We conclude the section by presenting some methods to estimate the nuisance functions $\mu$ and $e$. Each nuisance component has its own difficulties for estimation. First, as assumed in Assumption \ref{(A4)}, the dose-probability curve $\mu$ has a hormetic form, so it is desirable to obtain an estimate $\widehat{\mu}$ that maintains the same form. Some examples of the estimation techniques are (semi-)parametric models with a quadratic term of the dose level, and carefully designed nonparametric methods such as multidimensional monotone BART \citep{Hugh2022}. Second, the generalized propensity score $e$ is a conditional density due to the continuous nature of $A$, which is often regarded as a statistically more difficult object than the propensity score under discrete treatment. Investigators may use nonparametric methods to estimate the conditional density of $A$ given $\bX$ \citep{Li2008, Zhu2015, Schuster2020}. 

% If nonparametric methods are used to estimate nuisance functions, we recommend using the cross-fitting procedure to avoid the potential risk of having biased PDI estimates (see Section \ref{sec:supp:CF}). Additionally, the cross-fitting helps characterize the excess risk of the estimated PDI; see the next section for details. However, in practice, nonparametrically estimated nuisance functions can be unstable due to limited sample size and may yield no or little improvement compared to well-designed parametric methods. For this reason, in the simulation and warfarin application, we utilized parametrically estimated nuisance functions. 

We conclude the section by presenting some methods for estimating nuisance functions $\mu^*$ and $e^*$. First, as assumed in Assumption \ref{(A4)}, the dose-probability curve $\mu^*$ has a hormetic form, so it is desirable to obtain an estimator of $\mu^*$ that maintains the same form. Some examples of estimation techniques are (semi-)parametric models with a quadratic term of the dose level, and carefully designed nonparametric methods such as multidimensional monotone Bayesian additive regression trees \citep{Hugh2022}. Second, the generalized propensity score $e^*$ is a conditional density due to the continuous nature of $A$, which is often regarded as a statistically more difficult object than the propensity score under discrete treatment. Again, one may use (semi-)parametric models to estimate the conditional density of $A$ given $\bX$ or nonparametric methods \citep{Li2008, Zhu2015, Schuster2020}. 

The nuisance functions can also be estimated nonparametrically, but this requires a more careful estimation procedure. Suppose nonparametric methods are naively used to estimate the nuisance functions by utilizing the entire dataset, denoted as $\mathcal{D}=\{1,\ldots,N\}$, to (i) estimate the nuisance functions and (ii) estimate the PDI by solving the ERM, using the nuisance functions estimated in step (i). In this setup, each observation is used twice---once in step (i) and again in step (ii).  Unfortunately, this ``double-dipping'' approach can lead to inconsistent PDI estimates unless the nuisance functions are estimated within function spaces of well-controlled complexity, such as the Donsker class or the Vapnik-Chervonenkis class.

To address this issue, we recommend using the cross-fitting procedure \citep{Schick1986, Victor2018}, which mitigates the risk of non-diminishing bias in PDI estimators. The cross-fitting procedure consists of the following steps: First, split the dataset $\mathcal{D}$ into non-overlapping split samples $\{ \mathcal{D}_1,\ldots,\mathcal{D}_S \}$; second, use a subset of data to estimate the nuisance functions; and third, use the remaining subsample to solve the ERM for estimating the PDI. Algorithm \ref{alg-CF} below summarizes the cross-fitting procedure. While any finite value of $S$ can be used in principle, one can select $S=2$ for simplicity and computational efficiency by following \citet{Bach2024}.

\begin{algorithm}[!htp]
\caption{Cross-fitting}
\label{alg-CF}

\begin{algorithmic}
\REQUIRE Split the data into equally sized $S$ folds, say $\mathcal{D}_1,\ldots,\mathcal{D}_S$
\FOR{$s=1,\ldots,S$}

\STATE Obtain $(\widehat{\mu}\LOO, \widehat{e}\LOO)$, estimators of the nuisance functions $\mu^*$ and $e^*$ using $\mathcal{D}_s^c$.

\STATE Choose hyperparameters from the cross-validation procedure (see Algorithm \ref{alg-CV} of the Appendix) using $\mathcal{D}_s$ where the loss function is evaluated with $(\widehat{\mu}\LOO, \widehat{e}\LOO)$.

\STATE Obtain a PDI estimator $(\widehat{f}_L\HOO,\widehat{f}_U\HOO)$ using the hyperparameters obtained in the previous step and $\mathcal{D}_s$.

\ENDFOR

\RETURN Aggregated PDI estimator $\widehat{f}_L = S^{-1} \sum_{s=1}^{S} \widehat{f}_L\HOO$ and $\widehat{f}_U = S^{-1} \sum_{s=1}^{S} \widehat{f}_U\HOO$.

\end{algorithmic}
\end{algorithm}

The cross-fitting procedure enables the characterization of the theoretical properties of the estimated PDI (e.g., excess risk) by ensuring that the nuisance functions are estimated on a separate portion of the data from where the PDI is evaluated. This approach mitigates the ``double-dipping'' issue by eliminating the dependence between the nuisance function estimation and PDI estimation steps; see the next section for further details. However, in practice, nonparametrically estimated nuisance functions can be unstable due to limited sample size and may yield little or no improvement compared to well-designed parametric methods. For this reason, in both the simulation study and data applications, we used parametrically estimated nuisance functions without the cross-fitting procedure, resulting in well-performing PDI estimates.

\section{Theoretical Properties}						\label{sec:theory}

We examine theoretical properties of the PDI estimator obtained from the ERM with cross-fitting. Let $\EXP\LD (\cdot) $ be the expectation operator that regards the sample $\mathcal{D}$ as given. Note that, for a generic function estimated from $\mathcal{D}_s^c$, denoted by $\widehat{g}\LOO$, we have $\EXP\LD \big\{ \widehat{g}\LOO (\bO) \big\} = \int \widehat{g} \LOO (\bo) \, dP(\bo)$. Similarly, we denote $\risk\LD(f_L,f_U \con \mu\LOO, e\LOO \big) = \EXP \LD \big\{ \loss(\bO, f_L,f_U \con \mu\LOO, e\LOO \big) \big\}$ and $\risk_\sur\LD$ analogously.

First, we show that using the surrogate indicator function is a reasonable choice when $\epsilon$ is small under an additional condition. We begin by introducing regularity conditions.
\begin{itemize}
\item[\HT{(RC1)}] The estimated propensity score $\widehat{e}\LOO$ is lower bounded by a constant $c_e>0$, i.e., $\widehat{e}\LOO(A \cond \bX) > c_e $ for all $(A,\bX)$. Additionally, the estimated dose curve $\widehat{\mu}\LOO(a,\bX)$ is bounded in the unit interval, i.e., $\widehat{\mu}\LOO(a,\bX) \in [0,1]$.

\item[\HT{(RC2)}] For any sequence $a_N=o(1)$, there exists a finite non-negative number $b$ and $b_0$ satisfying $\EXP\LD \big[ \ind \big\{ \hf_L\HOO (\bX) - \hf_U \HOO (\bX) \in (0,a_N) \big\} \big] \leq a_N b_0 N^{b}$.

\end{itemize}

Regularity Condition \HL{(RC1)} states that the estimated propensity score and dose-probability curve satisfy the positivity and boundedness conditions, respectively. In practice, \HL{(RC1)} can be empirically verified by investigating $\widehat{e}\LOO $ and $\widehat{\mu}\LOO $ over the observed samples, or it can be enforced by trimming or applying a link function. Of note, these conditions are commonly made for tractable inference of causal effects (e.g., Assumption 5.1 of \citet{Victor2018}).

Regularity Condition \HL{(RC2)} means that the probability of $\hf_L\HOO$ and $\hf_U\HOO$ being closer than a small bandwidth $a_N$ is proportional to the bandwidth with a factor $b_0 N^{b}$. Similar conditions have been used in the literature (e.g., (3.3) of \citet{QM2011}, (16) of \citet{LvdL2016}, (A5) and (A6) of \citet{Shi2020}, (A5) of \citet{Cai2023}). To better understand the condition, one can consider a derivative of the left-hand side with respect to $a_N$ at 0, which reduces the condition that the derivative value is upper bounded by a polynomial function $b_0 N^{b}$. This implies that the probability density function of a random variable $\hf_L\HOO - \hf_U\HOO$, which can be seen as the degree of non-monotonicity, is effectively bounded by a polynomial function. A sufficient condition of \HL{(RC2)} is that $\hf_L\HOO(\bX) \leq \hf_U\HOO(\bX)$ uniformly holds over $\bX$. This monotonicity condition can be empirically checked by investigating the empirical distribution of $\hf_L\HOO (\bX) - \hf_U\HOO (\bX)$ for the observed $\bX$ values. If the monotonicity condition appears to be violated, potentially undermining the validity of \HL{(RC2)}, one can make \HL{(RC2)} more plausible by using $\widetilde{Q}_{\kappa}$ in \eqref{eq-monotonicity} with a large $\kappa$ to enforce the constraint $\hf_L\HOO \leq \hf_U\HOO$.

The following lemma shows that using the surrogate indicator function is a reasonable choice when $\epsilon$ is small.

\begin{lemma}			\label{lemma-surloss}
Let $(\hf_L\HOO , \hf_U\HOO)$ be the PDI estimator obtained from the ERM \eqref{eq-otherERM} with the cross-fitting procedure. \\
\HT{(Result 1)} Suppose Assumptions \ref{(A1)}-\ref{(A5)} and Regularity Condition \HL{(RC1)}. Then, we have the following results for some constant $c_0$: 
\begin{align*}
\big| \risk_\sur \LD(f_L^*, f_U^* \con \widehat{\mu}\LOO, \widehat{e}\LOO) - \risk \LD(f_L^*, f_U^* \con \widehat{\mu}\LOO, \widehat{e}\LOO) \big| \leq c_0 \epsilon \ .
\end{align*}
\HT{(Result 2)} Suppose Assumptions \ref{(A1)}-\ref{(A5)} and Regularity Conditions \HL{(RC1)}-\HL{(RC2)} are satisfied. Then, we have the following results for some constant $c_0$: 
\begin{align*}
\big| \risk_\sur \LD(\hf_L \HOO, \hf_U \HOO \con \widehat{\mu}\LOO, \widehat{e}\LOO) - \risk \LD( \hf_L \HOO , \hf_U \HOO \con \widehat{\mu}\LOO, \widehat{e}\LOO) \big| \leq c_0 \epsilon N^{b} \ .
\end{align*}
\end{lemma}
\noindent \HL{(Result 1)} of Lemma \ref{lemma-surloss} states that, by viewing the estimated nuisance components as fixed functions, the surrogate risk function evaluated at the optimal PDI is not far from the original risk function evaluated at the optimal PDI. \HL{(Result 2)} of the Lemma states that a similar result is established for the PDI estimator obtained from the ERM under the additional condition \HL{(RC2)}.

Next, we present the result of the convergence of the estimated PDI in terms of the excess risk bound. In order to do so, we introduce additional regularity conditions.
\begin{itemize}[leftmargin=1cm]

\item[\HT{(RC3)}] The support of $\bX$ is compact, and the density of $\bX$ is bounded above by a constant.

\item[\HT{(RC4)}] Let $f_L^\dagger$ and $f_U^\dagger$ be the optimal PDI given $\widehat{\mu}\LOO$ and $\widehat{e}\LOO$, i.e., 
\begin{align*}
(f_L^\dagger, f_U^\dagger) \in \argmin_{(f_L, f_U) \in \Fprod} \risk \LD( f_L, f_U \con \widehat{\mu}\LOO, \widehat{e}\LOO ) \ .
\end{align*}
Then, the optimal PDI bounds $(f_L^*,f_U^*)$ and $(f_L^\dagger,f_U^\dagger)$ belong to a Besov space with smoothness parameter $\beta>0$, which is $\mathcal{B}_{1,\infty}^\beta (\R^d) = \{ f \in L_{\infty}(\R^d) \cond \sup_{t>0} t^{-\beta} \{ \omega_{r, L_1(\R^d) } (f,t) \}	< \infty	, r > \beta \}$ where $\omega_r$ is the modulus of continuity of order $r$. 

% \item[\HT{(RC5)}] Let $L$ the surrogate loss function at the true and estimated nuisance components, i.e., $L=\loss_\sur(\cdot \con \mu^*, e^*)$ or $L=\loss_\sur(\cdot \con \widehat{\mu}\LOO, \widehat{e}\LOO)$. Then, it satisfies the following conditions:	
% \begin{itemize}[leftmargin=1cm]
% \item[\HT{(RC5-1)}] For all $(\bo,\ell, u)$, there exists a constant $B>0$ satisfying $0 \leq L (\bo,\ell, u ) \leq B$.
% \item[\HT{(RC5-2)}] $L (\bo,\ell, u)$ is locally Lipschitz continuous with respect to $(\ell,u)$ for $\ell \leq u$.
% \item[\HT{(RC5-3)}] For all $(t,\bo)$, we have $L( \bo, \ell^W, u^W) \leq L(\bo, \ell, u)$ where $ \ell^W = \ell \cdot \ind \big\{ | \ell | \leq c_0 \big\} + \text{sign}(\ell) c_0 \cdot \ind \big\{ c_0 < | \ell | \big\}$.
% \item[\HT{(RC5-4)}] $\EXP \big[ \big\{ L \big( \bO, \widehat{f}_L^W, \widehat{f}_U^W \big) - L \big( \bO, f_L^*, f_U^* \big) \big\}^2 \big] \leq V \cdot \big[ \EXP \big\{ L \big( \bO, {f}_L^W, {f}_U^W \big) - L \big( \bO, f_L^*, f_U^* \big) \big\} \big]^v$ is satisfied for constant $v \in [0,1]$, $V \geq B^{2-v}$, and for all $f_L,f_U \in \HH$. 
% \end{itemize}

\item[\HT{(RC5)}] There exist positive constants $C$, $r_\mu$, and $r_e$ satisfying $ \EXP \LD \big[ \big\{ \widehat{\mu}\LOO(A , \bX) - \mu^*(A , \bX) \big\}^2 \big] \leq C \cdot N^{- 2 r_\mu}$, and $\EXP \LD \big[ \big\{ \widehat{e}\LOO(A \cond \bX) - e^*(A \cond \bX) \big\}^2 \big] \leq C \cdot N^{ - 2 r_e}$ with probability not less than $1-\Delta_N$ where $\Delta_N \rightarrow 0$ as $N \rightarrow \infty$. 
\end{itemize}
\noindent Regularity Condition \HL{(RC3)} states that the covariate does not concentrate to a certain point. Regularity Condition \HL{(RC4)} states that the optimal PDI corresponding to the true nuisance functions $(\mu^*,e^*)$ and the one corresponding to the estimated nuisance functions $(\widehat{\mu}\LOO,\widehat{e}\LOO)$ have smoothness of order $\beta$ in a Besov space. Regularity Condition \HL{(RC5)} implies that the estimated nuisance functions converge to their truth as $N$ grows. Regularity Condition \HL{(RC5)} can be seen as requirements on the smoothness of the nuisance functions $\mu^*$ and $e^*$. For instance, if $\mu^*$ belongs to the H\"older space with smoothness exponent $\delta_{\mu}$, the kernel density estimator for $\mu^*$ achieves a convergence rate of $O_P(N^{-2\delta_{\mu}/(2\delta_{\mu}+d')})$ in terms of mean squared error, where $d' = d + 1$ is the dimension of $(A, \bX)$. This rate is minimax optimal, and similar rates can be attained by other nonparametric methods; for further details, see \citet{Stone1980} and Chapter 1 of \citet{Tsybakov2009}. Therefore, \HL{(RC5)} would hold for any $r_{\mu}$ smaller than $ \delta_{\mu}/(2\delta_{\mu}+d')$. A similar interpretation applies to $e^*$.

Under these Regularity Conditions, the convergence of the PDI estimator can be characterized in terms of excess risk. Theorem \ref{thm-ExcessRisk} formally states the result.
\begin{THEOREM}						\label{thm-ExcessRisk}	
Let $(\hf_L\HOO, \hf_U\HOO)$ be the PDI estimator obtained from ERM with the cross-fitting procedure. Suppose that Assumptions \ref{(A1)}-\ref{(A5)} and Regularity Conditions \HL{(RC1)}-\HL{(RC5)} hold. Then, for all fixed $\epsilon>0$, $d/(d+\tau) < p < 1$, $\tau>0$, and $\lambda>0$, we obtain the following result with probability $P_{\bO}^{N}$ not less than $1-3e^{-\tau}-\Delta_N$:
\begin{align*}
&
\risk \LD\big( \hf_L\HOO , \hf_U\HOO \con \mu^*, e^* \big) 
- \risk \LD\big( f_L^*, f_U^* \con \mu^*, e^* \big)
\\
&
\leq
c_1 \lambda\gamma^{-d} + c_2 \gamma^\beta 
+ c_3 \big\{ \gamma^{(1-p)(1+\epsilon)d} \lambda^p N \big\}^{-\frac{1}{2-p}} 
+ c_4 N^{-1/2}\tau^{1/2} + c_5 N^{-1} \tau
+ c_6 \epsilon N^b + c_7 N^{-r_e-r_\mu}
\end{align*}
and
\begin{align*}
&
\risk \LD\big( \hf_L\HOO , \hf_U\HOO \con \widehat{\mu}\LOO, \widehat{e}\LOO \big) 
- \risk \LD\big( f_L^*, f_U^* \con \mu^*, e^* \big)
\\
&
\leq
c_1 \lambda\gamma^{-d} + c_2 \gamma^\beta 
+ c_3 \big\{ \gamma^{(1-p)(1+\epsilon)d} \lambda^p N \big\}^{-\frac{1}{2-p}} 
+ c_4 N^{-1/2}\tau^{1/2} + c_5 N^{-1} \tau
+ c_6 \epsilon N^b + c_8 N^{-r_e-r_\mu}
\end{align*}
where constants $c_1,\ldots,c_8$ do not depend on $N$.
\end{THEOREM} 
Theorem \ref{thm-ExcessRisk} states that the estimated PDI with the cross-fitting procedure converges to the optimal PDI in terms of the excess risk bound. This convergence is subject to several regularity conditions, including the boundedness of the estimated nuisance components and the density of $\bX$, the smoothness of the optimal PDI, the condition outlined in Lemma \ref{lemma-surloss}, and certain properties of the surrogate loss functions. If the hyperparameters are chosen with specific rates, specifically $\gamma \asymp N^{-1/(2\beta + d)}$, $\lambda \asymp N^{-(\beta+d)/(2\beta + d)}$, and $\epsilon \asymp N^{-\beta/(2\beta+d) - b}$ where $b$ is the parameter satisfying \HL{(RC2)} in Lemma \ref{lemma-surloss}, the excess risk bound has a rate of $O_P \big( N^{-r_e - r_\mu} + N^{-\beta/(2\beta+d) } \big)$, which has two leading terms. The first term is proportional to the product of the convergence rates of the nuisance components. The first term is proportional to the product of the convergence rates of the nuisance components, which can achieve an $o_P(N^{-1/2})$ rate when both $e^*$ and $\mu^*$ are estimated at $o_P(N^{-1/4})$ rates. These rates can be attained using various machine learning techniques, such as lasso \citep{Belloni2011, Belloni2013}, random forests \citep{Wager2016, Syrgkanis2020}, neural networks \citep{ChenWhite1999, Farrell2021}, and boosting \citep{Luo2016}, provided that $e^*$ and $\mu^*$ exhibit sufficient smoothness. If at least one of the true nuisance functions were known, as in experimental settings where $e^*$ was known, the first term would be zero. The second term is from the approximation error of using the ERM over the Gaussian RKHS. When the optimal PDI is sufficiently smooth compared to the dimension of the covariate, i.e., $\beta/d \simeq \infty$, the second term approaches $O_P(N^{-1/2})$.

While Theorem \ref{thm-ExcessRisk} characterizes the convergence of the PDI estimator in terms of excess risk, it does not ensure that the PDI estimator converges to the optimal PDI. To address this limitation, we directly analyze the convergence rate of the PDI estimator directly. We begin by making the following regularity condition to establish this convergence result:

    \begin{itemize}[leftmargin=1.5cm]
        \item[\HL{(RC6)}] There exists a constant $\mathfrak{b}>0$ and $\underline{L} >0$ such that $
        f_U^*(\bX)  - f_L^*(\bX) \geq 2 \mathfrak{b}$,  and
    \begin{align*}
        &
        \bigg| 
        \frac{\partial}{\partial a} \mu^*(a,\bX) 
        \bigg|
        \geq \underline{L} \quad \text{ for all } 
        a \in [f_L^*(\bX) - \mathfrak{b}, f_L^*(\bX) + \mathfrak{b} ] \cup 
        [f_U^*(\bX) - \mathfrak{b}, f_U^*(\bX) + \mathfrak{b} ] \ .
    \end{align*} 
    \end{itemize}

Regularity Condition \HL{(RC6)} consists of two parts. The first part means that the optimal PDI has a positive length. The second part requires that the dose-response curve $\mu^*(\cdot,\bX)$ exhibits sufficient variation near the boundary points of the optimal PDI. Roughly speaking, the second part ensures that differences in the dose-response curve correspond to differences in the dose level, provided that the dose level is near the boundary points of the optimal PDI.     This assumption is comparable to the conditions required in M-estimation, where the gradient of the moment equation must be non-zero near the true parameter value (e.g., Theorem 5.21 in \citet{Vaart1998}).

Theorem \ref{thm-convergence f} below establishes the convergence rate of the PDI estimator with respect to the $L_2(P)$-norm under the regularity conditions introduced above.
\begin{theorem} \label{thm-convergence f}
    Suppose that the conditions of Theorem \ref{thm-ExcessRisk} hold, and Regularity Condition \HL{(RC6)} also holds. Then, there exists a sequence $\delta_N$ such that $\delta_N \rightarrow 0$ as $N \rightarrow \infty$, and the following result holds with probability $P_{\bO}^{N}$ not less than $1-3e^{-\tau}-\Delta_N-\delta_N$: \\
    For all fixed $\epsilon>0$, $d/(d+\tau) < p < 1$, $\tau>0$, and $\lambda>0$, we have
\begin{align*}
    &
    \EXP \LD
    \Big[ \big\{ \widehat{f}_L\HOO(\bX) - f_L^*(\bX) \big\}^2
    +
    \big\{ \widehat{f}_U\HOO(\bX) - f_U^*(\bX) \big\}^2 \Big]
    \\
    &
    \leq 
    c_1' \lambda\gamma^{-d} + c_2' \gamma^\beta 
+ c_3' \big\{ \gamma^{(1-p)(1+\epsilon)d} \lambda^p N \big\}^{-\frac{1}{2-p}} 
+ c_4' N^{-1/2}\tau^{1/2} + c_5' N^{-1} \tau
+ c_6' \epsilon N^b + c_7' N^{-r_e-r_\mu}  
\end{align*}
where constants $c_1',\ldots,c_7'$ do not depend on $N$.  
\end{theorem}

\noindent As in Theorem \ref{thm-ExcessRisk}, the convergence rate established in Theorem \ref{thm-convergence f} follows the same rate, differing only in the constant terms. Consequently, the interpretation of Theorem \ref{thm-ExcessRisk} similarly applies to the convergence rate of the PDI estimator. In particular, if $\gamma \asymp N^{-1/(2\beta + d)}$, $\lambda \asymp N^{-(\beta+d)/(2\beta + d)}$, and $\epsilon \asymp N^{-\beta/(2\beta+d) - b}$, the convergence rate of the PDI estimator has a rate of $O_P \big( N^{-r_e - r_\mu} + N^{-\beta/(2\beta+d) } \big)$.

\section{Simulation}						\label{sec:Simulation}

We conducted a simulation study to assess the finite-sample performance of the proposed method based on the following data generating process with $N \in \{500,1000,1500,2000\}$ observations. Each observation consisted of a 10-dimensional covariates $\bX$, a dose level $A$, and a discretized outcome $R$. We generated the first four covariates, $(X_{1},X_{2},X_{3},X_{4})$, from $\text{Unif}(0,1)$, the next three, $(X_{5}, X_{6}, X_{7})$, from $N(0,1)$, and the remaining three, $(X_{8},X_{9},X_{10})$, from $\text{Ber}(0.5)$, with each covariate generated independently. We then generated $A$ from $N \big(
\sum_{j=1}^{10} X_{j}/15 + 0.2, 0.1^2 \big)$, and $R$ from $\text{Ber}( \nu(A,\bX) )$ where $\nu(a,\bx)
=
\text{expit}(
-(300+30\sum_{j=1}^{10}x_j)(a-0.45)^2+0.1\sum_{j=1}^{10}x_{j}+2.5)$. We considered the range of the probability level $\alpha$ in $\{0.6, 0.65, 0.7\}$.

Using the generated data, we first estimated the PDI based on the approach proposed in Sections \ref{sec:lossft} and \ref{sec:ERM}. Specifically, we estimated the propensity score and the dose-probability curve using parametric models that are widely used in practice. For the propensity score, we fitted a linear regression model of $A$ regressed on $\bX$ positing a normal error and obtained the density estimate based on the regression model, i.e., $\widehat{e}(A \cond \bX) = \phi ( A \con \widehat{\eta}_A(\bX), \widehat{\sigma}_A^2 )$ where $\phi(\cdot \con \eta , \sigma^2 )$ is the density of $N(\eta , \sigma^2 )$, and $\widehat{\eta}_A(\bX)$ and $\widehat{\sigma}_A^2$ are the estimated mean function and variance obtained from the regression model, respectively. We obtained the estimated dose-probability curve $\widehat{\mu} (A,\bX)$ based on a logistic regression model of $R$ regressed on $(A, A^2, \bX)$. We remark that this parametric model is misspecified because the coefficient of $A^2$ in the logistic regression model is assumed to be constant, while it varies with $\bX$ in the true model. However, we intentionally considered this misspecified model to reflect its widespread use in practice.

For the surrogate loss function bandwidth parameter, we chose $\epsilon=10^{-3}$. The kernel bandwidth parameter $\gamma$ was determined using the median heuristic described in Appendix \ref{sec:supp:alg-CV}. The monotonicity-inducing regularization parameter $\kappa$ in \eqref{eq-monotonicity} was selected as $\kappa=0$, which reduced to the optimization problem \eqref{eq-otherERM}; we remark that even though the monotonicity-inducing term is not introduced, the proportion of resulting in non-monotonic bounds was very small; see the results below for details. The regularization parameter $\lambda$ and the internal division ratio parameter $\rho$ for initial points were selected through 5-fold cross-validation from the sets $\lambda \in \{ 2^{-4}, 2^{0}, 2^{4} \}$ and $\rho \in \{0, 0.5, 1\}$, respectively. The described PDI estimator is referred to as \HT{(Direct)}. 

For comparison, we additionally considered the following two estimators. 
% First, we implemented the constant-width PDI estimator in Remark \ref{remark-2}, which is referred to as \HT{(D-CW)}. 
First, we obtained the indirect PDI estimator by following Section \ref{sec:indirect} based on the dose-probability curve estimator used in \HL{(Direct)}; this indirect estimator is referred to as \HT{(Ind-Para)}. Furthermore, we obtained an additional indirect PDI estimator where the dose-probability curve was estimated by an ensemble learner of random forest and gradient boosting \citep{Friedman2001}; this PDI estimator is referred to as \HT{(Ind-Ens)}. For indirect methods, we performed grid search to find an interval that made the estimated outcome regression greater than or equal to the probability level $\alpha$, i.e., $[\hf_L(\bX), \hf_U(\bX)] = \{ a \cond \widehat{\mu}(a,\bX) \geq \alpha \}$. If such an interval did not exist, we took the dose level maximizing the estimated dose-probability curve as the indirect rule recommendation, i.e., $\hf_L(\bX)=\hf_U(\bX) = \argmax_{a} \widehat{\mu} (a,\bX)$. Lastly, for reference, we considered the optimal PDI, which is referred to as (True). We evaluated the performance of these PDI estimators using a test dataset $\mathcal{D}_{\text{test}}$ having $N \in \{500,1000,1500,2000\}$ observations generated from the same distribution as the training data. We repeated the simulation 1000 times for each $N$. 

To evaluate the performance of a PDI estimator $(\hf_L, \hf_U)$, we considered the following criteria. First, we obtained the empirical excess risk (EER) over the test set, which is defined by:
\begin{align*}
\text{EER}
=
\frac{1}{| \mathcal{D}_{\text{test}} | } \sum_{i \in \mathcal{D}_{\text{test}} }
\big\{ \mathcal{L}(\bO_i, \hf_L, \hf_U \con \mu^*,e^*) - 
\mathcal{L}(\bO_i, f_L^*,f_U^* \con \mu^*,e^*)
\big\}
\ . 
\end{align*}
Note that the EER is an empirical analogue of the excess risk in Theorem \ref{thm-ExcessRisk}. Of note, the EER of the optimal PDI serves as the reference value of zero. Second, we calculated the proportion in which a method in view does not produce a valid PDI interval. 
Third, we focused on the mean absolute error (MAE) and the root mean squared error (RMSE), which are defined as follows:
\begin{align*}
&
\text{MAE} = \frac{1}{| \mathcal{D}_{\text{test}} | } \sum_{i \in \mathcal{D}_{\text{test}} } \big\{ | f_L^* (\bX_i) - \hf_L(\bX_i) | + | f_U^* (\bX_i) - \hf_U(\bX_i) | \big\} \ ,
\\
&\text{RMSE} = \bigg[ \frac{1}{| \mathcal{D}_{\text{test}} | } \sum_{i \in \mathcal{D}_{\text{test}} } \big[ \{ f_L^* (\bX_i) - \hf_L(\bX_i) \}^2 + \{ f_U^* (\bX_i) - \hf_U(\bX_i) \}^2 \big]
\bigg]^{1/2}
\ .
\end{align*}
For these four evaluation metrics (EER, proportion of invalid PDI, MAE, and RMSE), a lower value indicates better performance.

Lastly, we also evaluated the performance of the PDI estimators in terms of widely used classification performance measures. Specifically, for each PDI estimator, we defined the contingency table counts, true positives, true negatives, false positives, and false negatives as follows:
\begin{align*}
&
{\rm TP} 
= 
\sum_{i \in \mathcal{D}_{\rm test}} \ind \big\{ R_i =1 , A_i \in [\widehat{f}_L(\bX_i),\widehat{f}_U(\bX_i)] \big\} 
\ , 
\\
&
{\rm TN} 
= 
\sum_{i \in \mathcal{D}_{\rm test}} \ind \big\{ R_i =0 , A_i \notin [\widehat{f}_L(\bX_i),\widehat{f}_U(\bX_i)] \big\} 
\ , 
\nonumber
\\
&
{\rm FP} 
= 
\sum_{i \in \mathcal{D}_{\rm test}} \ind \big\{ R_i =1 , A_i \notin [\widehat{f}_L(\bX_i),\widehat{f}_U(\bX_i)] \big\} 
\ , 
\nonumber
\\
&
{\rm FN} 
= 
\sum_{i \in \mathcal{D}_{\rm test}} \ind \big\{ R_i =0 , A_i \in [\widehat{f}_L(\bX_i),\widehat{f}_U(\bX_i)] \big\} \ .
\nonumber
\end{align*}
Using these contingency table counts, we calculated accuracy, F1 score, Matthew's correlation coefficient (MCC) \citep{MCC}, recall, and Cohen's kappa \citep{Cohen1960}, which are defined as follows:
\begin{align*}
&
\text{Accuracy}
&&
=
\frac{ {\rm TP} + {\rm TN} }{{\rm TP} + {\rm TN} + {\rm FP} + {\rm FN}}
\ , 
\\
&
\text{F1}
&&
=
\frac{ 2{\rm TP} }{2{\rm TP} + {\rm FP} + {\rm FN}}
\ ,
\\
&
\text{MCC}
&&
=
\frac{{\rm TP} \times {\rm TN} 
-
{\rm FP} \times {\rm FN} 
}{
\{ (
{\rm TP} + 
{\rm FP} ) \times
(
{\rm TP} + 
{\rm FN} )
\times
(
{\rm TN} + 
{\rm FP} )
\times(
{\rm TN} + 
{\rm FN} ) \}^{1/2} } \ ,
\\
&
\text{Recall}
&&
= 
\frac{ {\rm TP} }{{\rm TP}+{\rm FN}}
\ , 
\\
&
\text{Cohen's kappa}
&&
=
\frac{2\times ({\rm TP} \times {\rm TN}-{\rm FN}\times {\rm FP})}{({\rm TP}+{\rm FP})\times ({\rm FP}+{\rm TN})+({\rm TP}+{\rm FN})\times ({\rm FN}+{\rm TN})} \ .
\end{align*}
A larger value of these classification measures indicates better performance with fewer type I and/or II errors.

The results of the simulation are summarized in Table \ref{tab:Simulation}. First, the EER of \HL{(Direct)} is the lowest among the competing estimators, likely due to its construction based on empirical risk minimization. Moreover, it decreases as the sample size $N$ increases, consistent with the result of Theorem \ref{thm-ExcessRisk}. Second, we focus on the proportion of invalid PDI estimates. \HL{(Direct)} generates invalid PDI intervals at only negligible proportions, while the two indirect methods show higher proportions to produce such intervals. Among these, \HL{(Ind-Ens)} is particularly susceptible to this issue. As $\alpha$ increases, the proportion of invalid intervals rises. This occurs because higher $\alpha$ values produce narrower PDIs, thereby increasing the risk of violating the monotonicity constraints. Third, in terms of the MAE and RMSE, \HL{(Direct)} uniformly outperforms the competing methods. We note that the EER, MAE, and RMSE cannot be calculated in real-world applications, as these measures require knowledge of the optimal PDI, which is not available. Lastly, in terms of classification performance measures, \HL{(Direct)} uniformly exhibits higher accuracy, F1 score, MCC, and Cohen's kappa compared to the other methods. We further note that \HL{(True)} generally yields the highest classification performance measures. Consequently, these measures (which can be calculated in real-world applications without knowledge of the optimal PDI) can serve as practical proxies for the EER, MAE, and RMSE in such scenarios.

Combining all numerical results, we conclude that the proposed direct method \HL{(Direct)} outperforms the other methods by always producing valid intervals, resulting in a smaller bias and fewer type I and II errors.

\begin{table}[!htp]

\renewcommand{\arraystretch}{1.05} \centering
\setlength{\tabcolsep}{4pt}
\footnotesize
\begin{tabular}{|c|c|c||c|c|c|c||c|c|c|c|c|}
\hline
$\alpha$ & Estimator & $N$ & EER & \begin{tabular}[c]{@{}c@{}}Invalid PDI\\ (\%)\end{tabular} & \begin{tabular}[c]{@{}c@{}}MAE\\ ($\times1000$)\end{tabular} & \begin{tabular}[c]{@{}c@{}}RMSE\\ ($\times1000$)\end{tabular} & Accuracy & F1 & MCC & Recall & \begin{tabular}[c]{@{}c@{}}Cohen's\\ kappa\end{tabular} \\ \hline
\multirow{13}{*}{0.6} & \multirow{4}{*}{(Ind-Para)} & 500 & 3.278 & 0.031 & 16.798 & 15.425 & 0.897 & 0.855 & 0.775 & 0.852 & 0.774 \\ \cline{3-12} 
& & 1000 & 0.366 & 0.003 & 11.914 & 10.810 & 0.899 & 0.858 & 0.780 & 0.856 & 0.780 \\ \cline{3-12} 
& & 1500 & 0.083 & 0.000 & 9.461 & 8.562 & 0.901 & 0.861 & 0.784 & 0.858 & 0.784 \\ \cline{3-12} 
& & 2000 & 0.037 & 0.000 & 8.407 & 7.595 & 0.902 & 0.862 & 0.786 & 0.859 & 0.786 \\ \cline{2-12} 
& \multirow{4}{*}{(Ind-Ens)} & 500 & 75.191 & 0.749 & 20.894 & 22.913 & 0.891 & 0.844 & 0.762 & 0.827 & 0.760 \\ \cline{3-12} 
& & 1000 & 30.929 & 0.307 & 18.150 & 18.963 & 0.893 & 0.846 & 0.765 & 0.829 & 0.764 \\ \cline{3-12} 
& & 1500 & 19.378 & 0.192 & 17.177 & 17.706 & 0.894 & 0.848 & 0.767 & 0.828 & 0.767 \\ \cline{3-12} 
& & 2000 & 13.396 & 0.132 & 16.287 & 16.533 & 0.895 & 0.849 & 0.769 & 0.829 & 0.768 \\ \cline{2-12} 
& \multirow{4}{*}{(Direct)} & 500 & \textbf{0.184} & \textbf{0.001} & \textbf{13.895} & \textbf{12.977} & \textbf{0.899} & \textbf{0.858} & \textbf{0.779} & \textbf{0.856} & \textbf{0.779} \\ \cline{3-12} 
& & 1000 & \textbf{0.070} & \textbf{0.000} & \textbf{10.045} & \textbf{9.314} & \textbf{0.900} & \textbf{0.860} & \textbf{0.783} & \textbf{0.859} & \textbf{0.783} \\ \cline{3-12} 
& & 1500 & \textbf{0.041} & \textbf{0.000} & \textbf{8.088} & \textbf{7.432} & \textbf{0.902} & \textbf{0.862} & \textbf{0.786} & \textbf{0.860} & \textbf{0.786} \\ \cline{3-12} 
& & 2000 & \textbf{0.016} & \textbf{0.000} & \textbf{7.267} & \textbf{6.645} & \textbf{0.902} & \textbf{0.863} & \textbf{0.787} & \textbf{0.861} & \textbf{0.787} \\ \cline{2-12} 
& (True) & & 0.000 & 0.000 & 0.000 & 0.000 & 0.903 & 0.864 & 0.788 & 0.861 & 0.788 \\ \hline \hline
\multirow{13}{*}{0.65} & \multirow{4}{*}{(Ind-Para)} & 500 & 7.222 & 0.071 & 17.543 & 16.179 & 0.893 & 0.847 & 0.766 & 0.827 & 0.765 \\ \cline{3-12} 
& & 1000 & 0.658 & 0.006 & 12.394 & 11.259 & 0.896 & 0.850 & 0.771 & 0.831 & 0.770 \\ \cline{3-12} 
& & 1500 & 0.100 & 0.000 & 9.833 & 8.902 & 0.897 & 0.853 & 0.775 & 0.832 & 0.774 \\ \cline{3-12} 
& & 2000 & 0.047 & 0.000 & 8.724 & 7.880 & 0.898 & 0.853 & 0.776 & 0.833 & 0.775 \\ \cline{2-12} 
& \multirow{4}{*}{(Ind-Ens)} & 500 & 179.484 & 1.792 & 24.247 & 27.649 & 0.883 & 0.826 & 0.741 & 0.786 & 0.738 \\ \cline{3-12} 
& & 1000 & 90.584 & 0.904 & 20.814 & 23.002 & 0.885 & 0.831 & 0.747 & 0.790 & 0.744 \\ \cline{3-12} 
& & 1500 & 62.496 & 0.623 & 19.398 & 21.273 & 0.886 & 0.832 & 0.750 & 0.790 & 0.747 \\ \cline{3-12} 
& & 2000 & 46.383 & 0.462 & 18.242 & 19.790 & 0.888 & 0.834 & 0.752 & 0.793 & 0.749 \\ \cline{2-12} 
& \multirow{4}{*}{(Direct)} & 500 & \textbf{0.275} & \textbf{0.002} & \textbf{14.348} & \textbf{13.476} & \textbf{0.895} & \textbf{0.850} & \textbf{0.770} & \textbf{0.831} & \textbf{0.769} \\ \cline{3-12} 
& & 1000 & \textbf{0.090} & \textbf{0.000} & \textbf{10.184} & \textbf{9.474} & \textbf{0.897} & \textbf{0.852} & \textbf{0.774} & \textbf{0.835} & \textbf{0.773} \\ \cline{3-12} 
& & 1500 & \textbf{0.028} & \textbf{0.000} & \textbf{8.108} & \textbf{7.502} & \textbf{0.898} & \textbf{0.854} & \textbf{0.777} & \textbf{0.835} & \textbf{0.776} \\ \cline{3-12} 
& & 2000 & \textbf{0.026} & \textbf{0.000} & \textbf{7.220} & \textbf{6.662} & \textbf{0.899} & \textbf{0.855} & \textbf{0.778} & \textbf{0.836} & \textbf{0.777} \\ \cline{2-12} 
& (True) & & 0.000 & 0.000 & 0.000 & 0.000 & 0.899 & 0.855 & 0.779 & 0.835 & 0.778 \\ \hline \hline
\multirow{13}{*}{0.7} & \multirow{4}{*}{(Ind-Para)} & 500 & 15.677 & 0.155 & 18.581 & 17.232 & 0.888 & 0.835 & 0.753 & 0.796 & 0.750 \\ \cline{3-12} 
& & 1000 & 1.437 & 0.014 & 13.094 & 11.929 & 0.890 & 0.838 & 0.757 & 0.800 & 0.755 \\ \cline{3-12} 
& & 1500 & 0.171 & 0.001 & 10.409 & 9.442 & 0.891 & 0.840 & 0.761 & 0.800 & 0.758 \\ \cline{3-12} 
& & 2000 & 0.096 & 0.001 & 9.235 & 8.356 & 0.892 & 0.841 & 0.762 & 0.801 & 0.760 \\ \cline{2-12} 
& \multirow{4}{*}{(Ind-Ens)} & 500 & 409.985 & 4.097 & 29.186 & 33.773 & 0.869 & 0.798 & 0.711 & 0.731 & 0.703 \\ \cline{3-12} 
& & 1000 & 228.890 & 2.286 & 24.775 & 28.383 & 0.874 & 0.806 & 0.721 & 0.740 & 0.714 \\ \cline{3-12} 
& & 1500 & 173.203 & 1.730 & 22.839 & 26.199 & 0.875 & 0.809 & 0.724 & 0.742 & 0.717 \\ \cline{3-12} 
& & 2000 & 133.861 & 1.337 & 21.328 & 24.364 & 0.877 & 0.812 & 0.728 & 0.746 & 0.721 \\ \cline{2-12} 
& \multirow{4}{*}{(Direct)} & 500 & \textbf{0.499} & \textbf{0.004} & \textbf{14.799} & \textbf{14.005} & \textbf{0.890} & \textbf{0.839} & \textbf{0.759} & \textbf{0.802} & \textbf{0.756} \\ \cline{3-12} 
& & 1000 & \textbf{0.067} & \textbf{0.000} & \textbf{10.587} & \textbf{9.892} & \textbf{0.891} & \textbf{0.841} & \textbf{0.761} & \textbf{0.804} & \textbf{0.759} \\ \cline{3-12} 
& & 1500 & \textbf{0.041} & \textbf{0.000} & \textbf{8.328} & \textbf{7.760} & \textbf{0.893} & \textbf{0.842} & \textbf{0.763} & \textbf{0.804} & \textbf{0.761} \\ \cline{3-12} 
& & 2000 & \textbf{0.020} & \textbf{0.000} & \textbf{7.392} & \textbf{6.872} & \textbf{0.893} & \textbf{0.843} & \textbf{0.764} & \textbf{0.805} & \textbf{0.762} \\ \cline{2-12} 
& (True) & & 0.000 & 0.000 & 0.000 & 0.000 & 0.893 & 0.843 & 0.765 & 0.803 & 0.762 \\ \hline
\end{tabular}
\caption{Summary of the Simulation Results. Each number shows an average of over 1000 simulation repetitions. A lower value indicates better performance for the first four evaluation metrics, while a higher value indicates better performance for the latter five metrics. The best performance among the three competing methods is highlighted in bold text. The evaluation metrics for \protect\HL{(True)} is aggregated across all $N$.}
\label{tab:Simulation}

\end{table}

\section{Application}				 \label{sec:Data}

We applied our method to explore the two-sided PDI in the contexts of warfarin dosing and the Job Corps program.

\subsection{Warfarin Dosing}

We first applied our method to study the two-sided PDI of warfarin, an anti-coagulant medicine. Warfarin is a medicine for preventing harmful blood clots from forming and growing bigger. It is particularly useful for patients suffering from conditions that can cause fatal blood clots such as stroke and heart attack. However, excessive doses of warfarin can lead to severe complications, including life-threatening bleeding and tissue death. Therefore, it is essential to prescribe warfarin in a suitable therapeutic dose to ensure its benefits to patients. 
As recommended by the American Heart Association, warfarin should be prescribed to keep a patient's international normalized ratio (INR) within a desired range, typically between 2 and 3 \citep{Warfarin_AHA_2014}. The INR level increases with the warfarin dose, leading to a dose-probability curve with an inverted U-shaped pattern, reflecting the hormetic response \citep{Blann2003}.

We used the dataset presented in \cite{Warfarin2009} to estimate the PDI of warfarin. Specifically, each patient's target INR range, the reported INR value, and the warfarin dose measured in mg/week were considered as the desired range $\thr$, the observed outcome $Y$, and the observed dose level $A$, respectively. Accordingly, $R = \ind (Y \in \thr)$ indicates whether the INR of a patient was within the desired range. Furthermore, we included the following pharmacogenetic and clinical variables as pre-treatment covariates $\bX$: gender (male/female), weight (in kg), height (in cm), age (nine discretized levels), and Amiodarone receipt status (took Amiodarone/did not took Amiodarone). We restricted our analysis to Asians because the targeted INR ranges for Asians were significantly different from those for other races; see Section \ref{sec:supp:Data-Warfarin} of the Appendix for details. Lastly, we discarded two observations with outlying dose levels greater than 80mg/week to guarantee the positivity assumption \ref{(A3)}. Consequently, we used 1407 patients in the analysis. 

In our analysis, we implemented the proposed direct method, which corresponds to \HL{(Direct)} in the simulation study, as follows. We split the sample into two sets, with $1000$ patients in the training set $\mathcal{D}_{\text{train}}$ and $407$ patients in the test set $\mathcal{D}_{\text{test}}$. We repeated the split $100$ times, and for each iteration, we used the training set to estimate the PDI, and the test set to evaluate the estimated PDI. Using the training set, we estimated the propensity score based on the linear regression model of $\log(A)$ on $\bX$ with normal error and the dose-probability curve based on the logistic regression model of $R$ on $A$, $A^2$, and $\bX$, which are denoted by $\widehat{e}$ and $\widehat{\mu}$, respectively. We note that $\log(A)$ was used in the propensity score model because it suggested better model diagnostics; see Section \ref{sec:supp:Data-Warfarin} for details. We then estimate the PDI from the ERM \eqref{eq-ERM} where the hyperparameters were chosen based on the same procedure as in the simulation study. The probability level $\alpha$ was chosen from $\{0.6,0.65,0.7 \}$. For comparison, we also implemented the two indirect methods \HL{(Ind-Para)} and \HL{(Ind-Ens)} considered in the simulation study.

In Table \ref{tab:Data}, we report the proportion of producing an invalid PDI and the classification performance measures that were used in the simulation. First, we find that the proportion of invalid PDI estimates follows a similar pattern to the simulation study: invalid PDI estimates are more frequent with the indirect methods than with the direct method. Second, the classification performance measures suggest that the direct method yields more accurate estimates. Specifically, the direct method uniformly shows higher values. Combining all results from both simulation and data analysis, the direct method can be a valuable tool for practitioners seeking to determine the therapeutic dose of warfarin.

\begin{table}[!htp]

\renewcommand{\arraystretch}{1.02} \centering
\setlength{\tabcolsep}{5pt}
\footnotesize
\begin{tabular}{|c|c||c||c|c|c|c|c|}
\hline
$\alpha$ & Estimator & Invalid PDI (\%) & Accuracy & F1 & MCC & Recall & Cohen's kappa \\ \hline
\multirow{3}{*}{0.6} & (Ind-Para) & 0.020 & 0.730 & 0.834 & 0.147 & 0.913 & 0.130 \\ \cline{2-8} 
 & (Ind-Ens) & 1.401 & 0.655 & 0.768 & 0.089 & 0.774 & 0.085 \\ \cline{2-8} 
 & (Direct) & \textbf{0.000} & \textbf{0.743} & \textbf{0.845} & \textbf{0.168} & \textbf{0.939} & \textbf{0.136} \\ \hline \hline
\multirow{3}{*}{0.65} & (Ind-Para) & 0.052 & 0.706 & 0.811 & 0.161 & 0.846 & 0.156 \\ \cline{2-8} 
 & (Ind-Ens) & 1.845 & 0.605 & 0.715 & 0.071 & 0.676 & 0.068 \\ \cline{2-8} 
 & (Direct) & \textbf{0.000} & \textbf{0.724} & \textbf{0.826} & \textbf{0.178} & \textbf{0.879} & \textbf{0.169} \\ \hline \hline
\multirow{3}{*}{0.7} & (Ind-Para) & 0.216 & 0.667 & 0.769 & 0.176 & 0.744 & 0.174 \\ \cline{2-8} 
 & (Ind-Ens) & 2.961 & 0.564 & 0.660 & 0.099 & 0.574 & 0.090 \\ \cline{2-8} 
 & (Direct) & \textbf{0.012} & \textbf{0.684} & \textbf{0.786} & \textbf{0.177} & \textbf{0.782} & \textbf{0.176} \\ \hline
\end{tabular}
\caption{Summary of the Warfarin Dosing Analysis. Each number shows an average of over 100 data splits. A Lower value indicates better performance for the proportion of invalid PDI, while a higher value indicates better performance for the latter five metrics. The best performance among the three competing methods is highlighted in bold text.}
\label{tab:Data}

\end{table} 

We conclude the data analysis by providing the summary of the PDI estimates obtained from the proposed direct method at $\alpha = 0.7$. Table \ref{tab:Data-2} summarizes the relationship between the PDI estimates and two covariates, age and gender. In terms of age, we find that the PDI interval is wider for younger patients and narrower for older patients. This result is consistent with previous medical studies which have established a negative correlation between age and the therapeutic dose range of warfarin \citep{Warfarin_Age2014, Warfarin_Age2018}. % , Warfarin_Age2015, Warfarin_Age2016 
In terms of gender, we find that the PDI interval is wider for female patients, but the relationship between gender and the therapeutic warfarin dose range remains uncertain based on previous medical studies. Some studies did not find significant differences in the therapeutic warfarin dose range between male and female patients \citep{Warfarin_Gender1, Warfarin_Gender2}, while others suggested that gender is an important factor, especially for Asian patients \citep{Warfarin_Gender_Significant1, Warfarin_Gender_Significant2}. Our analysis, which only included Asian patients, seems to support the latter view. However, more research is needed to establish the role of gender in determining the therapeutic dose of warfarin.

\begin{table}[!htp]

\renewcommand{\arraystretch}{1.02} \centering
\setlength{\tabcolsep}{3.75pt}
\footnotesize
\begin{tabular}{|c|ccccccc|}
\hline
\multirow{2}{*}{Gender} & \multicolumn{7}{c|}{Age} \\ \cline{2-8} 
& \multicolumn{1}{c|}{Under 29} & \multicolumn{1}{c|}{30-39} & \multicolumn{1}{c|}{40-49} & \multicolumn{1}{c|}{50-59} & \multicolumn{1}{c|}{60-69} & \multicolumn{1}{c|}{70-79} & Over 80 \\ \hline
Female & \multicolumn{1}{c|}{(10.2, 73.0)} & \multicolumn{1}{c|}{(11.3, 71.7)} & \multicolumn{1}{c|}{(11.9, 71.1)} & \multicolumn{1}{c|}{(12.7, 70.2)} & \multicolumn{1}{c|}{(13.6, 69.2)} & \multicolumn{1}{c|}{(14.6, 
 68.2)} & (15.4, 67.5) \\ \hline
Male & \multicolumn{1}{c|}{(12.3, 70.3)} & \multicolumn{1}{c|}{(13.3, 69.4)} & \multicolumn{1}{c|}{(14.2, 68.6)} & \multicolumn{1}{c|}{(15.1, 67.7)} & \multicolumn{1}{c|}{(16.1, 66.6)} & \multicolumn{1}{c|}{(17.3, 65.3)} & (19.0, 63.7) \\ \hline
\end{tabular}
\caption{Average of the PDI Estimates at $\alpha=0.7$ Obtained from 100 Data Split (unit: mg/week).}
\label{tab:Data-2}
\vspace*{-0.5cm}

\end{table}

\subsection{The Job Corps Program} \label{sec:Data_JC}

We next applied the proposed method in a social science context, focusing on formulating the two-sided PDI for the Job Corps program \citep{JobCorps2008}. Administered by the U.S. Department of Labor, Job Corps has been the nation’s largest training program for disadvantaged youths; as of 2023, the program has served over two million participants and continues to support more than 40,000 youths annually across centers nationwide \citep{JobCorps2024, JobCorps2025}. The Job Corps program offers intensive vocational training, academic education, and a variety of support services aimed at helping participants secure employment. 

Not surprisingly, several studies have identified a causal relationship between participation in Job Corps and employment outcomes. For example, \citet{JobCorps2001} and \citet{JobCorps2008} found that Job Corps has a positive long-term effect on employment, suggesting that participation increases the likelihood of securing a job. However, they also observed a negative short-term effect, likely due to the time commitment required by the program, which may delay immediate job opportunities. These findings suggest that the relationship between participation duration and employment outcomes may follow an inverted U-shaped pattern, where moderate participation maximizes job prospects, while both insufficient and prolonged participation can reduce employment opportunities.

We used the dataset in \citet{Huber2020} to estimate the PDI of the Job Corps participation duration for employment. We considered the following demographic variables as pre-treatment covariates, which were measured at the time of the survey: gender (male/female), age (16 to 24), race (white/black/etc), ethnicity (Hispanic/non-Hispanic), native English speaker status (yes/no), marital or cohabitation status (yes/no), parental status (yes/no), prior work experience (yes/no), average weekly gross earnings (in USD), health status (good/bad), and emotional problem status (has a problem/no problem). 
In the follow-up survey conducted after 1 year, the total hours spent in the Job Corps program during the first year were recorded, which we denote as $\widetilde{A}$. The treatment variable $A$ was then defined as the total hours spent in the Job Corps program raised to the power of 0.35, i.e., $A=\widetilde{A}^{0.35}$. A power of 0.35 was selected to transform the treatment variable, ensuring it more closely resembles a normal distribution; see Appendix \ref{sec:supp:Data-JC} for details. In the follow-up survey conducted after 2 years, the employment status in the second year was recorded and defined as $R=Y$, where $Y$ indicates that a participant was employed two years after the survey.
Following \citet{Flores2012} and \citet{Huber2020}, we restricted the sample to observations with positive treatment intensity, i.e., Job Corps participants. Furthermore, we included only individuals with a high school diploma or general education diploma, as Assumption \ref{(A4)} is less likely to hold for non-high school graduates within the considered range of $\alpha$; see Appendix \ref{sec:supp:Data-JC} for further details. As a consequence, we used 894 participants in the analysis. 

Like the warfarin analysis in the previous section, we implemented the proposed method \HL{(Direct)} as follows. The sample was split into two subsets 100 times: a training set with 800 participants, used to estimate the PDI, and a test set with 94 participants, used to evaluate the performance of the estimated PDI. Using the training set, we estimated the propensity score based on a linear regression model of $A$ on $\bX$ with normal error; note that $A$ was already transformed to resemble a normal distribution. The dose-probability curve was estimated based on a logistic regression model of $R$ on $(A,A^2,\bX)$. These estimated models are denoted by $\widehat{e}$ and $\widehat{\mu}$, respectively. The remaining implementation details were consistent with those in the simulation study and the warfarin analysis, and therefore, we omit them here.
 
Table \ref{tab:Data JC} provide the summary of the performance of the three competing PDI estimators.  First, we observe that invalid PDI estimates are more frequent with the indirect methods than with the direct method, which is consistent with the findings from the simulation study and the warfarin analysis. Second, the direct method generally exhibits higher classification performance measures, with the exception of MCC and Cohen's kappa at $\alpha=0.6$ and $0.65$, suggesting that our proposed approach is more accurate than the other two. As a result, the direct method appears to be an effective approach for determining the optimal Job Corps participation duration for employment in the following year.

\begin{table}[!htp]

\renewcommand{\arraystretch}{1.02} \centering
\setlength{\tabcolsep}{5pt}
\footnotesize

\begin{tabular}{|c|c||c||c|c|c|c|c|}
\hline
$\alpha$              & Estimator  & Invalid PDI (\%) & Accuracy       & F1             & MCC            & Recall         & Cohen's kappa  \\ \hline
\multirow{3}{*}{0.6}  & (Ind-Para) & \textbf{0.000}   & 0.815          & 0.897          & \textbf{0.075} & 0.976          & \textbf{0.054} \\ \cline{2-8} 
                      & (Ind-Ens)  & 1.266            & 0.723          & 0.830          & 0.006          & 0.842          & 0.005          \\ \cline{2-8} 
                      & (Direct)  & \textbf{0.000}   & \textbf{0.816} & \textbf{0.897} & 0.072          & \textbf{0.978} & 0.046          \\ \hline \hline
\multirow{3}{*}{0.65} & (Ind-Para) & 0.011            & 0.807          & 0.891          & \textbf{0.084} & 0.960          & \textbf{0.065} \\ \cline{2-8} 
                      & (Ind-Ens)  & 1.585            & 0.698          & 0.808          & 0.032          & 0.797          & 0.029          \\ \cline{2-8} 
                      & (Direct)  & \textbf{0.000}   & \textbf{0.810} & \textbf{0.893} & 0.079          & \textbf{0.967} & 0.059          \\ \hline \hline
\multirow{3}{*}{0.7}  & (Ind-Para) & 0.117            & 0.788          & 0.878          & 0.076          & 0.929          & 0.066          \\ \cline{2-8} 
                      & (Ind-Ens)  & 2.149            & 0.671          & 0.786          & 0.038          & 0.754          & 0.036          \\ \cline{2-8} 
                      & (Direct)  & \textbf{0.000}   & \textbf{0.799} & \textbf{0.886} & \textbf{0.087} & \textbf{0.945} & \textbf{0.073} \\ \hline
\end{tabular}

\caption{Summary of the Job Corps Analysis. Each number shows an average of over 100 data splits. A Lower value indicates better performance for the proportion of invalid PDI, while a higher value indicates better performance for the latter five metrics. The best performance among the three competing methods is highlighted in bold text.}
\label{tab:Data JC}

\end{table}

We conclude the data analysis by providing the summary of the PDI estimates, measured in hours per year, obtained from the proposed direct method at $\alpha = 0.7$. Table \ref{tab:Data-2 JC} summarizes the relationship between the PDI estimates and three covariates: ethnicity, native English speaker status, and gender. Regarding ethnicity, Hispanic participants are associated with wider PDI intervals. Similarly, native English speakers exhibit wider PDI intervals compared to non-native speakers. Lastly, male participants have wider PDI intervals than female participants. These findings suggest that male, Hispanic, and native English-speaking participants are more likely to secure employment regardless of the duration of their participation in the Job Corps program. In contrast, female, non-Hispanic, and non-native English-speaking participants exhibit narrower PDI intervals, indicating that their employment outcomes are more sensitive to the length of program participation and may require more strategic planning to improve their employment prospects. Our analysis indicates that these covariates are critical factors driving heterogeneity in labor market outcomes, aligning with previous research in economics that highlights the heterogeneous causal effect of the Job Corps program across gender, race, and ethnic groups \citep{JAE2014, Strittmatter2019}.

\begin{table}[!htp]

\renewcommand{\arraystretch}{1.02} \centering
\setlength{\tabcolsep}{6pt}
\footnotesize 

\begin{tabular}{|cc|cc|cc|}
\hline
\multicolumn{2}{|c|}{Ethnicity}                        & \multicolumn{2}{c|}{Non-Hispanic}                    & \multicolumn{2}{c|}{Hispanic}                        \\ \hline
\multicolumn{2}{|c|}{Native English Speaker Status}    & \multicolumn{1}{c|}{Non-native}     & Native         & \multicolumn{1}{c|}{Non-native}     & Native         \\ \hline
\multicolumn{1}{|c|}{\multirow{2}{*}{Gender}} & Male   & \multicolumn{1}{c|}{(63.8, 2443.8)} & (21.9, 2959.0) & \multicolumn{1}{c|}{(32.8, 2782.4)} & (18.0, 3058.8) \\ \cline{2-6} 
\multicolumn{1}{|c|}{}                        & Female & \multicolumn{1}{c|}{(81.6, 2298.0)} & (40.7, 2675.2) & \multicolumn{1}{c|}{(47.9, 2596.3)} & (27.0, 2875.7) \\ \hline
\end{tabular}

\caption{Average of the PDI Estimates at $\alpha=0.7$ Obtained from 100 Data Split  (unit: hour/year).}
\label{tab:Data-2 JC}
\vspace*{-0.5cm}

\end{table}

\section{Concluding Remarks} \label{sec:Conclusion}

In this paper, we present a method for estimating the personalized two-sided PDI. To directly estimate the PDI from an ERM, we develop a loss function that is robust to misspecification of the nuisance components (the propensity score and the dose-probability curve), and design a computationally tractable surrogate loss function for obtaining PDI estimators over the product RKHS. We show that the excess risk bound of our estimated PDI converges to the true optimal PDI at a rate that depends on the estimation error of the nuisance parameters and the approximation error of the PDI using functions within the product RKHS. The simulation results show that our direct method can yield PDIs that are more likely to be valid and more accurate in terms of bias and classification performance measures when compared to indirect methods. 

The proposed methodology offers several opportunities for extension. First, our framework is well-suited to settings with a moderate number of covariates, as illustrated by the two real-world applications in Section \ref{sec:Data}, involving 5 and 11 variables, respectively. To increase its versatility, the method could be extended to accommodate high-dimensional covariates, expanding its applicability to a broader range of scenarios. Second, many real-world datasets are collected in online settings, where the optimal decision rule must be updated sequentially as new data arrive. While our framework is developed for an offline setting, which aligns with the structure of the two real-world datasets analyzed, it would be valuable to extend the framework to an online setting and investigate the corresponding regret bounds. Third, while the ERM in \eqref{eq-ERM-surloss} is solved within the RKHS framework, alternative function spaces---such as those generated by (deep) neural networks \citep{DL2015, DLbook2016}, random forests \citep{Breiman2001}, or generalized additive models \citep{gam}---could be used in principle. However, establishing the convergence analysis of the PDI estimates in these alternative spaces requires substantial effort, as it involves distinct approximation theories that differ fundamentally from the RKHS approximation theory. Furthermore, the parameterization of these spaces differs from that of the RKHS, which may render the proposed DC algorithm incapable of preserving convexity under their parameterization. Therefore, we defer the exploration of these directions to future work. Lastly, we also note that our method relies on the hormesis, i.e., Assumption \ref{(A4)}. This assumption restricts the shape of the dose-probability curve, but can be empirically verified in certain cases, such as the data examples in Section \ref{sec:Data}. However, in cases where this assumption is difficult to verify due to insufficient observations or heterogeneous data sources, an investigator should be aware of this assumption and, if applicable, consider focusing on uncovering the mechanism between treatment and outcome through randomized experiments. 

We conclude the paper by revisiting the key distinction between the two approaches for personalized dose recommendations in continuous treatment. The first approach focuses on optimal single-level dose rules designed to maximize the outcome, while the second centers on therapeutic dose intervals that aim to achieve a desired range of outcomes. The former tends to be more interpretable, and its characterization and estimation are generally more straightforward, contributing to its popularity. However, despite the greater attention given to the single-level dose approach, the interval-based method is equally important in many applications, as demonstrated by the motivating examples in the introduction. From a policymaker's perspective, recommending a single fixed dose may be impractical due to uncertainties about its feasibility. Instead, proposing a dose interval is often a more feasible approach, as any value within the range can achieve the desired outcome while addressing concerns about implementation. Additionally, a single-level dose rule does not provide insight into whether the optimized outcome is truly preferred, whereas our interval-based approach inherently accounts for such considerations. As a result, we expect the interval-based approach to complement single-level dose recommendations by offering greater flexibility and practical advantages in a variety of contexts.

\newpage 

\appendix

\section*{Appendix}

\section{Details of the Main Paper}			\label{sec:supp:details}

\subsection{Existence of the Optimal PDI}		\label{sec:supp:Existence}

We discuss the existence of the optimal PDI, $(f_L^*,f_U^*)$, under Assumptions \ref{(A4)} and \ref{(A5)}. It is trivial that $a_{\bX}$ in \ref{(A4)} belongs to the level set $\big\{ a \cond \mu^*(a,\bX) \geq \alpha \big\}$, which is uniquely determined based on Assumption \ref{(A5)}. We now show that $\big\{ a \cond \mu^*(a,\bX) \geq \alpha \big\}$ is not a point, but an interval. Let $L_\mu$ be the Lipschitz constant of $\mu^*( \cdot, \bX)$, i.e., $\big| \mu^*(a_1,\bX) - \mu^*(a_2, \bX) \big| \leq L_\mu \big| a_1-a_2 \big|$. Since $\mu^*(a_{\bX},\bX) \geq \alpha+c_\mu$, we find $\mu^*(a_{\bX} - c_\mu/L_\mu, \bX) > \alpha$ and $\mu^*(a_{\bX} + c_\mu/L_\mu, \bX) > \alpha$, indicating that $[a_{\bX} - c_\mu/L_\mu , a_{\bX} + c_\mu/L_\mu] \subseteq \big\{ a \cond \mu^*(a,\bX) \geq \alpha \big\}$. As a result, we find
\begin{align*}
& 
f_L^* (\bX) = \inf_{a} \big\{ a \cond \mu^*(a,\bX) \geq \alpha \big\} \leq a_{\bX} - \frac{c_\mu}{L_\mu}
\ , &&
f_U^* (\bX) = \sup_{a} \big\{ a \cond \mu^*(a,\bX) \geq \alpha \big\} \geq a_{\bX} + \frac{c_\mu}{L_\mu} \ ,
\end{align*}
indicating $f_L^*(\bX) < f_U^*(\bX)$. 

% As alternative sufficient conditions, \citet{Chen2022} imposed $\mu^*(a,\bX)$ satisfies the following three conditions: (i) $\mu^*(0,\bX) \leq \alpha$ and $\mu^*(1,\bX) \leq \alpha$; (ii) there exists $a_{\bX}$ satisfying $\mu^*(a_{\bX}, \bX) \geq \alpha$; (iii) $\mu^*(a,\bX)$ crosses $\alpha$ at most twice as $a$ goes from 0 to 1. We argue that Assumptions \ref{(A4)} and \ref{(A5)} are more clear than the conditions in \citet{Chen2022} despite conceptual similarity.

\subsection{Details of the AIPW Loss Function} \label{sec:supp:TwoLoss}

Recall that
\begin{align*}
& \mathcal{L}_{\text{AIPW}}(\bO,f_L,f_U \con \mu, e)
\\
& = 
\alpha
\bigg[
\frac{ \{\mu(A,\bX)-R\} \ind \big\{ A \in [ f_L(\bX), f_U(\bX)] \big\}
}{ e(A \cond \bX) }
+ \int \big\{ 1 - \mu(a,\bX) \big\} \ind \big\{ a \in [ f_L(\bX), f_U(\bX)] \big\} \, da \bigg]
\\
&
+ 
(1-\alpha)
\bigg[
\frac{ \big\{ R - \mu(A,\bX) \big\} \ind \big\{ A \notin [ f_L(\bX), f_U(\bX)] \big\}}{ e(A \cond \bX) }
+ \int \mu(a,\bX) \ind \big\{ a \notin [ f_L(\bX), f_U(\bX)] \big\} \, da \bigg] \ ,
\end{align*}

We first introduce the surrogate loss function of $\mathcal{L}_{\text{AIPW}}$. 
\begin{align*}
& \mathcal{L}_{\text{AIPW}}(\bO,f_L,f_U \con \mu, e)
\\
& = 
\alpha
\bigg[
\frac{ \{\mu(A,\bX)-R\} \Psi_\epsilon \big( f_L(\bX) , A, f_U(\bX) \big) 
}{ e(A \cond \bX) }
+ \int \big\{ 1 - \mu(a,\bX) \big\} \ind \big\{ a \in [ f_L(\bX), f_U(\bX)] \big\} \, da \bigg]
\\
&
+ 
(1-\alpha)
\bigg[
\frac{ \big\{ R - \mu(A,\bX) \big\} \big\{ 1-\Psi_\epsilon \big( f_L(\bX) , A, f_U(\bX) \big) \big\}}{ e(A \cond \bX) }
+ \int \mu(a,\bX) \ind \big\{ a \notin [ f_L(\bX), f_U(\bX)] \big\} \, da \bigg] \ ,
\end{align*}
Here $\ind \big\{ A \in [f_L(\bX), f_U(\bX)] \big\}$ is replaced with $\Psi_\epsilon \big( f_L(\bX) , A, f_U(\bX) \big) $. 

We find the two IPW terms in $\mathcal{L}_{\text{AIPW}}$ are
\begin{align*}
& 
\frac{ \alpha \{\mu(A,\bX)-R\} \ind \big\{ A \in [ f_L(\bX), f_U(\bX)] \big\}
+
(1-\alpha) \big\{ R - \mu(A,\bX) \big\} \ind \big\{ A \notin [ f_L(\bX), f_U(\bX)] \big\}
}{ e(A \cond \bX) }
\\
& =
\frac{ \alpha \{\mu(A,\bX)-R\} \ind \big\{ A \in [ f_L(\bX), f_U(\bX)] \big\}
-
(1-\alpha) \{\mu(A,\bX)-R\} \big[ 1 - \ind\big\{ A \in [ f_L(\bX), f_U(\bX)] \big\} \big]
}{ e(A \cond \bX) }
\\
& = 
\frac{ \{\mu(A,\bX)-R\}\big[ -1 + \alpha + \ind\big\{ A \in [ f_L(\bX), f_U(\bX)] \big\} \big] }{ e(A \cond \bX) }
\\
& = 
\frac{\{\mu(A,\bX)-R\} \big[\ind\big\{ A \in [ f_L(\bX), f_U(\bX)] \big\} \big] }{ e(A \cond \bX) } + B_1(\bO)
\end{align*}
where $B_1(\bO) = \{\mu(A,\bX)-R\} (\alpha-1)/e(A \cond \bX)$.
The two outcome-integrated terms are
\begin{align*}
& 
\alpha \int \big\{ 1 - \mu(a,\bX) \big\} \ind \big\{ a \in [ f_L(\bX), f_U(\bX)] \big\} \, da
+ 
(1-\alpha) \int \mu(a,\bX) \ind \big\{ a \notin [ f_L(\bX), f_U(\bX)] \big\} \, da
\\
&
= 
\int \alpha \ind \big\{ a \in [ f_L(\bX), f_U(\bX)] \big\} + (1-\alpha) \mu(a,\bX) - \mu(a,\bX) \ind \big\{ a \in [ f_L(\bX), f_U(\bX)] \big\} \, da
\\
&
=
\int \big\{ \alpha - \mu(a,\bX) \big\} \ind \big\{ a \in [ f_L(\bX), f_U(\bX)] \big\} \, da 
+ \int (1-\alpha) \mu(a,\bX) \, da
\\
&
=
\int \big\{ \alpha - \mu(a,\bX) \big\} \ind \big\{ a \in [ f_L(\bX), f_U(\bX)] \big\} \, da 
+ B_2(\bO) \ .
\end{align*}
where $B_2(\bO) = \int (1-\alpha) \mu(a,\bX) \, da$. Therefore, $\mathcal{L}_\text{AIPW} - \mathcal{L}^{(1)}$ does not depend on $(f_L,f_U)$.

\subsection{Details of the Median Heuristic and Cross-validation} \label{sec:supp:alg-CV}

First, we provide details on the median heuristic \citep{Garreau2017}. The bandwidth parameter $\gamma$ is set as the median of the pairwise distances between the observed covariates, i.e., $\gamma = \median_{i,j} \big\| \bX_i - \bX_j \big\|$. This choice ensures that the entries of the Gram matrix, given by $\exp\big( - \big\| \bX_i - \bX_j \big\|_2^2 / \gamma^2 \big)$, remain within a reasonable range, avoiding extreme values. This approach is referred to as the median heuristic.

Next, we summarize the cross-validation procedure for selecting $\lambda$ in Algorithm \ref{alg-CV}.

\begin{algorithm}[!htp]
\caption{Cross-validation}
\label{alg-CV}

\begin{algorithmic}
\REQUIRE Candidates of hyperparameters $\Theta$; Number of folds $M$; Data $\mathcal{D}$
\STATE Split the data into equally sized $L$ folds, say $\mathcal{D}_1,\ldots,\mathcal{D}_M$
\FOR{$t=1,\ldots,T$}

\STATE Let $\theta_t \in \Theta$ $(t=1,\ldots,T)$ be the $t$th hyperparameter candidate

\FOR{$m = 1,\ldots,M$}

\STATE Let $( \widehat{\bxi}_L , \widehat{\bxi}_U )$ be the solution to \eqref{eq-otherERM} obtained by using $\mathcal{D}_m^c$ and $\theta_t$

\STATE Let $\widehat{\loss}_{t,m}$ be the empirical loss evaluated over $\mathcal{D}_m$

\ENDFOR

\STATE Let $\widehat{\loss}_t$ be the average of $\widehat{\loss}_{t,1}, \ldots, \widehat{\loss}_{t,M}$

\ENDFOR

\RETURN Optimal hyperparameter $\theta^*= \argmin_{\theta_t} \widehat{\loss}_t$

\end{algorithmic}
\end{algorithm}

\subsection{Details on the Data Analysis}

\subsubsection{Warfarin Dosing} \label{sec:supp:Data-Warfarin}

\begin{figure}[!htb]
\centering
\includegraphics[width=0.75\textwidth]{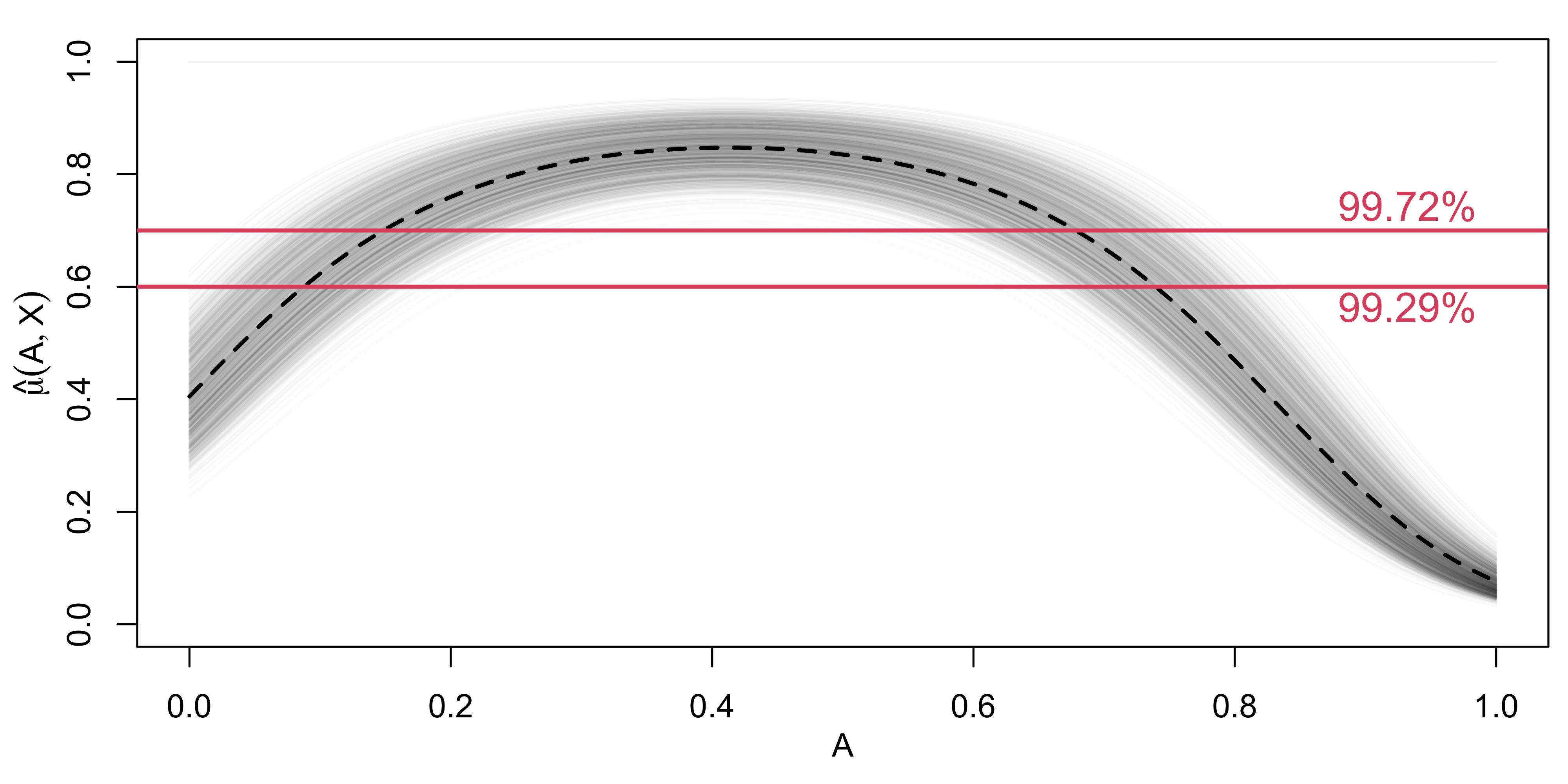}
\caption{A graphical summary of the estimated dose-probability curve. The gray curves show $\widehat{\mu}(a,\bX_i)$ for $a \in [0,1]$ and patient $i$'s covariates $\bX_i$. The black dashed line shows the mean of the dose-probability curve across all $N=1407$ patients. The red solid lines show the lower and upper ends of $\alpha=0.6$ and $0.7$, respectively. The red percentages represent the proportion of observations with a valid PDI estimate at $\alpha=0.6,0.7$.}
\label{Fig-OR}
\end{figure}

We first present a graphical summary of the estimated dose-probability curve. Figure \ref{Fig-OR} shows the estimated dose-probability curve,  which has an inverted U-shape for almost all observations. Additionally, visual inspection guarantees the existence of the PDI for $\alpha \in [0.6,0.7]$. Specifically, the PDIs are well-defined for 1397 observations (99.29\%) at $\alpha=0.6$ and 1403 observations (99.72\%) at $\alpha=0.7$. This suggests that Assumptions \ref{(A4)} and \ref{(A5)} are likely satisfied.

\begin{figure}[!htb]
\centering
\includegraphics[width=1\textwidth]{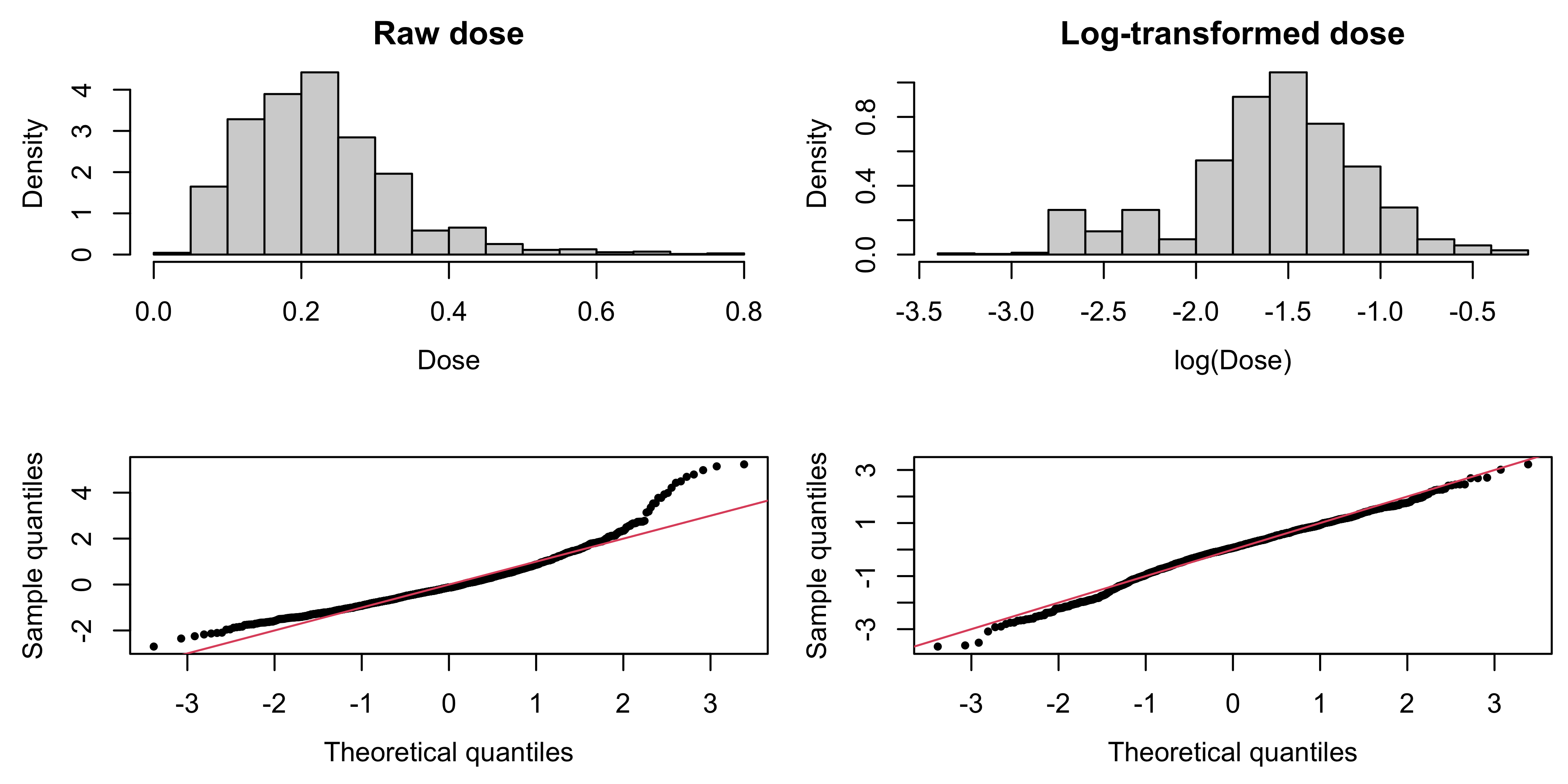}
\caption{Graphical summaries of propensity score model diagnosis. The left and right panels show the raw dose $A_i$ and the log-transformed dose $\log(A_i)$, respectively. The top panel shows histograms of the (log) dose levels of $N=1407$ patients, and the bottom panel shows the QQ plot obtained from the parametric propensity score model, respectively. The red solid lines in the bottom panel visually guide the QQ line for the standard normal distribution.}
\label{Fig-PS}
\end{figure}

Next, we validate the propensity score model. Figure \ref{Fig-PS} shows the empirical distribution of the observed dose level $A_i$ and its log-transformed value $\log(A_i)$, and the QQ plots of the residuals obtained from the parametric propensity score model where $A_i$ or $\log(A_i)$ is regressed on the covariates $\bX_i$, respectively. The visual diagnosis suggests that the propensity score model using the log-transformed dose is better than that using the raw dose. 

Lastly, we compare the desired ranges $\thr$ for Asians and those for other races. Table \ref{Tab-thr_range} shows the frequencies of $\thr$ for Asians and non-Asians who have a complete set of covariates. We find that the desired ranges for Asians are very different from those for non-Asians. Therefore, we only focus on the Asian subgroup to make $\thr$ homogeneous.

\begin{table}[!htp]
\renewcommand{\arraystretch}{1.1} \centering
\setlength{\tabcolsep}{4pt}
\footnotesize
\begin{tabular}{ccccccccccccc}
\hline
\multicolumn{1}{|c|}{Lower end of $\thr$} & \multicolumn{1}{c|}{0.8} & \multicolumn{1}{c|}{1.2} & \multicolumn{1}{c|}{1.25} & \multicolumn{1}{c|}{1.3} & \multicolumn{1}{c|}{1.5} & \multicolumn{3}{c|}{1.7} & \multicolumn{1}{c|}{1.75} & \multicolumn{3}{c|}{1.8} \\ \hline
\multicolumn{1}{|c|}{Upper end of $\thr$} & \multicolumn{1}{c|}{1.8} & \multicolumn{1}{c|}{2.2} & \multicolumn{1}{c|}{2.25} & \multicolumn{1}{c|}{2.3} & \multicolumn{1}{c|}{2.5} & \multicolumn{1}{c|}{2.7} & \multicolumn{1}{c|}{2.8} & \multicolumn{1}{c|}{3.3} & \multicolumn{1}{c|}{2.75} & \multicolumn{1}{c|}{2.2} & \multicolumn{1}{c|}{2.5} & \multicolumn{1}{c|}{2.8} \\ \hline
\multicolumn{1}{|c|}{Non-Asian} & \multicolumn{1}{c|}{1} & \multicolumn{1}{c|}{1} & \multicolumn{1}{c|}{6} & \multicolumn{1}{c|}{1} & \multicolumn{1}{c|}{4} & \multicolumn{1}{c|}{184} & \multicolumn{1}{c|}{0} & \multicolumn{1}{c|}{0} & \multicolumn{1}{c|}{4} & \multicolumn{1}{c|}{1} & \multicolumn{1}{c|}{3} & \multicolumn{1}{c|}{5} \\ \hline
\multicolumn{1}{|c|}{Asian} & \multicolumn{1}{c|}{0} & \multicolumn{1}{c|}{0} & \multicolumn{1}{c|}{0} & \multicolumn{1}{c|}{0} & \multicolumn{1}{c|}{631} & \multicolumn{1}{c|}{1} & \multicolumn{1}{c|}{236} & \multicolumn{1}{c|}{249} & \multicolumn{1}{c|}{0} & \multicolumn{1}{c|}{0} & \multicolumn{1}{c|}{0} & \multicolumn{1}{c|}{0} \\ \hline
\multicolumn{1}{l}{} & \multicolumn{1}{l}{} & \multicolumn{1}{l}{} & \multicolumn{1}{l}{} & \multicolumn{1}{l}{} & \multicolumn{1}{l}{} & \multicolumn{1}{l}{} & \multicolumn{1}{l}{} & \multicolumn{1}{l}{} & \multicolumn{1}{l}{} & \multicolumn{1}{l}{} & \multicolumn{1}{l}{} & \multicolumn{1}{l}{} \\ \cline{1-12}
\multicolumn{1}{|c|}{Lower end of $\thr$} & \multicolumn{2}{c|}{2} & \multicolumn{1}{c|}{2.1} & \multicolumn{1}{c|}{2.2} & \multicolumn{1}{c|}{2.25} & \multicolumn{1}{c|}{2.3} & \multicolumn{2}{c|}{2.5} & \multicolumn{1}{c|}{2.75} & \multicolumn{1}{c|}{3} & \multicolumn{1}{c|}{3} & \\ \cline{1-12}
\multicolumn{1}{|c|}{Upper end of $\thr$} & \multicolumn{1}{c|}{3} & \multicolumn{1}{c|}{3.5} & \multicolumn{1}{c|}{3.1} & \multicolumn{1}{c|}{3.2} & \multicolumn{1}{c|}{3.25} & \multicolumn{1}{c|}{3.3} & \multicolumn{1}{c|}{3} & \multicolumn{1}{c|}{3.5} & \multicolumn{1}{c|}{3.75} & \multicolumn{1}{c|}{3.5} & \multicolumn{1}{c|}{4} & \\ \cline{1-12}
\multicolumn{1}{|c|}{Non-Asian} & \multicolumn{1}{c|}{2805} & \multicolumn{1}{c|}{435} & \multicolumn{1}{c|}{1} & \multicolumn{1}{c|}{11} & \multicolumn{1}{c|}{1} & \multicolumn{1}{c|}{3} & \multicolumn{1}{c|}{2} & \multicolumn{1}{c|}{335} & \multicolumn{1}{c|}{1} & \multicolumn{1}{c|}{2} & \multicolumn{1}{c|}{10} & \\ \cline{1-12}
\multicolumn{1}{|c|}{Asian} & \multicolumn{1}{c|}{286} & \multicolumn{1}{c|}{1} & \multicolumn{1}{c|}{0} & \multicolumn{1}{c|}{0} & \multicolumn{1}{c|}{0} & \multicolumn{1}{c|}{0} & \multicolumn{1}{c|}{0} & \multicolumn{1}{c|}{5} & \multicolumn{1}{c|}{0} & \multicolumn{1}{c|}{0} & \multicolumn{1}{c|}{0} & \\ \cline{1-12}
\end{tabular}
\caption{Frequency Table of $\thr$ among Asians and Non-Asians.}
\label{Tab-thr_range}
\end{table}

\subsubsection{The Job Corps Program} \label{sec:supp:Data-JC}

\begin{figure}[!htb]
\centering
\includegraphics[width=1\textwidth]{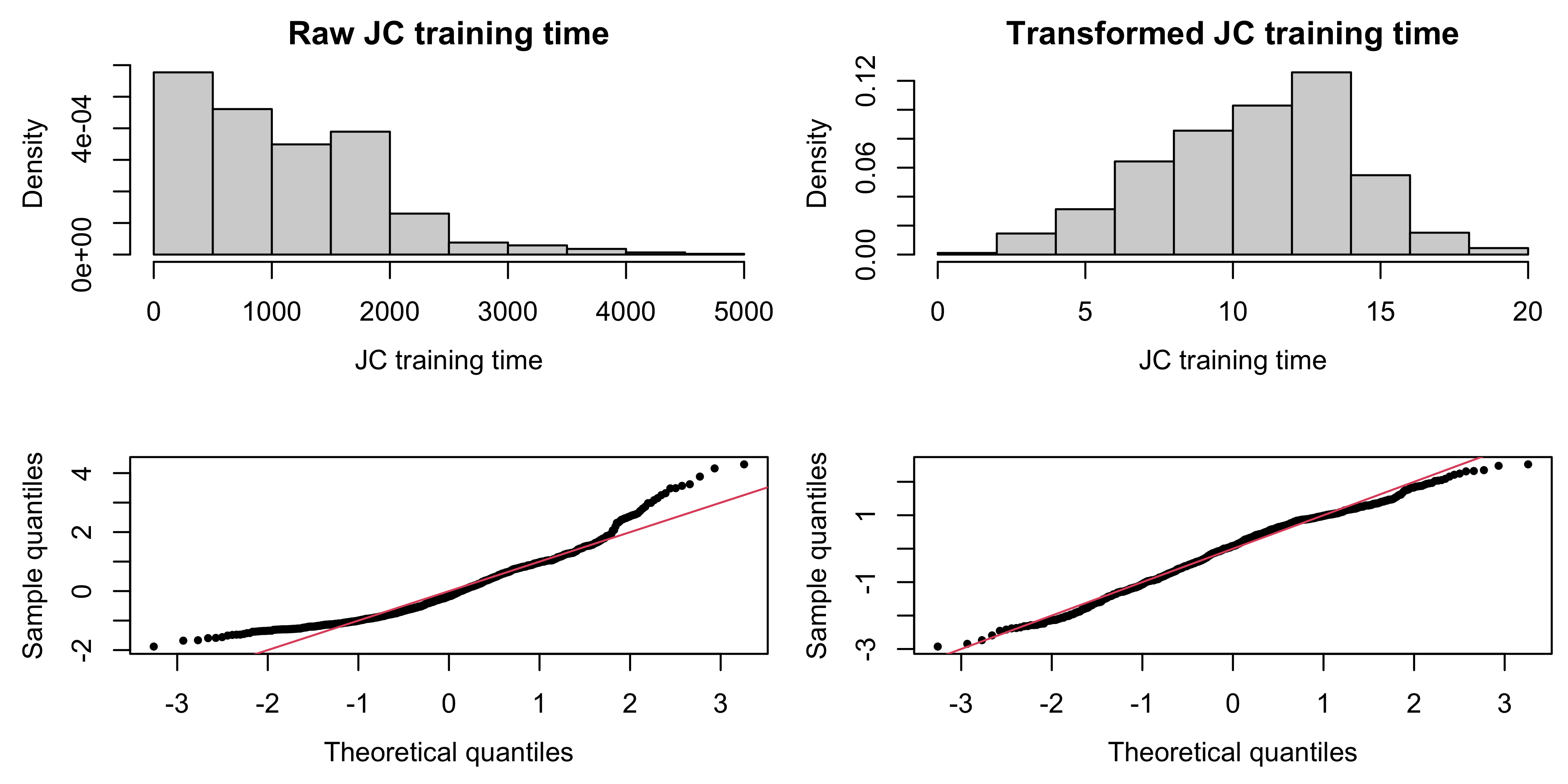}
\caption{Graphical summaries of propensity score model diagnosis. The left and right panels show the raw Job Corps training time $\widetilde{A}_i$ and the transformed Job Corps training time $A_i=\widetilde{A}_i^{0.35}$, respectively. The top panel shows histograms of the training time of $N=894$ patients, and the bottom panel shows the QQ plot obtained from the parametric propensity score model, respectively. The red solid lines in the bottom panel visually guide the QQ line for the standard normal distribution.}
\label{Fig-PS_JC}
\end{figure}

We first present the propensity score model. Figure \ref{Fig-PS_JC} shows the empirical distribution of the original Job Corps training time $\widetilde{A}_i$ and its transformed value $A_i=\widetilde{A}_i^{0.35}$, and the QQ plots of the residuals obtained from the parametric propensity score model where $\widetilde{A}_i$ or $A_i$ is regressed on the covariates $\bX_i$, respectively. The visual diagnosis suggests that the propensity score model using the transformed Job Corps training time is better than that using the raw dose.

\begin{figure}[!htb]
\centering
\includegraphics[width=\textwidth]{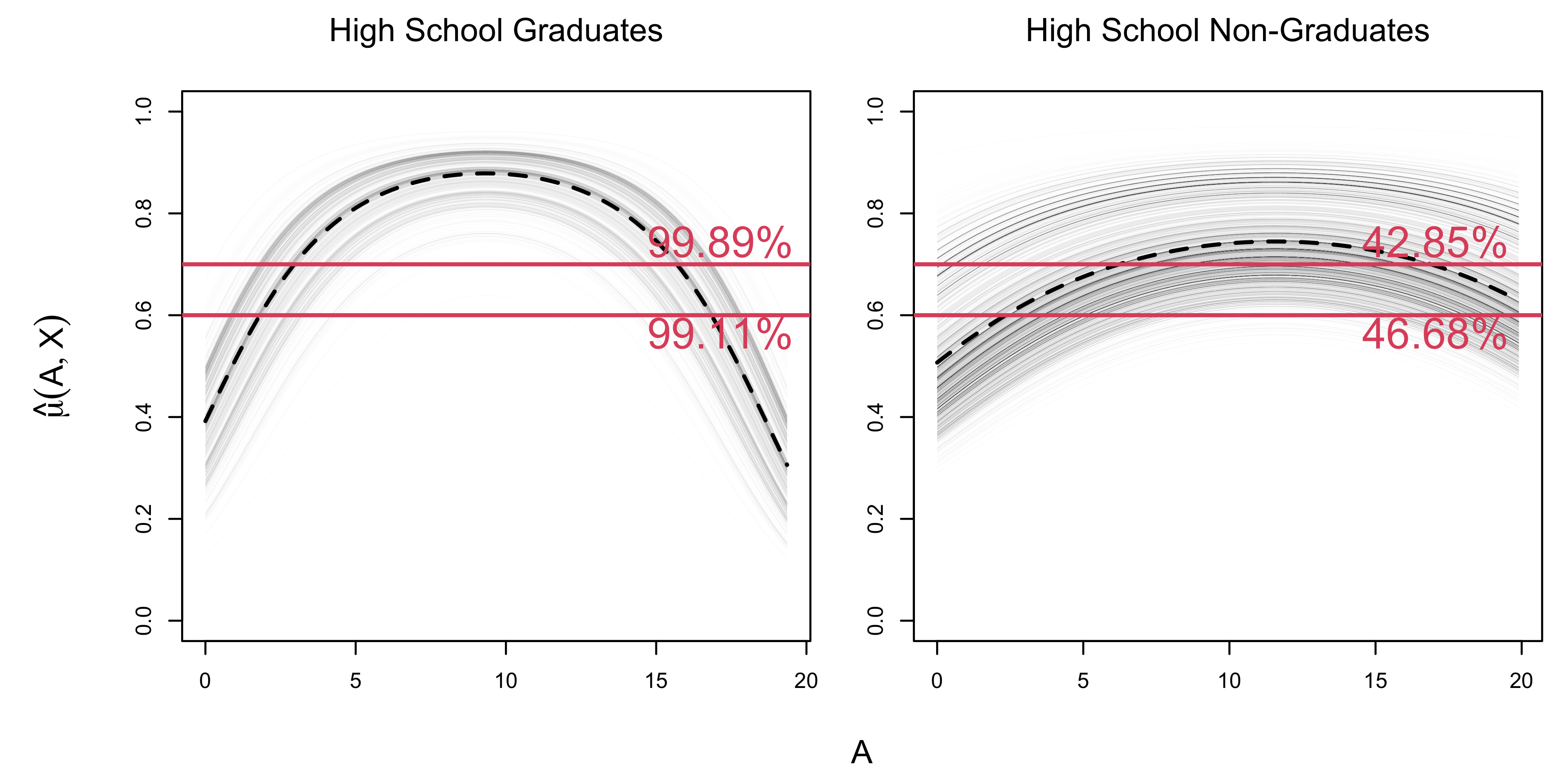}
\caption{A graphical summary of the estimated dose-probability curve for high school graduates (left) and high school non-graduates (right). The gray curves show $\widehat{\mu}(a,\bX_i)$ for $a \in [0,20]$ and patient $i$'s covariates $\bX_i$. The black dashed lines show the mean of the dose-probability curve across all high school graduates $(N=894)$ and high school non-graduates $(N=3106)$. The red solid lines show the lower and upper ends of $\alpha=0.6$ and $0.7$, respectively. The red percentages represent the proportion of observations with a valid PDI estimate at $\alpha=0.6,0.7$.}
\label{Fig-OR_JC}
\end{figure}

Next, we present a graphical summary of the estimated dose-probability curve. Figure \ref{Fig-OR_JC} shows the estimated dose-probability curves for high school graduates ($N=894$) and high school non-graduates $(N=3106)$. Although the estimated dose-probability curves for both groups exhibit an inverted U-shape, visual inspection suggests that the PDI is well-defined for high school graduates but not for high school non-graduates within $\alpha \in [0.6, 0.7]$. Specifically, for high school graduates, the PDIs are well-defined for 886 observations (99.11\%) at $\alpha=0.6$ and 893 observations (99.89\%) at $\alpha=0.7$. In contrast, for high school non-graduates, the PDIs are well-defined for only 1450 observations (46.68\%) at $\alpha=0.6$ and 1331 observations (42.85\%) at $\alpha=0.7$. This suggests that Assumptions \ref{(A4)} and \ref{(A5)} are likely satisfied for high school graduates but appear to be violated for high school non-graduates.

\newpage 

\section{Proof} \label{sec:supp:proof}

\subsection{Proof of Lemma \ref{lemma-minimizer}} \label{sec-Lemma1}

Let $(f_L^{\#}, f_U^{\#})$ be the minimzer of the risk function $\risk$. Based on the form of the loss function $\loss$, the minimizer should satisfy $f_L^{\#}(\bX) \leq f_U^{\#}(\bX)$ for all $\bX$. For a PDI candidate satisfying $f_L (\bX) \leq f_U (\bX)$, we find
\begin{align*}
&
\EXP \Big\{	\loss^{(1)} (\bO, f_L, f_U \con \mu, e) \, \Big| \,	 \bX \Big\}
\\
& =
\EXP \bigg[
\frac{ \big\{ \mu(A, \bX) - R \big\} 
\big[ \ind \{ A \in [ f_L(\bX), f_U(\bX)] \} \big]
}{e(A \cond \bX) } \, \bigg| \, \bX \bigg] 
+
\int_{f_L(\bX)}^{f_U(\bX)} \big\{ \alpha - \mu(a, \bX) \big\} \, da 
\\
& =
\EXP \bigg[
\frac{ \big\{ \mu(A,\bX) - \mu^*(A, \bX) \big\} 
\ind \{ A \in [ f_L(\bX), f_U(\bX)] \} 
}{e(A \cond \bX) } \, \bigg| \, \bX \bigg] 
+
\int_{f_L(\bX)}^{f_U(\bX)} \big\{ \alpha - \mu(a, \bX) \big\} \, da 
\\
& =
\int_{f_L(\bX)}^{f_U(\bX)} \big\{ \mu(a,\bX) - \mu^*(a,\bX) \big\} \frac{e^*(a \cond \bX)}{e(a \cond \bX)} \, da 
+
\int_{f_L(\bX)}^{f_U(\bX)} \big\{ \alpha - \mu(a, \bX) \big\} \, da
\ .
\numeq \label{eq-proof-lemma1}
\end{align*}
If $\mu=\mu^*$, \eqref{eq-proof-lemma1} reduces to
\begin{align*}
&
\EXP \Big\{	\loss^{(1)} (\bO, f_L, f_U \con \mu^*, e) \, \Big| \,	 \bX \Big\}
\\
& 
=
\int_{f_L(\bX)}^{f_U(\bX)} \underbrace{ \big\{ \mu^*(a,\bX) - \mu^*(a,\bX) \big\} }_{=0} \frac{e^*(a \cond \bX)}{e(a \cond \bX)} \, da 
+
\int_{f_L(\bX)}^{f_U(\bX)} \big\{ \alpha - \mu^*(a, \bX) \big\} \, da
\\
&
=
\int_{f_L(\bX)}^{f_U(\bX)} \big\{ \alpha - \mu^*(a, \bX) \big\} \, da \ .
\end{align*}
Likewise, if $e=e^*$, \eqref{eq-proof-lemma1} becomes
\begin{align*}
&
\EXP \Big\{	\loss^{(1)} (\bO, f_L, f_U \con \mu, e^*) \, \Big| \,	 \bX \Big\}
\\
&
=
\int_{f_L(\bX)}^{f_U(\bX)} \big\{ \mu(a,\bX) - \mu^*(a,\bX) \big\} 
\underbrace{ \frac{e^*(a \cond \bX)}{e^*(a \cond \bX)} }_{=1}
\, da 
+
\int_{f_L(\bX)}^{f_U(\bX)} \big\{ \alpha - \mu(a, \bX) \big\} \, da
\\
&
=
\int_{f_L(\bX)}^{f_U(\bX)} \big\{ \alpha - \mu^*(a, \bX) \big\} \, da \ .
\end{align*}
Therefore, the minimizer $(f_L^{\#},f_U^{\#})$ must have a form
\begin{align*}
& f_L^{\#}(\bX) = \inf_{\ell} \big\{ (\ell,u) \cond ^\forall a \in (\ell, u) \ \Rightarrow \ \mu^*(a,\bX) \geq \alpha \big\} \ ,
\\
& f_U^{\#}(\bX) = \sup_{u} \big\{ (\ell,u) \cond ^\forall a \in (\ell, u) \ \Rightarrow \ \mu^*(a,\bX) \geq \alpha \big\} \ , 
\end{align*}
which is equivalent to the definition of the PDI, i.e., $(f_L^*,f_U^*) = (f_L^{\#},f_U^{\#})$. This completes the proof.

\subsection{Proof of Lemma \ref{lemma-surloss}}

The proof is part of the proof Theorem \ref{thm-ExcessRisk}; see the derivation of \eqref{eq-RateofB} for details.

\subsection{Proof of Theorem \ref{thm-ExcessRisk}}		\label{sec:proof of Theorem}

Let $(f_L^\dagger, f_U^\dagger)$ and $(f_L^\#,f_U^\#)$ be the intermediate quantities that satisfy
\begin{align*}
    &
    (f_L^\dagger, f_U^\dagger)
    =
    \argmin_{(f_L,f_U)}
    \risk_\sur \LD(f_L,f_U \con \widehat{\mu}\LOO,\widehat{e}\LOO) 
    \ ,
    \\
    &
    (f_L^\#, f_U^\#)
    =
    \argmin_{(f_L,f_U)}
    \risk_\sur \LD(f_L,f_U \con \mu^*,e^*) \ .
\end{align*}

We decompose the difference between the two risk functions as follows.
\begin{align*}
&
\risk \LD\big( \widehat{f}_L\HOO , \widehat{f}_U\HOO \con \mu^*, e^* \big) 
- \risk \LD\big( f_L^*, f_U^* \con \mu^*, e^* \big)
\\
&
=
\risk \LD\big( \widehat{f}_L\HOO , \widehat{f}_U\HOO \con \mu^*, e^* \big) 
-
\risk \LD\big( \widehat{f}_L\HOO , \widehat{f}_U\HOO \con \widehat{\mu}\LOO, \widehat{e}\LOO \big) 
\\
& \hspace*{1cm}
+
\risk \LD\big( \widehat{f}_L\HOO , \widehat{f}_U\HOO \con \widehat{\mu}\LOO, \widehat{e}\LOO \big) 
-
\risk_\sur \LD\big( \widehat{f}_L\HOO , \widehat{f}_U\HOO \con \widehat{\mu}\LOO, \widehat{e}\LOO \big) 
\\
& \hspace*{1cm}
+
\risk_\sur \LD\big( \widehat{f}_L\HOO , \widehat{f}_U\HOO \con \widehat{\mu}\LOO, \widehat{e}\LOO \big) 
-
\risk_\sur \LD\big( f_L^\dagger, f_U^\dagger \con \widehat{\mu}\LOO, \widehat{e}\LOO \big) 
\\
& \hspace*{1cm}
+
\risk_\sur \LD\big( f_L^\dagger, f_U^\dagger \con \widehat{\mu}\LOO, \widehat{e}\LOO \big) 
-
\risk_\sur \LD\big( f_L^\#, f_U^\# \con \mu^*, e^* \big) 
\\
& \hspace*{1cm}
+
\underbrace{ 
\risk_\sur \LD\big( f_L^\#, f_U^\# \con \mu^*, e^* \big) 
- 
\risk_\sur \LD\big( f_L^*, f_U^* \con \mu^*, e^* \big) 
}_{\leq 0}
\\
& \hspace*{1cm}
+ 
\risk_\sur \LD\big( f_L^*, f_U^* \con \mu^*, e^* \big)
- \risk \LD\big( f_L^*, f_U^* \con \mu^*, e^* \big)  
\\
&
\leq
\risk \LD\big( \widehat{f}_L\HOO , \widehat{f}_U\HOO \con \mu^*, e^* \big) 
-
\risk \LD\big( \widehat{f}_L\HOO , \widehat{f}_U\HOO \con \widehat{\mu}\LOO, \widehat{e}\LOO \big) 
&&
\Leftarrow \quad (A)
\\
& \hspace*{1cm}
+
\risk \LD\big( \widehat{f}_L\HOO , \widehat{f}_U\HOO \con \widehat{\mu}\LOO, \widehat{e}\LOO \big) 
-
\risk_\sur \LD\big( \widehat{f}_L\HOO , \widehat{f}_U\HOO \con \widehat{\mu}\LOO, \widehat{e}\LOO \big) 
&&
\Leftarrow \quad (B)
\\
& \hspace*{1cm}
+
\risk_\sur \LD\big( \widehat{f}_L\HOO , \widehat{f}_U\HOO \con \widehat{\mu}\LOO, \widehat{e}\LOO \big) 
-
\risk_\sur \LD\big( f_L^\dagger, f_U^\dagger \con \widehat{\mu}\LOO, \widehat{e}\LOO \big) 
&&
\Leftarrow \quad (C)
\\
& \hspace*{1cm}
+
\risk_\sur \LD\big( f_L^\dagger, f_U^\dagger \con \widehat{\mu}\LOO, \widehat{e}\LOO \big) 
-
\risk_\sur \LD\big( f_L^\#, f_U^\# \con \mu^*, e^* \big) 
&&
\Leftarrow \quad (D)
\\  
& \hspace*{1cm}
+ 
\risk_\sur \LD\big( f_L^*, f_U^* \con \mu^*, e^* \big)
- \risk \LD\big( f_L^*, f_U^* \con \mu^*, e^* \big) \ .
&&
\Leftarrow \quad (E)
\end{align*}
Likewise, we have
\begin{align*}
&
\risk \LD\big( \widehat{f}_L\HOO , \widehat{f}_U\HOO \con \widehat{\mu}\LOO, \widehat{e}\LOO \big) 
- \risk \LD\big( f_L^*, f_U^* \con \mu^*, e^* \big)
\\
&
=
\risk \LD\big( \widehat{f}_L\HOO , \widehat{f}_U\HOO \con \widehat{\mu}\LOO, \widehat{e}\LOO \big) 
-
\risk_\sur \LD\big( \widehat{f}_L\HOO , \widehat{f}_U\HOO \con \widehat{\mu}\LOO, \widehat{e}\LOO \big) 
\\
& \hspace*{1cm}
+
\risk_\sur \LD\big( \widehat{f}_L\HOO , \widehat{f}_U\HOO \con \widehat{\mu}\LOO, \widehat{e}\LOO \big) 
-
\risk_\sur \LD\big( f_L^\dagger, f_U^\dagger \con \widehat{\mu}\LOO, \widehat{e}\LOO \big) 
\\
& \hspace*{1cm}
+
\risk_\sur \LD\big( f_L^\dagger, f_U^\dagger \con \widehat{\mu}\LOO, \widehat{e}\LOO \big) 
-
\risk_\sur \LD\big( f_L^\#, f_U^\# \con \mu^*, e^* \big) 
\\
& \hspace*{1cm}
+
\underbrace{ 
\risk_\sur \LD\big( f_L^\#, f_U^\# \con \mu^*, e^* \big) 
- 
\risk_\sur \LD\big( f_L^*, f_U^* \con \mu^*, e^* \big) 
}_{\leq 0}
\\
& \hspace*{1cm}
+ 
\risk_\sur \LD\big( f_L^*, f_U^* \con \mu^*, e^* \big)
- \risk \LD\big( f_L^*, f_U^* \con \mu^*, e^* \big)  
\\
&
\leq
\risk \LD\big( \widehat{f}_L\HOO , \widehat{f}_U\HOO \con \widehat{\mu}\LOO, \widehat{e}\LOO \big) 
-
\risk_\sur \LD\big( \widehat{f}_L\HOO , \widehat{f}_U\HOO \con \widehat{\mu}\LOO, \widehat{e}\LOO \big) 
&&
\Leftarrow \quad (B)
\\
& \hspace*{1cm}
+
\risk_\sur \LD\big( \widehat{f}_L\HOO , \widehat{f}_U\HOO \con \widehat{\mu}\LOO, \widehat{e}\LOO \big) 
-
\risk_\sur \LD\big( f_L^\dagger, f_U^\dagger \con \widehat{\mu}\LOO, \widehat{e}\LOO \big) 
&&
\Leftarrow \quad (C)
\\
& \hspace*{1cm}
+
\risk_\sur \LD\big( f_L^\dagger, f_U^\dagger \con \widehat{\mu}\LOO, \widehat{e}\LOO \big) 
-
\risk_\sur \LD\big( f_L^\#, f_U^\# \con \mu^*, e^* \big) 
&&
\Leftarrow \quad (D)
\\  
& \hspace*{1cm}
+ 
\risk_\sur \LD\big( f_L^*, f_U^* \con \mu^*, e^* \big)
- \risk \LD\big( f_L^*, f_U^* \con \mu^*, e^* \big) \ .
&&
\Leftarrow \quad (E)
\end{align*}

% We bound $(A1)+(A2)$, $(B1)+(B2)$, and $(C)$, respectively, in the following bullet points. As a consequence, we get the following result:
% \begin{align*}
% &
% \risk \big( \widehat{f}_L\HOO , \widehat{f}_U\HOO \con \mu^*, e^* \big) 
% - \risk \big( f_L^*, f_U^* \con \mu^*, e^* \big)
% \\
% &
% \leq
% \underbrace{
% C_1 \big\| \widehat{e}\LOO - e^* \big\|_{P,2} \big\|\widehat{\mu}\LOO - \mu^* \big\|_{P,2}	
% }_{\geq (A1)+(A2); \text{ see \eqref{eq-RateofA}}}
% +
% \underbrace{
% C_2 N^{-\beta/(2\beta+d) }
% }_{\geq (B1)+(B2); \text{ see \eqref{eq-RateofB}}}
% +
% \underbrace{
% C_3 N^{-\beta/(2\beta+d) } 
% }_{\geq (C) \text{ with high prob.; see \eqref{eq-RateofC}}}
% \\
% & 
% = O_P \Big( \big\| \widehat{e}\LOO - e^* \big\|_{P,2} \big\|\widehat{\mu}\LOO - \mu^* \big\|_{P,2}	 \Big)
% +
% O_P \Big( N^{-\beta/(2\beta+d) } \Big) \text{ with high probability} \ .
% \end{align*}

In \eqref{eq-RateofA}, \eqref{eq-RateofB}, \eqref{eq-RateofC-Full}, \eqref{eq-RateofD}, and \eqref{eq-RateofE}, we establish:
\begin{align*}
    &
    \eqref{eq-RateofA}:
    &&
    \text{With probability not less than $1-\Delta_N$, \quad }
    (A) 
    \lesssim 
    N^{-r_{\mu} - r_{e}} \ ;
    \\
    &
    \eqref{eq-RateofB}:
    &&
    (B)
    \lesssim
    N^{ b } \epsilon \ ;
    \\
    &
    \eqref{eq-RateofC-Full}:
    &&
    \text{With probability not less than $1-3e^{-\tau}, $}
    \\
    &
    &&
    (C)
    \lesssim
    c_1 \lambda\gamma^{-d} + c_2 \gamma^\beta 
+ c_3 \Big\{ \gamma^{(1-p)(1+\epsilon)d} \lambda^p N \Big\}^{-\frac{1}{2-p}} 
+ c_4 N^{-1/2}\tau^{1/2} + c_5N^{-1} \tau  
 \ ;
\\
    &
    \eqref{eq-RateofD}:
    &&
    \text{With probability not less than $1-\Delta_N$, \quad } (D) \lesssim C N^{-r_\mu-r_e} + C' \epsilon
    \leq 
    C N^{-r_\mu-r_e} + C' N^b \epsilon 
     \ ;
    \\
    &
    \eqref{eq-RateofE}:
    &&
    (E) \lesssim \epsilon
    \leq N^b \epsilon 
\ .
\end{align*}
In the rest of the proof, we obtain the rate of each term, which completes the proof.

\subsubsection{Upper bound of (A)} \label{sec-ub (A)}

For any measurable functions $(f_L,f_U): \mathcal{X}^{\otimes 2} \rightarrow \R^{\otimes 2}$, the difference between the loss function at $(\mu^*,e^*)$ and $(\mu', e')$ is
\begin{align*}
&
\loss ( \bO, f_L, f_U \con \mu^*, e^*) - \loss ( \bO, f_L, f_U \con \mu', e')
\\
&
=
\frac{ \ind \{ A \in [ f_L(\bX), f_U(\bX)] \} }{ e^*(A \cond \bX) e'(A \cond \bX) }
\left[
\begin{array}{l}
R \big\{ e^*(A \cond \bX) - e'(A \cond \bX) \big\} \\
+ \mu^*(A,\bX) e'(A \cond \bX) - \mu'(A,\bX) e^*(A \cond \bX)
\end{array}
\right]
\\
&
\hspace*{0.5cm}
+ \int \Big\{ \mu'(a,\bX) - \mu^*(a,\bX) \Big\} \Big\{ \ind \{ a \in [ f_L(\bX), f_U(\bX)] \} \Big\} \, da
&&
\hspace*{-2cm}
\text{ if $f_L(\bX) \leq f_U(\bX)$,}
\\
& = 0
&&
\hspace*{-2cm}
\text{ if $f_L(\bX) > f_U(\bX)$.}
\end{align*}
Note that for all $(f_L,f_U)$, we have
\begin{align*}
&
\EXP \left[
\frac{ \ind\big\{ f_L(\bX) \leq f_U(\bX) \big\} 
\big[ \ind \{ A \in [ f_L(\bX), f_U(\bX)] \} \big]
}{ e^*(A \cond \bX) e'(A \cond \bX) }
\left[
\begin{array}{l}
R \big\{ e^*(A \cond \bX) - e'(A \cond \bX) \big\} \\
+ \mu^*(A,\bX) e'(A \cond \bX) \\
- \mu'(A,\bX) e^*(A \cond \bX)
\end{array}
\right]\right]
\\
&
=
\EXP \left[
\frac{ \ind\big\{ f_L(\bX) \leq f_U(\bX) \big\} 
\big[ \ind \{ A \in [ f_L(\bX), f_U(\bX)] \} \big]
\big\{ \mu^*(A,\bX) - \mu'(A,\bX) \big\}
}{ e'(A \cond \bX) }
\right] \ , 	
\end{align*}
and
\begin{align*}
&
\EXP \Big[
\ind\big\{ f_L(\bX) \leq f_U(\bX) \big\} 
\int \Big\{ \mu'(a,\bX) - \mu^*(a,\bX) \Big\} \Big\{ \ind \{ a \in [ f_L(\bX), f_U(\bX)] \} \Big\} \, da
\Big]
\\
&
=
\EXP \left[
\frac{ \ind\big\{ f_L(\bX) \leq f_U(\bX) \big\} 
\big[ \ind \{ A \in [ f_L(\bX), f_U(\bX)] \} \big]
\big\{ \mu'(A,\bX) - \mu^*(A,\bX) \big\}
}{ e^* (A \cond \bX) }
\right] \ .
\end{align*}

As a consequence, $(A)$ is upper bounded as follows.
\begin{align*}
&
\Big|
\risk \big( \widehat{f}_L\HOO , \widehat{f}_U\HOO \con \mu^*, e^* \big) 
-
\risk \LD\big( \widehat{f}_L\HOO , \widehat{f}_U\HOO \con \widehat{\mu}\LOO, \widehat{e}\LOO \big) 
\Big|
\\
& = 
\Big|
\EXP\LD
\Big\{
\loss ( \bO, \widehat{f}_L \HOO, \widehat{f}_U\HOO \con \mu^*, e^*) - \loss ( \bO, \widehat{f}_L\HOO, \widehat{f}_U\HOO \con \widehat{\mu}\LOO, \widehat{e}\LOO)
\Big\}
\Big|
\\
&
=
\left|
\EXP\LD \left[
\begin{array}{l}
\displaystyle{
\frac{ \ind\big\{ \widehat{f}_L\HOO (\bX) \leq \widehat{f}_U\HOO (\bX) \big\} 
\ind \{ A \in [ \widehat{f}_L\HOO (\bX), \widehat{f}_U\HOO (\bX)] \} 
}{ \widehat{e}\LOO (A \cond \bX) e^* (A \cond \bX) }
}
\\
\hspace*{2cm}
\times \big\{ e^* (A \cond \bX) - \widehat{e}\LOO (A \cond \bX) \big\} \big\{ \mu^*(A,\bX) - \widehat{\mu}\LOO (A,\bX) \big\}
\end{array}
\right] 
\right|
\\
&
\leq
\Bigg\|
\frac{ \ind\big\{ \widehat{f}_L\HOO (\bX) \leq \widehat{f}_U\HOO (\bX) \big\} 
\ind \{ A \in [ \widehat{f}_L\HOO (\bX), \widehat{f}_U\HOO (\bX)] \} 
}{ \widehat{e}\LOO (A \cond \bX) e^* (A \cond \bX) } \Bigg\|_{P,\infty}
\\
& \hspace*{2cm} \times
\Big|
\EXP \Big[ \big\{ e^* (A \cond \bX) - \widehat{e}\LOO (A \cond \bX) \big\} \big\{ \mu^*(A,\bX) - \widehat{\mu}\LOO (A,\bX) \big\} \Big]
\Big|
\\
&
\precsim \big\| \widehat{e}\LOO - e^* \big\|_{P,2} \big\| \widehat{\mu}\LOO -\mu^* \big\|_{P,2} \ .
\end{align*}
The two inequalities are obtained by applying the H\"older's inequality. In particular, we used the assumption that the (estimated) propensity scores are away from zero. 
% Similarly, (A2) is upper bounded by the same cross-product, i.e., $(A2) \precsim \big\| \widehat{e}\LOO - e^* \big\|_{P,2} \big\| \widehat{\mu}\LOO -\mu^* \big\|_{P,2} $. 
Consequently, we have a constant $C_1$ satisfying
\begin{align}									\label{eq-RateofA}
(A) \leq C_{1} \big\| \widehat{e}\LOO - e^* \big\|_{P,2} \big\| \widehat{\mu}\LOO -\mu^* \big\|_{P,2}
\lesssim 
N^{-r_{\mu} - r_{e}}
\ ,
\end{align}
where the second inequality holds from \HL{(RC5)}. 

\subsubsection{Upper bound of (B)} \label{sec-ub (B)}

For any measurable functions $(f_L,f_U): \mathcal{X}^{\otimes 2} \rightarrow \R^{\otimes 2}$, we consider the following three cases where (i) $f_L(\bX) \leq f_U(\bX)$, (ii) $f_U(\bX) + \epsilon \leq f_L(\bX)$, and (iii) $f_L(\bX) \in ( f_U(\bX) , f_U(\bX) + \epsilon )$. \\

\noindent--- Case 1: $f_L(\bX) \leq f_U(\bX)$ \\[0.1cm]	

We find $\Phi_\epsilon \big( f_L(\bX) , f_U(\bX) \big)= \ind \big\{ f_L(\bX) \leq f_U(\bX) \big\} $. Thus,
\begin{align*}
&
\loss_\sur (\bO , f_L , f_U \con \widehat{\mu}\HOO, \widehat{e}\HOO) - \loss (\bO , f_L, f_U \con \widehat{\mu}\HOO, \widehat{e}\HOO)
\\
&
=
\ind \big\{ f_L(\bX) \leq f_U(\bX) \big\}
\frac{ \big\{ \widehat{\mu}\LOO(A,\bX) - R \big\}
\big[ \Psi_\epsilon \big( f_L(\bX), A, f_U(\bX) \big)
- 
\ind \big\{ A \in [f_L(\bX), f_U(\bX) ] \big\} \big]
}{\widehat{e}\LOO(A \cond \bX)}
\ .
\end{align*}
Since $\{ \widehat{\mu}\LOO(A,\bX) -R \}/ \widehat{e}\LOO(A \cond \bX)$ is uniformly bounded under Assumption \hyperlink{(R1)}{(R1)}, we have
\begin{align*}
&
\bigg|
\EXP\LD 
\Big[ \ind \big\{ f_L(\bX) \leq f_U(\bX) \big\}
\Big\{
\loss_\sur (\bO , f_L , f_U \con \widehat{\mu}\LOO , \widehat{e}\LOO) - \loss (\bO , f_L, f_U \con \widehat{\mu}\LOO, \widehat{e}\LOO)
\Big\}
\Big]
\bigg|
\\
&
\precsim
\bigg|
\EXP \Big[
\ind \big\{ f_L(\bX) \leq f_U(\bX) \big\}
\big[
\Psi_\epsilon \big( f_L(\bX), A, f_U(\bX) \big)
-
\ind \big\{ A \in [f_L(\bX), f_U(\bX) ] \big\}
\big]
\Big] \bigg| \ .
\end{align*}
Given $f_L \leq f_U$, we find
\begin{align*}
&
\Psi_\epsilon \big( f_L(\bX), A, f_U(\bX) \big)
-
\ind \big\{ A \in [f_L(\bX), f_U(\bX) ] \big\}
\\
& = 
\frac{A-f_L(\bX)+\epsilon}{\epsilon} \ind \big\{ A \in [ f_L(\bX) -\epsilon, f_L(\bX) ] \big\}
+
\frac{f_U(\bX)-A+\epsilon}{\epsilon} \ind \big\{ A \in [ f_U(\bX) , f_U(\bX) + \epsilon ] \big\}
 \ ,
\end{align*}
which implies
\begin{align*}
&
\EXP \big\{ 
\Psi_\epsilon \big( f_L(\bX), A, f_U(\bX) \big) 
-
\ind \big\{ A \in [f_L(\bX), f_U(\bX) ] \big\}
\cond \bX \big\}
\\
& =
\int_{f_L(\bX)-\epsilon}^{f_L(\bX)}
\frac{a-f_L(\bX)+\epsilon}{\epsilon} P(A=a \cond \bX) \, da
+
\int_{f_U(\bX)}^{f_U(\bX)+\epsilon}
\frac{f_U(\bX)-a+\epsilon}{\epsilon} P(A=a \cond \bX) \, da
\\
& \leq
c_e
\int_{f_L(\bX)-\epsilon}^{f_L(\bX)}
\frac{a-f_L(\bX)+\epsilon}{\epsilon} \, da
+
c_e
\int_{f_U(\bX)}^{f_U(\bX)+\epsilon}
\frac{f_U(\bX)-a+\epsilon}{\epsilon} \, da		
\\
& =
c_e \cdot \epsilon \ .
\end{align*}
The inequality is from the positivity assumption \ref{(A3)}, and the last equality is trivial. Combining the result, we get
\begin{align*}
&
\bigg|
\EXP\LD 
\Big[ \ind \big\{ f_L(\bX) \leq f_U(\bX) \big\}
\Big\{
\loss_\sur (\bO , f_L , f_U \con \widehat{\mu}\LOO , \widehat{e}\LOO) - \loss (\bO , f_L, f_U \con \widehat{\mu}\LOO, \widehat{e}\LOO)
\Big\}
\Big]
\bigg|
\\
&
\precsim
\bigg|
\EXP \Big[
\ind \big\{ f_L(\bX) \leq f_U(\bX) \big\}
\big[
\Psi_\epsilon \big( f_L(\bX), A, f_U(\bX) \big)
-
\ind \big\{ A \in [f_L(\bX), f_U(\bX) ] \big\}
\big]
\Big] \bigg|
\\
&
\leq c_e \cdot \epsilon \ .
\end{align*}

\noindent--- Case 2: $f_U(\bX) + \epsilon \leq f_L(\bX)$ \\[0.1cm]

We have $\loss_\sur (\bO , f_L, f_U \con \widehat{\mu}\LOO, \widehat{e}\LOO) - \loss (\bO , f_L, f_U \con \widehat{\mu}\LOO, \widehat{e}\LOO) = 0$ since $\loss = \loss_\sur \equiv C_\loss$. \\[0.3cm]	

\noindent--- Case 3: $f_L(\bX) \in ( f_U(\bX) , f_U(\bX) + \epsilon )$\\[0.1cm]

From the range of the loss functions, we have 
\begin{align*}
\big| \loss_\sur (\bO , {f}_L , f_U \con \widehat{\mu}\LOO, \widehat{e}\LOO ) - \loss (\bO , {f}_L , f_U \con \widehat{\mu}\LOO, \widehat{e}\LOO ) \big| \leq 2 C_\loss \ .
\end{align*}
Consequently, 
\begin{align*}
&
\bigg|
\EXP\LD 
\Big[ \ind \big\{ f_L(\bX) \in ( f_U(\bX) , f_U(\bX) + \epsilon ) \big\}
\\
&
\hspace*{2cm}\times 
\Big\{
\loss_\sur (\bO , f_L , f_U \con \widehat{\mu}\LOO , \widehat{e}\LOO) - \loss (\bO , f_L, f_U \con \widehat{\mu}\LOO, \widehat{e}\LOO)
\Big\}
\Big]
\bigg|
\\
& 
\leq 2 C_\loss \cdot \EXP\LD \big[ \ind \big\{ f_L (\bX) - f_U (\bX) \in (0,\epsilon) \big\} \big] \ .
\end{align*}

Combining three cases, we find 
\begin{align*}
&
\big| \risk_\sur \LD(f_L, f_U \con \widehat{\mu}\LOO, \widehat{e}\LOO )
-
\risk \LD(f_L, f_U \con \widehat{\mu}\LOO, \widehat{e}\LOO ) \big| 
\\
&
\leq c_e \cdot \epsilon + 0 + 2 C_\loss \cdot \EXP\LD \big[ \ind \big\{ f_L (\bX) - f_U (\bX) \in (0,\epsilon) \big\} \big] \ .
\end{align*}
At the optimal PDI, we have $\EXP\LD \big[ \ind \big\{ f_L^* (\bX) - f_U^* (\bX) \in (0,\epsilon) \big\} \big] = 0$. At the estimated PDI, this term is assumed to be upper bounded by a term having a rate $N^{b} \epsilon$. As a consequence, we get an upper bound of $(B)$ for some constant $C_2$:
\begin{align*}
&
\big| \risk_\sur \LD(f_L^*, f_U^* \con \widehat{\mu}\LOO, \widehat{e}\LOO )
-
\risk \LD(f_L^*, f_U^* \con \widehat{\mu}\LOO, \widehat{e}\LOO ) \big| 
\precsim \epsilon
\\
&
\big| \risk_\sur \LD(\widehat{f}_L\HOO, \widehat{f}_U\HOO \con \widehat{\mu}\LOO, \widehat{e}\LOO )
-
\risk \LD(\widehat{f}_L\HOO, \widehat{f}_U\HOO \con \widehat{\mu}\LOO, \widehat{e}\LOO ) \big| 
\precsim N^{b} \epsilon \ .
\end{align*}
The second result implies
\begin{align} \label{eq-RateofB}
& (B) \leq C_2 \cdot N^{ b } \epsilon
\ . 
\end{align}
Note that this provides the proof of Lemma \ref{lemma-surloss}. In addition, if $\epsilon \asymp N^{-\beta/(2\beta+d) - b}$, we get 
\begin{align*}					 
(B) \leq C_2 N^{-\beta/(2\beta+d) } \ .
\end{align*}

\subsubsection{Upper bound of (C)} \label{sec-ub (C)}

The outline of the proof is similar to that of Theorem 2 of \citet{Chen2016}. We remark that lemmas and theorems presented in \citet{SVM} and \citet{SVM2013} are essential for the proof. For notational brevity, let $L(\bO, \ell, u) = \loss_\sur (\bO, \ell, u\cdot \con \widehat{\mu}\LOO, \widehat{e}\LOO)$, $R (\ell, u) = \EXP\LD \big\{ \loss_\sur (\bO, \ell, u \con \widehat{\mu}\LOO, \widehat{e}\LOO) \big\}$, and the expectation involving $L$ is calculated conditioning on the estimation split data, i.e., $\EXP (L) := \EXP \LD \big\{ \loss_\sur(\cdot \con \widehat{\mu}\LOO,\widehat{e}\LOO) \big\}$. Let $n$ be the number of observations in the ERM dataset $\mathcal{D}_s$, which is proportional to $N$. 

We require $L$ to satisfy the following conditions:	
\begin{itemize}[leftmargin=1cm]
\item[\hypertarget{(C1)}{(C1)}] For all $(\bo,\ell, u)$, there exists a constant $B>0$ satisfying $0 \leq L (\bo,\ell, u ) \leq B$.
\item[\hypertarget{(C2)}{(C2)}] $L (\bo,\ell, u)$ is locally Lipschitz continuous with respect to $(\ell,u)$ for $\ell \leq u$.
\item[\hypertarget{(C3)}{(C3)}] For all $(t,\bo)$, we have $L( \bo, \ell^W, u^W) \leq L(\bo, \ell, u)$ where $ \ell^W = \ell \cdot \ind \big\{ | \ell | \leq c_0 \big\} + \text{sign}(\ell) c_0 \cdot \ind \big\{ c_0 < | \ell | \big\}$.
\item[\hypertarget{(C4)}{(C4)}] $\EXP \big[ \big\{ L \big( \bO, f_L^W, f_U^W \big) - L \big( \bO, f_L^\dagger, f_U^\dagger \big) \big\}^2 \big] \leq V \cdot \big[ \EXP \big\{ L \big( \bO, {f}_L^W, {f}_U^W \big) - L \big( \bO, f_L^\dagger, f_U^\dagger \big) \big\} \big]^v$ is satisfied for constant $v \in [0,1]$, $V \geq B^{2-v}$, and for all $f_L,f_U \in \HH$. 
\end{itemize}

With some algebra, we find that the surrogate loss function $L$ satisfies these conditions by (i) adding a sufficiently large constant to make $L$ non-negative and (ii) adding a very large constant for PDI candidates outside of the proper dose range $\R^2 \setminus [0,1]^2$. Then, we find each condition is satisfied as follows:
\begin{itemize}[leftmargin=1cm]
\item[\hyperlink{(C1)}{(C1)}] Trivial after adding a baseline constant.
\item[\hyperlink{(C2)}{(C2)}] The surrogate loss function is Lipschitz continuous for given $\epsilon$ due to its construction. 
\item[\hyperlink{(C3)}{(C3)}] Trivial after adding a penalizing constant over $\R^2 \setminus [0,1]^2$.
\item[\hyperlink{(C4)}{(C4)}] This holds with $v=0$ and $V=4B^2$.
\end{itemize}

Our ultimate goal is to show the following inequality holds 
\begin{align*}
&
\risk_\sur \LD\big( \widehat{f}_L\HOO , \widehat{f}_U\HOO \con \widehat{\mu}\LOO, \widehat{e}\LOO \big) 
-
\risk_\sur \LD\big( f_L^\dagger, f_U^\dagger \con \widehat{\mu}\LOO, \widehat{e}\LOO \big) 
\nonumber
\\
& =
\EXP \big\{ L \big( \bO, \widehat{f}_L^W, \widehat{f}_U^W \big) -   L \big( \bO, f_L^\dagger, f_U^\dagger \big) \big\}
\nonumber
\\
&
\leq
c_1 \lambda\gamma^{-d} + c_2 \gamma^\beta 
+ c_3 \Big\{ \gamma^{(1-p)(1+\epsilon)d} \lambda^p N \Big\}^{-\frac{1}{2-p}} 
+ c_4 N^{-1/2}\tau^{1/2} + c_5N^{-1} \tau \ .
\numeq \label{eq-RateofC-Full}
\end{align*}
with probability $P_{\bO}^{n}$ not less than $1-3e^{-\tau}$ where $c_1,\ldots,c_5$ does not depend on $N$. Then, taking $\gamma \asymp N^{-1/(2\beta + d)}$ and $\lambda \asymp N^{-(\beta+d)/(2\beta + d)}$, we get the following asymptotic rate with probability not less than $1-3 e^{-\tau}$:
\begin{align}						\label{eq-RateofC}
\risk_\sur \LD\big( \widehat{f}_L\HOO , \widehat{f}_U\HOO \con \widehat{\mu}\LOO, \widehat{e}\LOO \big) 
-
\risk_\sur \LD\big( f_L^*, f_U^* \con \widehat{\mu}\LOO, \widehat{e}\LOO \big) 
=
O_P \big( 	N^{- \beta/(2\beta + d)} 	\big) \ .
\end{align}
Since the estimated nuisance functions are random (depending on $\mathcal{D}_s^c$), we use $O_P$-notation instead of $O$-notation.

To show the result, we first present a formal result by extending Theorem 7.23 of \citet{SVM} to our setting. \\

\noindent\fbox{%
\parbox{\textwidth}{%
\begin{theorem}[Extension of Theorem 7.23. of \citet{SVM}] \label{thm-Thm723}
Let $L$ be a loss function having a non-negative value. Also, let $\HH$ be a Gaussian RKHS with bandwidth parameter $\gamma$ over $\mathcal{X} \subseteq \R^d$. Consider the ERM of the form
\begin{align}				\label{eq-ERM}
\big( \widehat{f}_L , \widehat{f}_U \big)
=
\argmin_{(f_L, f_U) \in \HH^{\otimes 2} }
\bigg[
\frac{1}{N} \sum_{i=1}^{N}
L
\big( \bO_i , f_L , f_U \big) 
+
\lambda
\Big\{ 
\big\| f_L \big\|_{\HH}^2
+
\lambda \big\| f_U \big\|_{\HH}^2
\Big\}
\bigg] \ .
\end{align}
Let $P_{\bO}$ and $P_{\bX}$ be distributions of $\bO$ and $\bX$, respectively. Furthermore, suppose Regularity Conditions \hyperlink{(R2)}{(R2)}, \hyperlink{(R3)}{(R3)}, and , \hyperlink{(R5)}{(R5)} hold. Assume that for fixed $n \geq 1$, there exist constant $p \in (0,1)$, and $\mathfrak{a} \geq B$ such that
\begin{align} \label{eq-Thm723}
\EXP_{\mathcal{O} \sim P_{\bO}^{|\mathcal{D}_s|} }
\Big[
e_n \big( \text{identity map}: \HH^{\otimes 2} \rightarrow L_2^{\otimes 2} ( \mu ) \big)
\Big]
\leq \mathfrak{a} \cdot i^{-\frac{1}{2p}} 
\ , 
\quad i \geq 1 \ .
\end{align}
Finally, fix $(f_{L,0},f_{U,0}) \in \HH^{\otimes 2}$ and a constant $B_0 \geq B$ such that $\big\| L (O , f_{L,0},f_{U,0}) \big\|_{\infty} \leq B_0$. Then, for all fixed $\tau>0$, and $\lambda>0$, the ERM \eqref{eq-ERM} using $\HH $ and $L$ satisfies	
\begin{align*}
&
\lambda \big\| \widehat{f}_L \big\|_{\HH}^2 + \lambda \big\| \widehat{f}_U \big\|_{\HH}^2 
+ R \big( \widehat{f}_L^W, \widehat{f}_U^W \big) - R \big( f_L^\dagger, f_U^\dagger \big)
\nonumber
\\
&
\leq
9
\big\{ \lambda \big\| f_{L,0} \big\|_{\HH}^2 + \lambda \big\| f_{U,0} \big\|_{\HH}^2 
+ R \big( f_{L,0}, f_{U,0} \big) - R \big( f_L^\dagger, f_U^\dagger \big) \big\}
\nonumber
\\
&
\hspace*{1cm}
+
K_0 \bigg\{ \frac{\mathfrak{a}^{2p} }{\lambda^p n} \bigg\}^{\frac{1}{2-p-v + vp}}
+
3 \bigg( \frac{72 V \tau}{n} \bigg)^{\frac{1}{2-v}}
+
\frac{15 B_0 \tau}{n} \ .
\end{align*}	
with probability $P_{\bO}^{n}$ not less than $1-3e^{-\tau}$, where where $K_0 \geq 1$ is a constant only depending on $p$, $c_0$, $B$, $v$, and $V$. 
\end{theorem} 
}%
}\\[0.5cm]

We establish Theorem \ref{thm-Thm723} following the proof of Theorem 7.23 of \citet{SVM}. When $a^{2p} > \lambda^p n $, we find
\begin{align*}
&
\lambda \| \widehat{f}_L \|_\HH^2
+
\lambda \| \widehat{f}_U \|_\HH^2
+
R (\widehat{f}_L^W, \widehat{f}_U^W)
-
R(f_L^\dagger, f_U^\dagger)
\\
&
\leq
\lambda \| \widehat{f}_L \|_\HH^2
+
\lambda \| \widehat{f}_U \|_\HH^2
+
\widehat{R} (\widehat{f}_L^W, \widehat{f}_U^W)
+
B
\\
&
\leq
\widehat{R} (0 , 0) + B
\leq 2B
\leq 2B \bigg(
\frac{a^{2p}}{\lambda^p n}
\bigg)^{\frac{1}{1-(1-p)(1-v)}} \ ,
\end{align*}
where $\widehat{R}(\ell, u)$ is the empirical risk $\sum_{i \in \mathcal{D}_s} L (\bO_i , \ell, u) / n$. Therefore, the claim holds whenever $K_0 \leq 2B$.

Next, we consider $a^{2p} \leq \lambda^p n $. Let $r^*$ be
\begin{align*}
r^* 
:=
\inf_{f_L , f_U \in \HH} 
\Big\{
\underbrace{
\lambda \| f_L \|_\HH^2
+
\lambda \| f_U \|_\HH^2
}_{\Gamma(f_L,f_U)}
+
R (f_L^W, f_U^W)
-
R(f_L^\dagger, f_U^\dagger)
\Big\} \ ,
\end{align*}
and we further define the following sets for $r> r^*$:
\begin{align*}
&
(\HH_L \otimes \HH_U)_r
:= 
\big\{ f_L, f_U \in \HH \cond 
\Gamma(f_L,f_U) + R(f_L^W, f_U^W) - R(f_L^\dagger, f_U^\dagger) \leq r
\big\}
\\
&
\mathcal{L}_r :=
\big\{
L ( \bO, f_L^W , f_U^W ) - L (\bO, f_L^\dagger, f_U^\dagger ) \cond
(f_L, f_U) \in (\HH_L \otimes \HH_U)_r 
\big\}\ .
\end{align*}
Let $B_\HH$ be the closed unit ball of $\HH$. For $\bflu \in (\HH_L \otimes \HH_U)_r$, we have
\begin{align*}
\Gamma(f_L, f_U) 
&
= 
\lambda \| f_L \|_\HH^2
+
\lambda \| f_U \|_\HH^2
\leq
\lambda \| f_L \|_\HH^2
+
\lambda \| f_U \|_\HH^2
+
R(f_L^W, f_U^W) - R(f_L^\dagger, f_U^\dagger) \leq r \ .
\end{align*}
This implies $\bflu \in (\HH_L \otimes \HH_U)_r$ also belong to $ ( (r/\lambda)^{1/2} B_\HH)^{\otimes 2}$.

To proceed, we adopt the approach used in the proof of Lemma 7.17 of \citet{SVM} to our setting. Given that $L \circ (\HH_L \otimes \HH_U)_r^W \subseteq L_2(P_{\bO})$, we have the relationship for (dyadic) entropy number $e_n$:
\begin{align*}
e_n \big( \mathcal{L}_r , \| \cdot \|_{L_2(P_{\bO})} \big) 
\leq
e_n \big( L \circ (\HH_L \otimes \HH_U)_r , \| \cdot \|_{L_2(P_{\bO})} \big) \ . 
\end{align*}
Now let $\{ \bflu_1,\ldots,\bflu_{2^{n-1}} \}$ $(\bflu=(f_{L,i}, f_{U,i}))$ be an $\epsilon$-net of $(\HH_L \otimes \HH_U)_r$ with respect to $\| \cdot \|_{L_2(P_{\bX})}$ where
\begin{align*}
\| \bflu \|_{L_2(P_{\bX})}^2
= 
\int \{ f_L(\bX) \}^2 + \{ f_U(\bX) \}^2 \, dP_{\bX} \ .
\end{align*}
See Section A.5.6 of \citet{SVM} for details on the dyadic entropy number. This means that, for $\bflu \in (\HH_L \otimes \HH_U)_r$, there exists an $i \in \{1,\ldots,2^{M-1} \}$ such that $\| \bflu - \bflu \|_{L_2(P_{\bX})} \leq \epsilon$. Thus, 
\begin{align*}
\Big\|
L \circ \bflu^W - L \circ \bflu^W 
\Big\|_{L_2(P_{\bO})}^2
& =
\int \Big|
L \big( \bO, f_L^W(\bX) , f_U^W(\bX) \big)
-
L \big( \bO, f_{L,i}^W(\bX) , f_{U,i}^W(\bX) \big)
\Big|^2
\, dP_{\bO}
\\
&
\leq
L_c ^2
\Big[
\int \Big| 
f_L^W(\bX) - f_{L,i}^W(\bX)
\Big|^2 \, dP_{\bX}
+
\int \Big| 
f_U^W(\bX) - f_{U,i}^W(\bX)
\Big|^2 \, dP_{\bX}
\Big]
\\
&
\leq
4 L_c^2 \epsilon^2 \ .
\end{align*}
Therefore, $\{ L \circ \bflu_1^W,\ldots,L \circ \bflu_{2^{n-1}}^W \}$ is an $(2L_c\epsilon)$-net of $L \circ (\HH_L \otimes \HH_U)_r$. As a consequence,
\begin{align*}
e_n \big( \mathcal{L}_r , \| \cdot \|_{L_2(P_{\bO})} \big) 
\leq 
2L_c
e_n \big( (\HH_L \otimes \HH_U)_r , \| \cdot \|_{L_2(P_{\bX})} \big) \ .
\end{align*}
Taking expectation (only over the data used in ERM which is $\mathcal{D}_s$), we have
\begin{align*}	
&
\EXP_{\mathcal{O} \sim P_{\bO}^{|\mathcal{D}_s|} } 
\Big[
e_n \big( \text{identity map}: \mathcal{L}_r \rightarrow L_2( \mathcal{O} ) \big)
\Big]
\\			
&
\leq
2 L_c 
\cdot
\EXP_{\mathcal{X} \sim P_{\bX}^{|\mathcal{D}_s|}} 
\Big[
\underbrace{
e_n \big( \text{identity map}: (\HH_L \otimes \HH_U)_r \rightarrow L_2^{\otimes 2} ( \mathcal{X} ) \big)
}_{ = \text{\HT{(Term 1)}} }
\Big] \ .
\end{align*}
Here, $L_2^{\otimes 2}(\mathcal{X}) = \{ (f,g) \cond \int f^2(\bX) \, dP_{\bX} + \int g^2(\bX) \, dP_{\bX} < \infty \}$.

The quantity \HL{(Term 1)} can be further upper bounded using Theorem 7.34 of \citet{SVM}, which we state below for completeness:\\

\noindent\fbox{%
\parbox{\textwidth}{%
\textbf{Theorem 7.34. }(\textit{Entropy Numbers for Gaussian Kernels}; \citet{SVM}) Let $\mu$ be a distribution on $\R^d$ having tail exponent $\tau \in (0, \infty]$, and $\HH_{\gamma}(\R^d)$ be the RKHS endowed with the Gaussian kernel with bandwidth $\gamma$. Then, for all $\epsilon>0$ and $d/(d+\tau) < p < 1$, there exists a constant $c \geq 1$ such that
\begin{align*}
e_n (\text{identity map}: \HH_{\gamma}(\R^d) \rightarrow L_2(\mu) ) \leq c_{\epsilon,p} \gamma^{- \frac{(1-p)(1+\epsilon) d}{2p}} i^{-\frac{1}{2p}} 
\end{align*}
for all $i \geq 1$ and $\gamma \in (0,1]$.
}%
}\\[0.5cm]

Since $\HH$ is induced by the Gaussian kernel, we have
\begin{align}						\label{eq-SVM-(1)RHS}
\text{\HL{(Term 1)}} &= e_n \big( \text{identity map}: (\HH_L \otimes \HH_U)_r \rightarrow L_2^{\otimes 2} ( \mathcal{X} ) \big)
\nonumber
\\
&
\leq
e_n \big( \text{identity map}: ( (r/\lambda)^{1/2} B_\HH)^{\otimes 2} \rightarrow \ell_\infty^{\otimes 2} ( (r/\lambda)^{1/2} B_\HH) \big)
\nonumber
\\
&
\leq
\bigg( \frac{r}{\lambda} \bigg)^{1/2}
e_n \big( \text{identity map}: (B_\HH)^{\otimes 2} \rightarrow \ell_\infty^{\otimes 2} ( B_\HH) \big) \ .
\end{align}
From Lemma 6.21 and Excercise 6.8 of \citet{SVM}, we have
\begin{align} \label{eq-upperbound}
\nonumber
& 
e_n \big( \text{identity map}: (B_\HH)^{\otimes 2} \rightarrow \ell_\infty^{\otimes 2} ( B_\HH) \big) \lesssim n^{-1/q} \text{ for some $q>0$}
\\
&
\Leftrightarrow \quad 
\log \mathcal{N} ( \epsilon, (B_\HH)^{\otimes 2}, \| \cdot \|_\infty ) \lesssim (1/\epsilon)^{q} \text{ for some $q>0$}
\end{align}
where $\mathcal{N} (\epsilon, \F ,\| \cdot \|)$ is the $\epsilon$-covering number of a space $\F $ with respect to a metric $\| \cdot \|$; see Definition 6.19 of \citet{SVM} for details. Therefore, to derive the upper bound of the right-hand side \eqref{eq-SVM-(1)RHS}, it suffices to characterize an upper bound of $\log \mathcal{N} ( \epsilon, (B_\HH)^{\otimes 2}, \| \cdot \|_\infty )$. In order to do so, let $C_b^m( S ) $ be all continuous functions from $S \subseteq \R^d$ with bounded $m$-th derivative. We further introduce Theorem 2.7.1 of \citet{VW1996} and an extension of Lemma 7 of \citet{Haussler1992} which are provided below for completeness: \\

\noindent\fbox{%
\parbox{\textwidth}{%
\textbf{Theorem 2.7.1.} (\citet{VW1996}) Let $\mathcal{X}$ be a bounded, convex subset of $\R^d$ with nonempty interior. There exists a constant $K$ depending only on $m$ and $d$ such that 
\begin{align} \label{eq-VW271}
\log \mathcal{N}( \epsilon, C_b^m(S), \| \cdot \|_\infty ) \lesssim (1/\epsilon)^{d/m} \ .
\end{align} 
}%
}\\[0.5cm]

\noindent\fbox{%
\parbox{\textwidth}{%
\textbf{Extension of Lemma 7} (\citet{Haussler1992}) 
\begin{align} \label{eq-Haussler}
\log \mathcal{N}(\epsilon, C_b^m(S)^{\otimes 2} , \| \cdot \|_{\infty}) 
\leq 
2 \log \mathcal{N}(\epsilon, C_b^m(S) , \| \cdot \|_{\infty})  \ .
\end{align}
}%
}\\[0.5cm]

We briefly show \eqref{eq-Haussler}. Let $U$ be an $\epsilon$-net of $C_b^m(S)$ having $n_\epsilon$ elements. Let $\mathcal{U} = U^{\otimes 2} = \{(u_1,u_2) \cond u_1, u_2 \in U \}$ having $n_\epsilon^2$ by construction. It suffices to show that $\mathcal{U}$ is an $\epsilon$-net of $C_b^m(S)^{\otimes 2}$. We arbitrarily take any function $(g_1,g_2) \in C_b^m(S)^{\otimes 2}$. Then, we can find $u_j \in U$ such that $\| g_j - u_j \|_{\infty} \leq \epsilon$, which implies
\begin{align*}
\big\| (g_1,g_2) - (u_1,u_2) \big\|_{\infty}
= \max_{i=1,2} \big\| g_i - u_i \big\|_{\infty}
\leq \epsilon \ .
\end{align*}
Therefore, $\mathcal{N}(\epsilon, C_b^m(S)^{\otimes 2} , \| \cdot \|_{\infty}) \leq | \mathcal{U} | = n_\epsilon^2 = \mathcal{N}(\epsilon, C_b^m(S) , \| \cdot \|_{\infty}) $.

Combining equations \eqref{eq-SVM-(1)RHS}-\eqref{eq-Haussler}, we establish that $\text{\HL{(Term 1)}} \lesssim (r/\gamma)^{1/2} n^{-m/d}$, which is essentially the same as the result of Theorems 6.26, 6.27, 7.33 of \citet{SVM}. Therefore, the remainder of the proof of Theorem 7.34 of \citet{SVM} applies to our setting. Therefore, from Theorem 7.34 of \citet{SVM}, we get
\begin{align*}
\text{\HL{(Term 1)}}
&
=
e_n \big( \text{identity map}: (B_\HH)^{\otimes 2} \rightarrow \ell_\infty^{\otimes 2} ( B_\HH) \big)
\\
&
\leq 
e_n \big( \text{identity map}: \HH^{\otimes 2} \rightarrow L_2^{\otimes 2} ( \mu ) \big)
\\
&
\leq
c_{\epsilon,p} \gamma^{ - \frac{(1-p)(1+\epsilon)d}{2p} } i^{-\frac{1}{2p}} \ ,
\end{align*}
for all $i \geq 1$ and $\gamma \in (0, 1]$. which satisfies condition \eqref{eq-Thm723} with $\mathfrak{a}=c_{\epsilon,p} \gamma^{ - \frac{(1-p)(1+\epsilon)d}{2p} }$. This further implies the following result:
\begin{align*}	
&
\EXP_{\mathcal{O} \sim P_{\bO}^{|\mathcal{D}_s|} }
\Big[
e_n \big( \text{identity map}: \mathcal{L}_r \rightarrow L_2( \mathcal{O} ) \big)
\Big]
\\
&
\leq
2 L_c 
\cdot
\EXP_{\mathcal{X} \sim P_{\bX}^{ |\mathcal{D}_s| } }
\Big[
\underbrace{
e_n \big( \text{identity map}: (\HH_L \otimes \HH_U)_r \rightarrow L_2^{\otimes 2} ( \mathcal{X} ) \big)
}_{\text{\HL{(Term 1)}}}
\Big]
\\
&
\leq 2L_c \bigg( \frac{r}{\lambda} \bigg)^{1/2} \cdot \mathfrak{a} \cdot i^{-\frac{1}{2p}} \quad \text{for any $i \geq 1$} \ .
\end{align*}

The rest of the proof of Theorem \ref{thm-Thm723} is the same as that in \citet{SVM}, which results in
\begin{align}
&
\lambda \big\| \widehat{f}_L \big\|_{\HH}^2 + \lambda \big\| \widehat{f}_U \big\|_{\HH}^2 
+ R \big( \widehat{f}_L^W, \widehat{f}_U^W \big) - R \big( f_L^\dagger, f_U^\dagger \big)
\nonumber
\\
&
\leq
9
\big\{ \underbrace{ \lambda \big\| f_{L,0} \big\|_{\HH}^2 + \lambda \big\| f_{U,0} \big\|_{\HH}^2 
+ R \big( f_{L,0}, f_{U,0} \big) - R \big( f_L^\dagger, f_U^\dagger \big)}_{=B(f_{L,0},f_{U,0})} \big\}
\nonumber
\\
&
\hspace*{1cm}
+
K_0 \bigg\{ \frac{ \gamma^{ - (1-p)(1+\epsilon) d} }{\lambda^p n} \bigg\}^{\frac{1}{2-p}}
+
3 \bigg( \frac{72 V \tau}{n} \bigg)^{\frac{1}{2-v}}
+
\frac{15 B_0 \tau}{n}
\nonumber 
\\
&
\stackrel{\text{take $v=0$}}{=} 
B(f_{L,0},f_{U,0})
+
K_0 \bigg\{ \frac{ \gamma^{ - (1-p)(1+\epsilon) d} }{\lambda^p n} \bigg\}^{\frac{1}{2-p}}
+
36 \sqrt{2} B \sqrt{ \frac{\tau}{n} }
+
15 B \frac{ \tau}{n}
\ .
\label{eq-SVM-proof2}
\end{align}

Since the above result holds for any $(f_{L,0}, f_{U,0}) \in \HH^{\otimes 2}$, we can further bound $B(f_{L,0},f_{U,0})$ by carefully choosing $(f_{L,0},f_{U,0})$ in a way presented in \citet{SVM2013}. We first define a function $Q_{r,\gamma} : \R^d \rightarrow \R$ as
\begin{align}								\label{eq-Qfunction}
Q_{r,\gamma} (\bz)
=
\sum_{j=1}^r
{r \choose j} (-1)^{1-j} \frac{1}{j^d} \bigg( \frac{2}{\gamma^2} \bigg)^{\frac{d}{2} } \mathcal{K}_{j\gamma/\sqrt{2} }(\bz)
\ , \
\mathcal{K}_\gamma(\bz) 
=
\exp \big\{ - \gamma^{-2} \big\| \bz \big\|_2^2 \big\} \ .
\end{align}
for $r \in \{1,2,\ldots\}$ and $\gamma > 0$. If $(f_L^\dagger, f_U^\dagger) \in L_2(\R^d) \cap L_\infty(\R^d)$, we can define $(f_{L,0}, f_{U,0})$ by convolving $Q_{r,\gamma}$ with $(f_L^\dagger, f_U^\dagger)$ as follows \citep{SVM2013}.
\begin{align*}
f_{L,0} (\bX)
= \big( Q_{r,\gamma} * f_L^* \big)(\bX)
= \int_{\R^d} Q_{r,\gamma}(\bX - \bz) f_L^*(\bz) \, d\bz \ .
\end{align*}
Next, we introduce Theorems 2.2 and 2.3 of \citet{SVM2013}.\\

\noindent\fbox{%
\parbox{\textwidth}{%
\textbf{Theorem 2.2. of \citet{SVM2013}} Let us fix some $q \in [1,\infty)$. Furthermore, assume that $P_{\bX}$ is a distribution on $\R^d$ that has a Lebesgue density $p_{\bX} \in L_p(\R^d)$ for some $p \in [1,\infty]$. Let $f: \R^d \rightarrow \R$ be such that $f \in L_q(\R^d) \cap L_\infty(\R^d)$. Then, for $r \in \{1,2,\ldots\}$, $\gamma > 0$, and $s \geq 1$ with $1=s^{-1} + p^{-1}$, we have
\begin{align*}
\big\| Q_{r,\gamma} * f - f \big\|_{L_q(P_{\bX})}^q 
\leq
C_{r,q} \cdot \big\| p_{\bX} \big\|_{L_p(\R^d)} \cdot \omega_{r,L_{qs}(\R^d)}^q ( f, \gamma/2)
\end{align*}
where $C_{r,q}$ is a constant only depending on $r$ and $q$.
}%
}\\[0.5cm]

\noindent\fbox{%
\parbox{\textwidth}{%
\textbf{Theorem 2.3. of \citet{SVM2013}} Let $f \in L_2(\R^d)$, $\HH$ be the RKHS of the Gaussian kernel with parameter $\gamma>0$, and $Q_{r,\gamma}$ be defined by \eqref{eq-Qfunction} for a fixed $r \in \{1,2,\ldots\}$. Then we have $Q_{r,\gamma} * f \in \HH$ with
\begin{align*}
\big\| Q_{r,\gamma} * f \big\|_{\HH} \leq \big( \gamma \sqrt{\pi} \big)^{ -\frac{d}{2} } \big( 2^r -1 \big) \big\| f \big\| _{L_2(\R^d) } \ .
\end{align*}
Moreover, if $f \in L_\infty(\R^d)$, we have $| Q_{r,\gamma} * f | \leq (2^r - 1) \big\| f \big\|_{L_\infty(\R^d) }$.
}%
}\\[0.5cm]

\noindent

As a result, we obtain
\begin{align}						\label{eq-SVM-proof3}
&
\lambda \big\| f_{L,0} \big\|_{\HH}^2 + \lambda \big\| f_{U,0} \big\|_{\HH}^2 
+ R \big( f_{L,0}, f_{U,0} \big) - R \big( f_L^\dagger, f_U^\dagger \big)
\\
\nonumber
&
=
\lambda \big\| Q_{r,\gamma} * f_L^\dagger \big\|_{\HH}^2 + \lambda \big\| Q_{r,\gamma} * f_U^\dagger \big\|_{\HH}^2 
+ R \big( Q_{r,\gamma} * f_L^\dagger , Q_{r,\gamma} * f_U^\dagger \big) - R \big( f_L^\dagger, f_U^\dagger \big)
\\
\nonumber
& 
\leq
\lambda \big( \gamma \sqrt{\pi} \big)^{ - d } \big( 2^r -1 \big)^2 \big\{ \big\| f_L^\dagger \big\| _{L_2(\R^d) }^2 + \big\| f_U^\dagger \big\| _{L_2(\R^d) }^2 \big\}
+ R \big( Q_{r,\gamma} * f_L^\dagger , Q_{r,\gamma} * f_U^\dagger \big) - R \big( f_L^\dagger, f_U^\dagger \big)
\\
\nonumber
& 
\leq
\lambda \big( \gamma \sqrt{\pi} \big)^{ - d } \big( 2^r -1 \big)^2 \big\{ \big\| f_L^\dagger \big\| _{L_2(\R^d) }^2 + \big\| f_U^\dagger \big\| _{L_2(\R^d) }^2 \big\} 
\nonumber
\\
\nonumber
& \hspace*{1cm}
+ B' \cdot \big\| Q_{r,\gamma} * f_L^\dagger - f_L^\dagger \big\|_{L_1(P_{\bX})}
+ B' \cdot \big\| Q_{r,\gamma} * f_U^\dagger - f_U^\dagger \big\|_{L_1(P_{\bX})}
\\
\nonumber
& 
\leq
\lambda \big( \gamma \sqrt{\pi} \big)^{ - d } \big( 2^r -1 \big)^2 \big\{ \big\| f_L^\dagger \big\| _{L_2(\R^d) }^2 + \big\| f_U^\dagger \big\| _{L_2(\R^d) }^2 \big\} 
\nonumber
\\
\nonumber
& \hspace*{1cm}
+ B' \cdot C_{r,1} \cdot \big\| p_{\bX} \big\|_{L_\infty (\R^d)} \cdot \big\{ \omega_{r,L_1(\R^d)} ( f_L^\dagger, \gamma/2)+ \omega_{r,L_1(\R^d)} ( f_U^\dagger, \gamma/2) \big\}
\\
\nonumber
&
\leq
\lambda \big( \gamma \sqrt{\pi} \big)^{ - d } \big( 2^r -1 \big)^2 \big\{ \big\| f_L^\dagger \big\| _{L_2(\R^d) }^2 + \big\| f_U^\dagger \big\| _{L_2(\R^d) }^2 \big\} 
+ B' c' \cdot C_{r,1} \cdot \big\| p_{\bX} \big\|_{L_\infty (\R^d)} \gamma^{\beta} \ .
\end{align}
The first equality is from the construction of $(f_{L,0},f_{U,0})$. The first inequality is from Theorem 2.3 of \citet{SVM2013}. The second inequality is from Lipschitz continuity of $L$. The third inequality is from Theorem 2.2 of \citet{SVM2013} with $q=s=1$ and $p=\infty$. The last inequality holds for some constant $c'$ since $f_L^\dagger, f_U^\dagger \in \mathcal{B}_{1,\infty}^\beta (\R^d)$ implies $\omega_{r,L_1(\R^d)}(f^\dagger, \gamma/2) \leq c' \gamma^{\beta}$ from the definition of a Besov space.

Combining the results in \eqref{eq-SVM-proof2} and \eqref{eq-SVM-proof3}, we have the following result.
\begin{align*}
&
R \big( \widehat{f}_L^W, \widehat{f}_U^W \big) - R \big( f_L^\dagger, f_U^\dagger \big)
\\
&
\leq
9 \big[ \lambda \big( \gamma \sqrt{\pi} \big)^{ - d } \big( 2^r -1 \big)^2 \big\{ \big\| f_L^\dagger \big\| _{L_2(\R^d) }^2 + \big\| f_U^\dagger \big\| _{L_2(\R^d) }^2 \big\}
+ B' c' \cdot C_{r,1} \cdot \big\| p_{\bX} \big\|_{L_\infty(\R^d)} \gamma^{\beta} \big] 
\\
& \hspace*{2cm}
+
K_0 \bigg\{ \frac{ \gamma^{ - (1-p)(1+\epsilon) d} }{\lambda^p n} \bigg\}^{\frac{1}{2-p}}
+
36 \sqrt{2} B \sqrt{ \frac{\tau}{n} }
+
15 B \frac{ \tau}{n}
\\
&
\leq
c_1 \lambda\gamma^{-d} + c_2 \gamma^\beta 
+ c_3 \Big\{ \gamma^{(1-p)(1+\epsilon)d} \lambda^p N \Big\}^{-\frac{1}{2-p}} 
+ c_4 N^{-1/2}\tau^{1/2} + c_5N^{-1} \tau \ .
\end{align*}
The latter inequality is from the fact that $n$ is proportional to $N$. This establishes \eqref{eq-RateofC-Full}.

\subsubsection{Upper bound of (D)} \label{sec-ub (D)}

Since $(f_L^\#, f_U^\#)$ is the minimizer of $\risk_\sur \LD(f_L, f_U \con \mu^*, e^*)$, we find 
\begin{align*}
    &
    \risk_\sur\LD
    (f_L^\dagger, f_U^\dagger \con \widehat{\mu}\LOO, \widehat{e}\LOO)
    \\
    &
    =
    \risk_\sur\LD
    (f_L^\dagger, f_U^\dagger \con \widehat{\mu}\LOO, \widehat{e}\LOO)
    +
    \risk_\sur\LD
    (f_L^\dagger, f_U^\dagger \con \mu^*, e^*)
    -
    \risk_\sur\LD
    (f_L^\dagger, f_U^\dagger \con \mu^*, e^*)
    \\
    &
    \geq 
    \risk_\sur\LD
    (f_L^\dagger, f_U^\dagger \con \widehat{\mu}\LOO, \widehat{e}\LOO)
    +
    \risk_\sur\LD
    (f_L^\#, f_U^\# \con \mu^*, e^*)
    -
    \risk_\sur\LD
    (f_L^\dagger, f_U^\dagger \con \mu^*, e^*) \ .
    \numeq \label{eq-D-1}
\end{align*}
Likewise, $(f_L^\dagger, f_U^\dagger)$ is the minimizer of $\risk_\sur \LD(f_L, f_U \con \widehat{\mu}\LOO, \widehat{e}\LOO)$, we find 
\begin{align*}
    &
    \risk_\sur\LD
    (f_L^\#, f_U^\# \con \mu^*, e^*)
    \\
    &
    =
    \risk_\sur\LD
    (f_L^\#, f_U^\# \con \mu^*, e^*)
    +
    \risk_\sur\LD
    (f_L^\#, f_U^\# \con \widehat{\mu}\LOO, \widehat{e}\LOO)
    -
    \risk_\sur\LD
    (f_L^\#, f_U^\# \con \widehat{\mu}\LOO, \widehat{e}\LOO)
    \\
    &
    \geq 
    \risk_\sur\LD
    (f_L^\#, f_U^\# \con \mu^*, e^*)
    +
    \risk_\sur\LD
    (f_L^\dagger, f_U^\dagger \con \widehat{\mu}\LOO, \widehat{e}\LOO)
    -
    \risk_\sur\LD
    (f_L^\#, f_U^\# \con \widehat{\mu}\LOO, \widehat{e}\LOO) \ .
    \numeq \label{eq-D-2}
\end{align*}
Results \eqref{eq-D-1} and \eqref{eq-D-2} imply that
\begin{align*}
    &
    \risk_\sur\LD
    (f_L^\dagger, f_U^\dagger \con \widehat{\mu}\LOO, \widehat{e}\LOO)
    -
    \risk_\sur\LD
    (f_L^\dagger, f_U^\dagger \con \mu^*, e^*)
    \\
    &
    \leq 
    \risk_\sur\LD
    (f_L^\dagger, f_U^\dagger \con \widehat{\mu}\LOO, \widehat{e}\LOO)
    -
    \risk_\sur\LD
    (f_L^\#, f_U^\# \con \mu^*, e^*)
    \\
    &
    \leq 
    \risk_\sur\LD
    (f_L^\#, f_U^\# \con \widehat{\mu}\LOO, \widehat{e}\LOO)
    -
    \risk_\sur\LD
    (f_L^\#, f_U^\# \con \mu^*, e^*) \ .
\end{align*}
Therefore, we find
\begin{align*}
    &
    \big| \risk_\sur\LD
    (f_L^\dagger, f_U^\dagger \con \widehat{\mu}\LOO, \widehat{e}\LOO)
    -
    \risk_\sur\LD
    (f_L^\#, f_U^\# \con \mu^*, e^*) \big| 
    \\ 
    &
    \leq 
    \max 
    \bigg\{ 
    \begin{array}{l} 
    \big| \risk_\sur\LD
    (f_L^\#, f_U^\# \con \widehat{\mu}\LOO, \widehat{e}\LOO)
    -
    \risk_\sur\LD
    (f_L^\#, f_U^\# \con \mu^*, e^*) \big| , \\
    \big| \risk_\sur\LD
    (f_L^\dagger, f_U^\dagger \con \widehat{\mu}\LOO, \widehat{e}\LOO)
    -
    \risk_\sur\LD
    (f_L^\dagger, f_U^\dagger \con \mu^*, e^*) \big| 
    \end{array}
    \bigg\} \ . 
    \numeq \label{eq-D-3}
\end{align*}
In addition, we remark that $f_L^\#(\bX) \leq f_U^\#(\bX)$ and $f_L^\dagger(\bX) \leq f_U^\dagger(\bX)$ for all $\bX$, because $\mathcal{L}_{\sur}$ is defined in a way that $\mathcal{L}_{\sur}(\bO,f_L,f_U \con \mu,e) \leq \mathcal{L}_{\sur}(\bO,f_L',f_U' \con \mu,e)$ for any pair $(f_L,f_U)$ and $(f_L',f_U')$ satisfying $f_L(\bX) \leq f_U(\bX)$ and $f_L(\bX) > f_U(\bX)$. Consequently, following the steps in Section \ref{sec-ub (B)}, we establish
\begin{align*}
    &
    \big| \risk_\sur\LD
    (f_L^\#, f_U^\# \con \widehat{\mu}\LOO, \widehat{e}\LOO) - \risk\LD
    (f_L^\#, f_U^\# \con \widehat{\mu}\LOO, \widehat{e}\LOO)
    \big| \leq c_e \cdot \epsilon
    \ ,
    \\
    &
    \big| \risk_\sur\LD
    (f_L^\#, f_U^\# \con \mu^*, e^*) - \risk\LD
    (f_L^\#, f_U^\# \con \mu^*, e^*)
    \big| \leq c_e \cdot \epsilon
    \ ,
    \\
    &
    \big| \risk_\sur\LD
    (f_L^\dagger, f_U^\dagger \con \widehat{\mu}\LOO, \widehat{e}\LOO) - \risk\LD
    (f_L^\dagger, f_U^\dagger \con \widehat{\mu}\LOO, \widehat{e}\LOO)
    \big| \leq c_e \cdot \epsilon
    \ ,
    \\
    &
    \big| \risk_\sur\LD
    (f_L^\dagger, f_U^\dagger \con \mu^*, e^*) - \risk\LD
    (f_L^\dagger, f_U^\dagger \con \mu^*, e^*)
    \big| \leq c_e \cdot \epsilon \ .
    \numeq \label{eq-D-4}
\end{align*}
Results \eqref{eq-D-3} and \eqref{eq-D-4} implies that
\begin{align*}
    &
    \big| \risk_\sur\LD
    (f_L^\dagger, f_U^\dagger \con \widehat{\mu}\LOO, \widehat{e}\LOO)
    -
    \risk_\sur\LD
    (f_L^\#, f_U^\# \con \mu^*, e^*) \big| 
    \\ 
    &
    \leq 
    \max 
    \bigg\{ 
    \begin{array}{l} 
    \big| \risk\LD
    (f_L^\#, f_U^\# \con \widehat{\mu}\LOO, \widehat{e}\LOO)
    -
    \risk\LD
    (f_L^\#, f_U^\# \con \mu^*, e^*) \big| , \\
    \big| \risk\LD
    (f_L^\dagger, f_U^\dagger \con \widehat{\mu}\LOO, \widehat{e}\LOO)
    -
    \risk\LD
    (f_L^\dagger, f_U^\dagger \con \mu^*, e^*) \big| 
    \end{array}
    \bigg\} 
    + 2 c_e \cdot \epsilon
    \ . 
    \numeq \label{eq-D-5}
\end{align*}
Following the derivation for \eqref{eq-RateofA}, we find the following result holds with probability at least $1-\Delta_N$:
\begin{align*}
    &
    \big| \risk\LD
    (f_L, f_U \con \widehat{\mu}\LOO, \widehat{e}\LOO)
    -
    \risk\LD
    (f_L, f_U \con \mu^*, e^*) \big|
    \\
    &
    \lesssim \big\| \widehat{e}\LOO - e^* \big\|_{P,2} \big\| \widehat{\mu}\LOO -\mu^* \big\|_{P,2}
    \\
    &
    \lesssim 
    N^{-r_{\mu} - r_{e}} \ .
\end{align*}
Therefore, \eqref{eq-D-5} reduces to
\begin{align*}
    \big| \risk_\sur\LD
    (f_L^\dagger, f_U^\dagger \con \widehat{\mu}\LOO, \widehat{e}\LOO)
    -
    \risk_\sur\LD
    (f_L^\#, f_U^\# \con \mu^*, e^*) \big|  
    \leq 
    C
    N^{-r_{\mu} - r_{e}}
    +
    C'
    \epsilon
    \numeq
    \label{eq-RateofD}
\end{align*}
for some constants $C$ and $C'$  with probability at least $1-\Delta_N$.

\subsubsection{Upper bound of (E)}

Following the steps in Section \ref{sec-ub (B)}, we establish
\begin{align*} 
    &
    \big| \risk_\sur\LD
    (f_L^*, f_U^* \con \mu^*,e^*) - \risk\LD
    (f_L^*, f_U^* \con \mu^*,e^*)
    \big| \leq c_e \cdot \epsilon  \ .
    \numeq \label{eq-RateofE}
\end{align*}

\subsection{Proof of Theorem \ref{thm-convergence f}}

Suppose that the result in Theorem \ref{thm-ExcessRisk} holds, which occurs with a probability not less than $1-3e^{-\tau}-\Delta_N$. We denote this event by $E_{\text{Thm \ref{thm-ExcessRisk}}}$. 

For readability, we restate the loss function in \eqref{eq-Loss}:
\begin{align*}
&
\loss ( \bO, f_L, f_U \con \mu^*, e^*)
=
\left\{
\begin{array}{ll}
\loss^{(1)} ( \bO, f_L, f_U \con \mu^*, e^*) & \text{if }f_L(\bX) \leq f_U(\bX) \\
C_\loss
& \text{if } f_L(\bX) > f_U(\bX)
\end{array}
\right. 
\ ,
\\
&
\loss^{(1)} ( \bO, f_L, f_U \con \mu^*, e^*)
=
\left[
\begin{array}{l}
\big\{ \mu(A, \bX) - R \big\} \ind \{ A \in [ f_L(\bX), f_U(\bX)] \} / e^*(A \cond \bX) 
\\
+
\int \big\{ \alpha - \mu^*(a, \bX) \big\} \ind \{ a \in [ f_L(\bX), f_U(\bX)] \} \, da 
\end{array}
\right]
\ ,
\end{align*}
where $C_\loss=2 \sup_{\bO, f_L, f_U} \big| \loss^{(1)}(\bO, f_L, f_U \con \mu^*, e^*) \big|$.

Suppose that $\Pr\LD \{ \widehat{f}_L\HOO(\bX) > \widehat{f}_U\HOO (\bX) \} = \int \ind \{ \widehat{f}_L\HOO(\bX) > \widehat{f}_U\HOO (\bX) \} \, dP_{\bX}  \rightarrow \underline{c}$ for a positive constant $\underline{c}$ as $N \rightarrow \infty$. Then, we find
\begin{align*}
    &
    \risk \LD( \widehat{f}_L\HOO, \widehat{f}_U\HOO \con \mu^*,e^*)
    -
    \risk \LD(  f_L^*,f_U^* \con \mu^*,e^*) 
    \\
    &
    = 
    \int 
    \big\{
    \loss (\bO, \widehat{f}_L\HOO, \widehat{f}_U\HOO \con \mu^*,e^*)
    -
    \loss (\bO, f_L^*,f_U^* \con \mu^*,e^*) 
    \big\}
    \, dP_{\bO}
    \\
    &
    =
    \int  
    \ind \big\{ \widehat{f}_L\HOO(\bX) > \widehat{f}_U\HOO (\bX) \big\}
    \big\{ 
    \loss (\bO, \widehat{f}_L\HOO, \widehat{f}_U\HOO \con \mu^*,e^*)
    -
    \loss (\bO, f_L^*,f_U^* \con \mu^*,e^*)
    \big\}
     \, dP_{\bO}
    \\
    &
    \hspace*{1cm}
    +
    \int 
    \ind \big\{ \widehat{f}_L\HOO(\bX) \leq  \widehat{f}_U\HOO (\bX) \big\}
    \big\{ 
    \loss (\bO, \widehat{f}_L\HOO, \widehat{f}_U\HOO \con \mu^*,e^*)
    -
    \loss (\bO, f_L^*,f_U^* \con \mu^*,e^*)
    \big\}
     \, dP_{\bO}
     \\
    &
    \stackrel{(a)}{\geq}
    \int  
    \ind \big\{ \widehat{f}_L\HOO(\bX) > \widehat{f}_U\HOO (\bX) \big\}
    \big\{ 
    \loss (\bO, \widehat{f}_L\HOO, \widehat{f}_U\HOO \con \mu^*,e^*)
    -
    \loss (\bO, f_L^*,f_U^* \con \mu^*,e^*)
    \big\}
     \, dP_{\bO} 
     \\
     &
     =
    \int  
    \ind \big\{ \widehat{f}_L\HOO(\bX) > \widehat{f}_U\HOO (\bX) \big\}
    \big\{ 
    C_\loss
    -
    \loss (\bO, f_L^*,f_U^* \con \mu^*,e^*)
    \big\}
     \, dP_{\bO} 
     \\ 
     &
     \stackrel{(b)}{=}
     0.5 C_{\loss}
    \int  
    \ind \big\{ \widehat{f}_L\HOO(\bX) > \widehat{f}_U\HOO (\bX) \big\} \, dP_{\bX}
    \\
    &
    \rightarrow 0.5 \underline{c} C_{\loss}
    > 0 \ .
\end{align*}
Inequality $\stackrel{(a)}{\geq}$ holds from the fact that $(f_L^*,f_U^*)$ is the minimizer of $\int \loss (f_L,f_U \con \mu^*,e^*)$ given that $f_L \leq f_U$. Equality $\stackrel{(b)}{=}$ holds from the fact that $C_\loss=2 \sup_{\bO, f_L, f_U} \big| \loss^{(1)}(\bO, f_L, f_U \con \mu^*, e^*) \big|$. Therefore, $\risk \LD( \widehat{f}_L\HOO, \widehat{f}_U\HOO \con \mu^*,e^*)
    -
    \risk \LD(  f_L^*,f_U^* \con \mu^*,e^*) $ converges to a positive value as $N$ grows to infinity. This contradicts the result of Theorem \ref{thm-ExcessRisk}. Therefore, we conclude that
    \begin{align*}
         &
         \int \ind \{ \widehat{f}_L\HOO(\bX) > \widehat{f}_U\HOO (\bX) \} \, dP_{\bX}  \rightarrow 0 \quad \text{ as } \quad 
         N \rightarrow \infty \ .
         \numeq \label{eq-converge1}
    \end{align*}
    We henceforth assume that $\widehat{f}_L\HOO \leq \widehat{f}_U\HOO$, which happens with probability tending to 1.

Provided that $f_L \leq f_U$, we find $\risk\LD(f_L,f_U \con \mu^*,e^*)$ is represented as follows:
\begin{align*}
    &
    \risk\LD(f_L,f_U \con \mu^*,e^*)
    \\
    &
    =
    \EXP \LD \big\{ \loss^{(1)} (\bO,f_L,f_U \con \mu^*,e^*) \big\}
    \\
    &
    =
    \EXP \LD 
    \left[ 
    \begin{array}{l}         
    \{ \mu^*(A,\bX)-R \} \ind \{ A \in [ f_L(\bX), f_U(\bX) ] \} /e^*(A \cond \bX)
    \\
    + \int \{ \alpha - \mu^*(a,\bX) \} \ind \{ a \in [ f_L(\bX), f_U(\bX) ] \} \, da
    \end{array}
    \right] 
    \\
    &
    =
    \EXP \LD 
    \left[ 
    \EXP \LD 
    \left[ 
    \begin{array}{l}         
    \{ \mu^*(A,\bX)-R \} \ind \{ A \in [ f_L(\bX), f_U(\bX) ] \} /e^*(A \cond \bX)
    \\
    + \int \{ \alpha - \mu^*(a,\bX) \} \ind \{ a \in [ f_L(\bX), f_U(\bX) ] \} \, da
    \end{array}
    \Bigg| A,\bX
    \right] 
    \right] 
    \\
    &
    =
    \EXP \LD 
    \left[ 
    \int \{ \alpha - \mu^*(a,\bX) \} \ind \{ a \in [ f_L(\bX), f_U(\bX) ] \} \, da
    \right] \ .
    \numeq \label{eq-risk1}
\end{align*}
Based on equation \eqref{eq-risk1}, we find the excess risk is simplified as follows, given that $\widehat{f}_L\HOO \leq \widehat{f}_U\HOO$:
\begin{align*}
    &
    \risk \LD( \widehat{f}_L\HOO, \widehat{f}_U\HOO \con \mu^*,e^*)
    -
    \risk \LD(  f_L^*,f_U^* \con \mu^*,e^*) 
    \\
    &
    =
    \EXP \LD 
    \left[ 
    \int \{ \alpha - \mu^*(a,\bX) \} 
    \bigg[ 
    \begin{array}{l}    
    \ind \{ a \in [ \widehat{f}_L\HOO(\bX), \widehat{f}_U\HOO(\bX) ] \} 
    \\
    -
    \ind \{ a \in [ f_L^*(\bX), f_U^*(\bX) ] \}
    \end{array}
    \bigg]
    \, da 
    \right] 
    \\
        & 
    =
    \EXP\LD \bigg[ 
    \int_{\widehat{f}_L\HOO(\bX)}^{f_L^*(\bX)} \{ \alpha - \mu^*(a,\bX) \} \, da
    +
    \int_{f_U^*(\bX)}^{\widehat{f}_U\HOO(\bX)} \{ \alpha - \mu^*(a,\bX) \} \, da
    \bigg] 
    \ . 
    \numeq \label{eq-risk2}
\end{align*}

From the proof of Lemma \ref{lemma-minimizer} in Section \ref{sec-Lemma1}, we find 
\begin{align*}
& f_L^{*}(\bX) = \inf \big\{ a \cond \mu^*(a,\bX) \geq \alpha \big\}
\ , \quad f_U^{*}(\bX) = \sup \big\{ a \cond \mu^*(a,\bX) \geq \alpha \big\} \ .
\end{align*}
Under Assumption \ref{(A5)}, we have that: 
\begin{align*}
 &
 \mu^*(f_L^*(\bX),\bX)=\mu^*(f_U^*(\bX),\bX)=\alpha
 \ ,
 \\
 &
 \mu^*(a,\bX) \geq \alpha \ , \quad \forall a \in [f_L^*(\bX), f_U^*(\bX)] 
 \ ,
 \\
 &
 \mu^*(a,\bX) < \alpha \ , \quad \forall a \notin [f_L^*(\bX), f_U^*(\bX)]  \ .
  \numeq \label{eq-result A5}
\end{align*}
From \eqref{eq-result A5}, we find that the following conditions for any $\widehat{f}_L\HOO(\bX)$ and $\widehat{f}_U\HOO(\bX)$:
\begin{align*}
    &
    \int_{\widehat{f}_L\HOO(\bX)}^{f_L^*(\bX)} \{ \alpha - \mu^*(a,\bX) \} \, da
    \geq 0 \ ,
    &&
    \int_{f_U^*(\bX)}^{\widehat{f}_U\HOO(\bX)} \{ \alpha - \mu^*(a,\bX) \} \, da
    \geq 0 \ .
    \numeq \label{eq-positive}
\end{align*}

In addition, Regularity Condition \HL{(RC6)} implies that 
\begin{align*}
    &
    \alpha - \mu^*(a,\bX)
    =
    \mu^*(f_L^*(\bX),\bX)
     - \mu^*(a,\bX)
    \geq 
    \underline{L} \big\{ f_L^*(\bX) - a \big\}
    \geq
    0
    \ ,
    &&
    \forall a \in [f_L^*(\bX)-\mathfrak{b},f_L^*(\bX)]
    \ ,
    \\
    &
    \alpha - \mu^*(a,\bX)
    =
    \mu^*(f_L^*(\bX),\bX)
     - \mu^*(a,\bX)
    \leq
    \underline{L} \big\{ f_L^*(\bX) - a \big\}
    \leq 0 
    \ ,
    &&
    \forall a \in [f_L^*(\bX),f_L^*(\bX)+\mathfrak{b}]
    \ ,
    \\
    &
    \alpha - \mu^*(a,\bX)
    =
    \mu^*(f_U^*(\bX),\bX)
     - \mu^*(a,\bX)
    \geq 
    \underline{L} \big\{ a-f_U^*(\bX) \big\}
    \geq
    0
    \ ,
    &&
    \forall a \in [f_U^*(\bX),f_U^*(\bX)+\mathfrak{b}]
    \ ,
    \\
    &
    \alpha - \mu^*(a,\bX)
    =
    \mu^*(f_U^*(\bX),\bX)
     - \mu^*(a,\bX)
    \leq
    \underline{L} \big\{ a-f_U^*(\bX) \big\}
    \leq 0 
    \ ,
    &&
    \forall a \in [f_U^*(\bX)-\mathfrak{b},f_U^*(\bX)] \ . 
\end{align*}
This implies that
\begin{align*}
 &   \int_{f_L^*(\bX)-\mathfrak{b}}^{f_L^*(\bX)}
    \big\{ \alpha - \mu^*(a,\bX) \big\} \, da
    \geq \underline{L} \mathfrak{b}^2 / 2 \ , 
&&
\int_{f_L^*(\bX)}^{f_L^*(\bX)+\mathfrak{b}}
    \big\{ \mu^*(a,\bX) - \alpha  \big\} \, da
    \geq \underline{L} \mathfrak{b}^2 / 2 \ , 
    \numeq \label{eq-result A6}
    \\
&   \int_{f_U^*(\bX)-\mathfrak{b}}^{f_U^*(\bX)}
    \big\{ \mu^*(a,\bX) - \alpha  \big\} \, da
    \geq \underline{L} \mathfrak{b}^2 / 2 \ , 
&&
\int_{f_U^*(\bX)}^{f_U^*(\bX)+\mathfrak{b}}
    \big\{ \alpha - \mu^*(a,\bX)   \big\} \, da
    \geq \underline{L} \mathfrak{b}^2 / 2 \ . 
\end{align*}

Let $E_L := \big\{ \bX \cond \widehat{f}_L^{(s)}(\bX) \notin [f_L^*(\bX)-\mathfrak{b},f_L^*(\bX)+\mathfrak{b}] \big\}$ or $E_U := \big\{ \bX \cond \widehat{f}_U^{(s)}(\bX) \notin [f_U^*(\bX)-\mathfrak{b},f_U^*(\bX)+\mathfrak{b}]  \big\} $ where a constant $\mathfrak{b}$ satisfies \eqref{eq-result A6}. We consider the following subevents of $E_L$:
\begin{itemize}
    \item (Case 1) \quad $\widehat{f}_L\HOO(\bX) < f_L^*(\bX) - \mathfrak{b}$: 

    We find 
    \begin{align*}
    &
    \int_{\widehat{f}_L\HOO(\bX)}^{f_L^*(\bX)} \{ \alpha - \mu^*(a,\bX) \} \, da
    +
    \int_{f_U^*(\bX)}^{\widehat{f}_U\HOO(\bX)} \{ \alpha - \mu^*(a,\bX) \} \, da
    \\
    &
    \geq 
    \int_{\widehat{f}_L\HOO(\bX)}^{f_L^*(\bX)} \{ \alpha - \mu^*(a,\bX) \} \, da
    \\
    &
    \geq 
    \int_{f_L^*(\bX)-\mathfrak{b}}^{f_L^*(\bX)} \{ \alpha - \mu^*(a,\bX) \} \, da
    \\
    &
    \geq 
    \underline{L} \mathfrak{b}^2/2 \ .
    \end{align*}
    The first and second inequalities are from \eqref{eq-result A5} and \eqref{eq-positive}. The third inequality is from \eqref{eq-result A6}. 

\item (Case 2) \quad $f_L^*(\bX) + \mathfrak{b} < \widehat{f}_L\HOO(\bX) \leq f_U^*(\bX)$:

We find 
\begin{align*}
    &
    \int_{\widehat{f}_L\HOO(\bX)}^{f_L^*(\bX)} \{ \alpha - \mu^*(a,\bX) \} \, da
    +
    \int_{f_U^*(\bX)}^{\widehat{f}_U\HOO(\bX)} \{ \alpha - \mu^*(a,\bX) \} \, da
    \\
    &
    =
    \int_{f_L^*(\bX)}^{\widehat{f}_L\HOO(\bX)} \{ \mu^*(a,\bX) - \alpha \} \, da
    +
    \int_{f_U^*(\bX)}^{\widehat{f}_U\HOO(\bX)} \{ \alpha - \mu^*(a,\bX) \} \, da
    \\
    &
    \geq
    \int_{f_L^*(\bX)}^{\widehat{f}_L\HOO(\bX)} \{ \mu^*(a,\bX) - \alpha \} \, da 
    \\
    &
    \geq
    \int_{f_L^*(\bX)}^{f_L^*(\bX) + \mathfrak{b}} \{ \mu^*(a,\bX) - \alpha \} \, da 
    \\
    &
    \geq 
    \underline{L} \mathfrak{b}^2/2 \ .
\end{align*}
The first four lines are from \eqref{eq-result A5} and \eqref{eq-positive}. The last line holds from \eqref{eq-result A6}. 

\item (Case 3) \quad $f_U^*(\bX)< \widehat{f}_L\HOO(\bX)$:

We find
\begin{align*} 
    &
    \int_{\widehat{f}_L\HOO(\bX)}^{f_L^*(\bX)} \{ \alpha - \mu^*(a,\bX) \} \, da
    +
    \int_{f_U^*(\bX)}^{\widehat{f}_U\HOO(\bX)} \{ \alpha - \mu^*(a,\bX) \} \, da
    \\
    &
    =
    \int_{f_L^*(\bX)}^{\widehat{f}_L\HOO(\bX)} \{  \mu^*(a,\bX) - \alpha \} \, da
    +
    \int_{f_U^*(\bX)}^{\widehat{f}_U\HOO(\bX)} \{ \alpha - \mu^*(a,\bX) \} \, da
    \\
    &
    =
    \int_{f_L^*(\bX)}^{f_U^*(\bX)} \{ \mu^*(a,\bX) - \alpha \} \, da
    +
    \int_{f_U^*(\bX)}^{\widehat{f}_L\HOO(\bX)}  \{ \mu^*(a,\bX) - \alpha \} \, da 
    +
    \int_{f_U^*(\bX)}^{\widehat{f}_U\HOO(\bX)} \{ \alpha - \mu^*(a,\bX) \} \, da
    \\
    &
    =
    \int_{f_L^*(\bX)}^{f_U^*(\bX)} \{ \mu^*(a,\bX) - \alpha \} \, da
    -
    \int_{f_U^*(\bX)}^{\widehat{f}_L\HOO(\bX)}  \{ \alpha - \mu^*(a,\bX) \} \, da 
    +
    \int_{f_U^*(\bX)}^{\widehat{f}_U\HOO(\bX)} \{ \alpha - \mu^*(a,\bX) \} \, da
    \\
    &
    =
    \int_{f_L^*(\bX)}^{f_U^*(\bX)} \{ \mu^*(a,\bX) - \alpha \} \, da 
    +
    \int_{\widehat{f}_L\HOO(\bX)}^{\widehat{f}_U\HOO(\bX)} \{ \alpha - \mu^*(a,\bX) \} \, da  
    \\
    &
    \geq 
    \int_{f_L^*(\bX)}^{f_U^*(\bX)} \{ \mu^*(a,\bX) - \alpha \} \, da 
    \\
    &
    \geq
    \int_{f_L^*(\bX)}^{f_L^*(\bX) + \mathfrak{b}} \{ \mu^*(a,\bX) - \alpha \} \, da 
    \\
    &
    \geq 
    \underline{L} \mathfrak{b}^2/2 \ .
\end{align*}
The equalities are obtained from straightforward algebra. The first and second inequalities are from \eqref{eq-result A5}. The last line holds from \eqref{eq-result A6}.

\end{itemize}

Therefore, $\bX \in E_L$ implies that 
\begin{align*}
    &
    \int_{\widehat{f}_L\HOO(\bX)}^{f_L^*(\bX)} \{ \alpha - \mu^*(a,\bX) \} \, da
    +
    \int_{f_U^*(\bX)}^{\widehat{f}_U\HOO(\bX)} \{ \alpha - \mu^*(a,\bX) \} \, da
    \geq 
    \underline{L} \mathfrak{b}^2/2 \ .
    \numeq
    \label{eq-EL}
\end{align*}
Likewise, one can also show that $\bX \in E_U$ implies that
\begin{align*}
    &
    \int_{\widehat{f}_L\HOO(\bX)}^{f_L^*(\bX)} \{ \alpha - \mu^*(a,\bX) \} \, da
    +
    \int_{f_U^*(\bX)}^{\widehat{f}_U\HOO(\bX)} \{ \alpha - \mu^*(a,\bX) \} \, da
    \geq 
    \underline{L} \mathfrak{b}^2/2 \ .
    \numeq
    \label{eq-EU}
\end{align*}

Therefore, we find

    \begin{align*}
    &
    \risk \LD( \widehat{f}_L\HOO, \widehat{f}_U\HOO \con \mu^*,e^*)
    -
    \risk \LD(  f_L^*,f_U^* \con \mu^*,e^*) 
    \\
    &
    =
    \EXP \LD 
    \left[ 
    \int_{\widehat{f}_L\HOO(\bX)}^{f_L^*(\bX)} \{ \alpha - \mu^*(a,\bX) \} \, da
    +
    \int_{f_U^*(\bX)}^{\widehat{f}_U\HOO(\bX)} \{ \alpha - \mu^*(a,\bX) \} \, da
    \right]
    \\
    &
    =
    \Pr\LD(\bX \in E_L \cup E_U)
    \times
    \EXP \LD 
    \left[ 
    \int_{\widehat{f}_L\HOO(\bX)}^{f_L^*(\bX)} \{ \alpha - \mu^*(a,\bX) \} \, da
    +
    \int_{f_U^*(\bX)}^{\widehat{f}_U\HOO(\bX)} \{ \alpha - \mu^*(a,\bX) \} \, da
    \, \bigg| \,  \bX \in E_L \cup E_U
    \right]
    \\
    &
    \quad 
    +
    \Pr\LD(\bX \in E_L^c \cap E_U^c)
    \times
    \EXP \LD 
    \left[ 
    \int_{\widehat{f}_L\HOO(\bX)}^{f_L^*(\bX)} \{ \alpha - \mu^*(a,\bX) \} \, da
    +
    \int_{f_U^*(\bX)}^{\widehat{f}_U\HOO(\bX)} \{ \alpha - \mu^*(a,\bX) \} \, da
    \, \bigg| \, \bX \in E_L^c \cap E_U^c
    \right]
    \\
    &
    \stackrel{(a)}{\geq }
    0.5
    \Pr\LD(\bX \in E_L \cup E_U)
    \times
    \underline{L} \mathfrak{b}^2
    \\
    &
    \quad 
    +
    \Pr\LD(\bX \in E_L^c \cap E_U^c)
    \times
    \EXP \LD 
    \left[ 
    \int_{\widehat{f}_L\HOO(\bX)}^{f_L^*(\bX)} \{ \alpha - \mu^*(a,\bX) \} \, da
    +
    \int_{f_U^*(\bX)}^{\widehat{f}_U\HOO(\bX)} \{ \alpha - \mu^*(a,\bX) \} \, da
    \, \bigg| \, \bX \in E_L^c \cap E_U^c
    \right]
    \\
    &
    \stackrel{(b)}{\geq }
    0.5
    \Pr\LD(\bX \in E_L \cup E_U)
    \times
    \underline{L} \mathfrak{b}^2 \ .
\end{align*}
Inequality $\stackrel{(a)}{\geq}$ holds from \eqref{eq-EL} and \eqref{eq-EU}. Inequality $\stackrel{(b)}{\geq}$ holds from \eqref{eq-positive}.

Suppose that $\Pr\LD(\bX \in E_L \cup E_U)  \rightarrow \underline{c}$ for a positive constant $\underline{c}$ as $N \rightarrow \infty$. Then, we find
    \begin{align*}
    &
    \risk \LD( \widehat{f}_L\HOO, \widehat{f}_U\HOO \con \mu^*,e^*)
    -
    \risk \LD(  f_L^*,f_U^* \con \mu^*,e^*) 
    \rightarrow 
    0.5 \underline{c} \underline{L} \mathfrak{b}^2 
    >
    0 \ .
    \end{align*}
    Again, this contradicts the result of Theorem \ref{thm-ExcessRisk}. Therefore, we conclude that 
    \begin{align*}
        \Pr\LD(\bX \in E_L \cup E_U)  \rightarrow 0 \quad \text{ as } \quad 
        N \rightarrow \infty
        \ .
        \numeq \label{eq-converge2}
    \end{align*}

    We now consider the following event $E_{f}$:
    \begin{align*}
        E_{f}
        :=
        \left\{ 
        \begin{array}{l}             
        \{ \widehat{f}_L^{(s)}(\bX) \leq \widehat{f}_U^{(s)} (\bX) \} \\
        \cap \ \{ \widehat{f}_L\HOO(\bX) \in [f_L^*(\bX)-\mathfrak{b},f_L^*(\bX)+\mathfrak{b}] \} \\
        \cap \ \{ \widehat{f}_U\HOO(\bX) \in [f_U^*(\bX)-\mathfrak{b},f_U^*(\bX)+\mathfrak{b}] \}
        \end{array}
        \right\}
        \ ,
    \end{align*}
    which occurs with probability not less than $1-\delta_N$ where $\delta_N \rightarrow 0$ as $N\rightarrow \infty$ from \eqref{eq-converge1} and \eqref{eq-converge2}.    
    
     Given $E_{\text{Thm \ref{thm-ExcessRisk}}} \cap E_f$, which happens with probability not less than $1-3e^{-\tau} - \Delta_N - \delta_N$, we find the following results:

\begin{itemize}

    \item (Case 1 for $f_L$) \quad $ \widehat{f}_L\HOO \in [f_L^* - \mathfrak{b} , f_L^*]$

    We find 
    \begin{align*}
    \int_{\widehat{f}_L\HOO(\bX)}^{f_L^*(\bX)}
    \big\{ \alpha - \mu^*(a,\bX) \big\}
    \, da
    &
    =
    \int_{\widehat{f}_L\HOO(\bX)}^{f_L^*(\bX)}
    \big\{ \mu^*(f_L^*(\bX),\bX) - \mu^*(a,\bX) \big\} \, da
    \\
    &
    \geq 
    \underline{L}
    \int_{\widehat{f}_L\HOO(\bX)}^{f_L^*(\bX)}
    \big\{ f_L^*(\bX) - a \big\} \, da
    \\
    &
    =
    0.5 
    \underline{L}
    \big\{ f_L^*(\bX) - \widehat{f}_L\HOO(\bX) \big\}^2 \ .
    \end{align*}
    The first line holds from \eqref{eq-result A5}. The second line holds from \eqref{eq-result A6}. The last line holds from straightforward algebra.

    \item (Case 2 for $f_L$) \quad $ \widehat{f}_L\HOO \in  [f_L^*  , f_L^* + \mathfrak{b}]$

    We find 
    \begin{align*}
    \int_{\widehat{f}_L\HOO(\bX)}^{f_L^*(\bX)}
    \big\{ \alpha - \mu^*(a,\bX) \big\}
    \, da
    &
    =
    \int_{f_L^*(\bX)}^{\widehat{f}_L\HOO(\bX)}
    \big\{ \mu^*(a,\bX) - \alpha  \big\}
    \, da
    \\
    &
    =
    \int_{f_L^*(\bX)}^{\widehat{f}_L\HOO(\bX)}
    \big\{ \mu^*(a,\bX) - \mu^*(f_L^*(\bX),\bX)  \big\}  \, da
    \\
    &
    \geq 
    \underline{L}
    \int_{f_L^*(\bX)}^{\widehat{f}_L\HOO(\bX)}
    \big\{ a - f_L^*(\bX)  \big\} \, da
    \\
    &
    =
    0.5 
    \underline{L}
    \big\{ f_L^*(\bX) - \widehat{f}_L\HOO(\bX) \big\}^2 \ .
    \end{align*}
    The first line holds from straightforward algebra.  
    The second line holds from \eqref{eq-result A5}. The second line holds from \eqref{eq-result A6}. The last line again holds from straightforward algebra.  

    \item (Case 1 for $f_U$) \quad $\widehat{f}_U\HOO \in [f_U^*,f_U^* + \mathfrak{b}]$

By following the analogous steps as in (Case 1 for $f_L$), one can establish
    \begin{align*}
        \int_{f_U^*(\bX)}^{\widehat{f}_U\HOO(\bX)}
        \big\{ \alpha - \mu^*(a,\bX) \big\} \, da 
        \geq 0.5 \underline{L} \big\{ f_U^*(\bX) - \widehat{f}_U\HOO(\bX) \big\}^2 
        \ .
    \end{align*}

    \item (Case 1 for $f_U$) \quad $\widehat{f}_U\HOO \in [f_U^*-\mathfrak{b},f_U^*]$

By following the analogous steps as in (Case 2 for $f_L$), one can establish
    \begin{align*}
        \int_{f_U^*(\bX)}^{\widehat{f}_U\HOO(\bX)}
        \big\{ \alpha - \mu^*(a,\bX) \big\} \, da 
        \geq 0.5 \underline{L} \big\{ f_U^*(\bX) - \widehat{f}_U\HOO(\bX) \big\}^2 
        \ .
    \end{align*}
\end{itemize}

Therefore,  given $E_{\text{Thm \ref{thm-ExcessRisk}}} \cap E_f$, we have
\begin{align*} 
    &
    \int_{\widehat{f}_L\HOO(\bX)}^{f_L^*(\bX)} \{ \alpha - \mu^*(a,\bX) \} \, da
    +
    \int_{f_U^*(\bX)}^{\widehat{f}_U\HOO(\bX)} \{ \alpha - \mu^*(a,\bX) \} \, da
    \\
    &
    \geq 
    0.5 \underline{L} \big[ \big\{ f_L^*(\bX) - \widehat{f}_L\HOO(\bX) \big\}^2
    +
    \big\{ f_U^*(\bX) - \widehat{f}_U\HOO(\bX) \big\}^2
    \big] \ .
    \numeq \label{eq-fconverge}
\end{align*}
Accordingly, we find the following representation of the excess risk in \eqref{eq-risk2}: 
\begin{align*}
    &
    \risk \LD( \widehat{f}_L\HOO, \widehat{f}_U\HOO \con \mu^*,e^*)
    -
    \risk \LD(  f_L^*,f_U^* \con \mu^*,e^*) 
    \\
    &
    =
    \EXP\LD \bigg[ 
    \int_{\widehat{f}_L\HOO(\bX)}^{f_L^*(\bX)} \{ \alpha - \mu^*(a,\bX) \} \, da
    +
    \int_{f_U^*(\bX)}^{\widehat{f}_U\HOO(\bX)} \{ \alpha - \mu^*(a,\bX) \} \, da
    \bigg] 
    \\
    &
    \geq 
    0.5 \underline{L}
    \times 
    \EXP \LD
    \Big[ \big\{ f_L^*(\bX) - \widehat{f}_L\HOO(\bX) \big\}^2
    +
    \big\{ f_U^*(\bX) - \widehat{f}_U\HOO(\bX) \big\}^2 \Big] \ .    
\end{align*}
The last line holds from \eqref{eq-fconverge}. Consequently, this concludes the following results hold with probability not less than $1-3e^{-\tau} - \Delta_N - \delta_N$:
\begin{align*}
    &
    \EXP \LD
    \Big[ \big\{ f_L^*(\bX) - \widehat{f}_L\HOO(\bX) \big\}^2
    +
    \big\{ f_U^*(\bX) - \widehat{f}_U\HOO(\bX) \big\}^2 \Big]
    \\
    &
    \leq 
    c_1' \lambda\gamma^{-d} + c_2' \gamma^\beta 
+ c_3' \big\{ \gamma^{(1-p)(1+\epsilon)d} \lambda^p N \big\}^{-\frac{1}{2-p}} 
+ c_4' N^{-1/2}\tau^{1/2} + c_5' N^{-1} \tau
+ c_6' \epsilon N^b + c_7' N^{-r_e-r_\mu}  
\end{align*}
where constants $c_1',\ldots,c_7'$ do not depend on $N$. Taking $\gamma \asymp N^{-1/(2\beta + d)}$ and $\lambda \asymp N^{-(\beta+d)/(2\beta + d)}$, we get the following asymptotic rate with probability not less than $1-3e^{-\tau} - \Delta_N - \delta_N$:
\begin{align}						\label{eq-RateofF}
\EXP \LD
    \Big[ \big\{ f_L^*(\bX) - \widehat{f}_L\HOO(\bX) \big\}^2
    +
    \big\{ f_U^*(\bX) - \widehat{f}_U\HOO(\bX) \big\}^2 \Big]
=
O_P \big( 	N^{- \beta/(2\beta + d)} + N^{-r_e-r\mu}	\big) \ .
\end{align}
This completes the proof.

\newpage

\bibliographystyle{apa}
\bibliography{TDI}

\end{document}

%% file: Commands.tex
\definecolor{mycolor}{RGB}{0,200,200} 

\newcommand{\indep}{\rotatebox[origin=c]{90}{$\models$} }

\newcommand{\bO}{\bm{O}}

\newcommand{\bX}{\bm{X}}

\newcommand{\bx}{\bm{x}}
\newcommand{\bo}{\bm{o}}
\newcommand{\bz}{\bm{z}}
\newcommand{\bxi}{\bm{\xi}}

\newcommand{\cond}{\, \vert \,}
\newcommand{\con}{;}
\newcommand{\pot}[2]{#1^{(#2)}}
\newcommand{\EXP}{\text{E}}
\renewcommand{\Pr}{\text{Pr}}

\newcommand{\ind}{\mathbbm{1}}
\newcommand{\T}{^{\intercal}}
\DeclareMathOperator*{\argmax}{arg\,max}
\DeclareMathOperator*{\argmin}{arg\,min}
\DeclareMathOperator*{\median}{median}

\newcommand{\R}{\mathbbm{R}}
\newcommand{\F}{\mathcal{F}}
\newcommand{\Fmnt}{\F_{\text{mnt}}}
\newcommand{\Fnmnt}{\F_{\text{nmnt}}}
\newcommand{\Fprod}{\F^{\otimes2}}

\newcommand{\thr}{\mathcal{T}}
\newcommand{\risk}{\mathcal{R}}
\newcommand{\loss}{\mathcal{L}}
\newcommand{\hf}{\widehat{f}}

\newcommand{\sur}{\text{sur}}
\newcommand{\HH}{\mathcal{H}}
\newcommand{\LOO}{^{(-s)}}
\newcommand{\HOO}{^{(s)}}
\newcommand{\LD}{^{\mathcal{D}}}
\newcommand{\bflu}{\bm{f}}

\hypersetup{
colorlinks=true,
linkcolor=blue,
filecolor=blue,
urlcolor=blue,
citecolor=blue
}

\newcommand{\HL}[1]{\hyperlink{#1}{#1}}
\newcommand{\HT}[1]{\hypertarget{#1}{#1}}

\newcommand\numeq{\addtocounter{equation}{1}\tag{\theequation}}